\documentclass[12pt]{article}
\usepackage{amsmath,amssymb}
\usepackage{bm}
\usepackage{graphicx}
\usepackage{setspace}
\usepackage{cite}
\usepackage{nameref,hyperref}

\begin{document}

\title{Coupled Hebbian learning and evolutionary dynamics in a formal model for structural synaptic plasticity}

\author{ \href{https://hpvladar.wordpress.com}{H.P. Vladar} and \href{https://www.parmenides-foundation.org/people/eoers-szathmary/}{E. Szathm\'ary} \\ \href{https://www.parmenides-foundation.org}{Parmenides Foundation}\\ Pullach/Munich, Germany}
\date{\today}
\maketitle

\begin{abstract}
Theoretical models of neuronal function consider different mechanisms through which networks learn, classify and discern inputs. A central focus of these models is to understand how associations are established amongst neurons, in order to predict spiking patterns that are compatible with empirical observations. Although these models have led to major insights and advances, they still do not account for the astonishing velocity with which the brain solves certain problems and what lies behind its creativity, amongst others features. We examine two important components that may crucially aid comprehensive understanding of said neurodynamical processes. First, we argue that once presented with a problem, different putative solutions are generated in parallel by different groups or local neuronal complexes, with the subsequent stabilization and spread of the best solutions. Using mathematical models we show that this mechanism accelerates finding the right solutions. This formalism is analogous to standard replicator-mutator models of evolution where mutation is analogous to the probability of neuron state switching (on/off). Although in evolution mutation rates are constant, we show that neuronal switching probability is determined by neuronal activity and their associative weights, described by the network of synaptic connections. The second factor that we incorporate is structural synaptic plasticity, i.e. the making of new and disbanding of old synapses, which we apply as a dynamical reorganization of synaptic connections. We show that Hebbian learning alone does not suffice to reach optimal solutions. However, combining it with parallel evaluation and structural plasticity opens up possibilities for efficient problem solving. In the resulting networks, topologies converge to subsets of fully connected components. Imposing costs on synapses reduces the connectivity, although the number of connected components remains robust. The average lifetime of synapses is longer for connections that are established early, and diminishes with synaptic cost.
\end{abstract}

\section{Introduction}

Many mechanisms of cognition, memory and other aspects of brain function remain unclear. It is acknowledged that associations build up by updating synapses between neurons that spike (nearly) synchronously to a given stimulus. In this way some neuronal circuits can predispose or anticipate a response to similar stimuli by retrieving information stored in synaptic weights. Synaptic weights may in turn be systematically altered by successful anticipation or recognition activity. At the same time, given the multidimensional space of alternative neuronal circuits and spiking sequences, undirected random variation in circuitry and spiking are extremely unlikely to produce better solutions for each new problem. 

The connectivity overall of the human brain is sparse where, roughly, $10^{11}$ neurons are estimated to connect through some $10^{15}$ synapses. Learning and cognition have been understood in terms of changes in associative weights on networks of fixed topology. However, the discovery that rewiring this network is not uncommon even in adult brains challenges the former views regarding the mechanisms of learning. This rewiring, known as structural synaptic plasticity (SSP), has been well documented experimentally \cite{Butz:2009qq,Holtmaat:2009le}. However, neither the full consequences nor the central role of SSP have been fully clarified. Yet, it is not only reasonable, but also supporting evidence exists, that SSP can encode information \cite{Chklovskii:2004eb}. Thus, associative weights and SSP are two mechanisms that have an effect on learning. These need not be mutually exclusive; rather, as we show in this article, they both seem to be necessary for different stages of learning, such as short and long term, respectively.

Our knowledge about what determines the establishment of new synapses is still limited, especially taking into account the sparseness and dimensions of the brain. Neither synaptic weights nor SSP explain on their own variability in circuitry associated with a particular stimulus. As such, they only show variability in time. If trial solutions to a problem (such as learning or recognizing a pattern) rely on serial evaluations, SSP is a poor candidate mechanism, even for long-term learning. Under serial evaluations the time for establishing new synapses would be prohibitively large to account for randomly testing connections amongst pairs of neurons.

Changeux \cite{Changeux:1973if,Changeux:1973wo} and Edelman \cite{Edelman:1987uz} proposed a selectionist \cite{Fernando:2012ju} framework for brain function. They noted that selection acts, through preferentially reinforcing and stabilizing some synaptic patterns over others, and through the elimination of dysfunctional neurons and neuronal connections. Although these ideas are correct, they are incomplete because they only consider the fate of initial topological variability in circuitry, thought to occur only during development. In their framework, selection acts on this standing variation, stabilizing functional circuits that remain unchanged throughout life, with later learning and problem solving resulting only from changing synaptic weights. In this sense, the role of selection is limited to establishing functional neuronal network at early stages. The ideas that we investigate in this article go beyond this view: we consider that selection of novel variation plays an active role in learning through life. 

Kilgard proposed a verbal model that accounts for circuitry variation during learning periods \cite{Kilgard:2012ci}. In his `expansion-renormalization model' he envisions that SSP accounts for such variation. The mechanism is as follows. When a cortical subnetwork is challenged by a novel task, new synapses are being generated in response, out of which only the functionally important ones are kept, while the obsolete ones are eliminated. This is like an iterated Changeux-type overproduction-selective stabilisation mechanism, and is being explicitly regarded as a Darwinian mechanism by the author. However, he fails to discuss particulars such as: what are the true units of variation, and how this mechanism quantitatively acts. Our ideas are conceptually similar, but we pin them down to specific `learning' units and develop quantitative models to understand how this variability is generated and how it affects learning. 

We note that there are at least two other sources of neuronal variability. The first one is the variance in spiking patterns and is due only to the stochastic behaviour of neurons (cf.\cite{Seung:2003we,Rolls:2012ix}). The second one, which is more fundamental, is due to SSP, which acts by rewiring the set of neurons in a complex. Selection is then able to act on the variation that is generated by the three mechanisms. We point out that the crucial one is SSP, but as we will explain throughout this article, the three mechanisms play different roles in learning.

\begin{figure}
\includegraphics[width=\columnwidth]{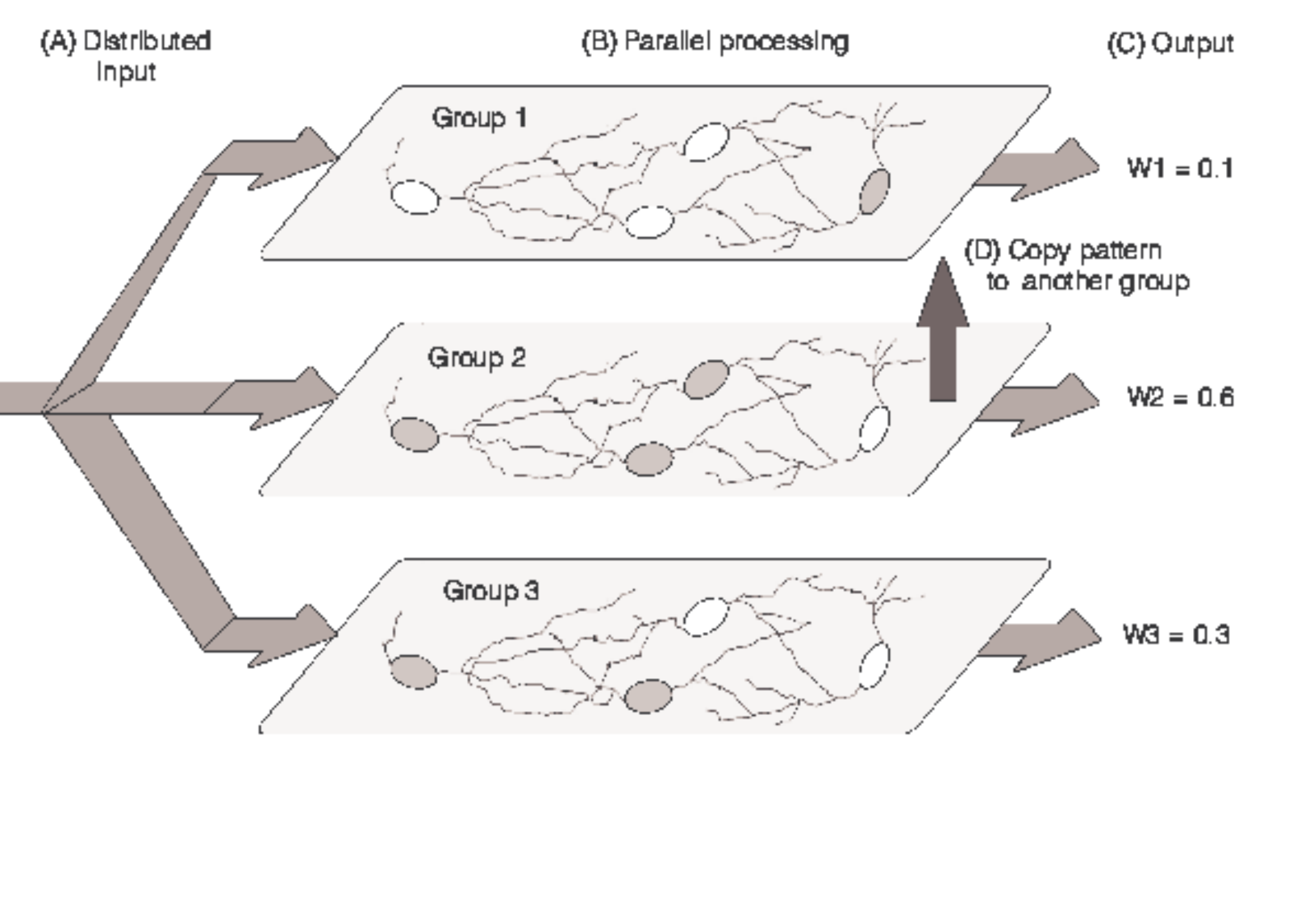}
\caption{\bf Replicative neurodynamics.}
(A) The input is fed into several local neuronal groups. (B) Each of these groups evaluate the input independently, thus trying in parallel distinct spiking patterns (represented by neurons in white and grey states), and (C) producing distinct outputs with corresponding reward/fitness values $W$. (D) Groups that result in higher fitness transmit their synaptic weights to other groups that performed poorly (connections amongst groups are assumed to exist but are not displayed on the figure, and not explicitly modelled). This parallel evaluation is repeated until an optimal solution spreads across all groups.
\label{Fig:Layers}
\end{figure}

 We assume that circuits that result in a sub-optimal solution relative to the rest of the circuits not only receive less reward, but also are more likely to be `overwritten' by transmitting the information in the form of synaptic weights and structure from other local complexes. During this transmission process, small variations are introduced to the new circuit through SSP. Iterating this mechanism results in the increase in the representation of the circuit that gives the best solution, gradually replacing other circuits until no better variants are further produced, and finally (and ideally) a solution is found.
Our central aim is to understand how different neuronal complexes might evaluate possible solutions in parallel and thus compete to converge to an optimal result during learning (Fig. \ref{Fig:Layers}). For this, we put together all these verbal ideas into a quantitative framework.
 
 We study the properties that need to be sought in order to understand more accurately how learning occurs. For this reason we build up from local mechanisms of neural learning. That is, we set our problem at a time scale that allows us follow whether neurons are found to be on or off. Each neuron is assumed to fire stochastically, but with a probability given by the input activity of other neurons in the complex. We will assume reinforcement learning, and as other works, employ simple measures such as distance between the output and the target. We emphasize that this is analogous to the gradient of a fitness landscape in evolution \cite{Crow:1970wa}. This analogy will allow us to tackle the problem with full force, partly by employing the mathematical models developed in evolutionary biology.
 
 Despite the high level of abstraction of our approach, we acknowledge that an ultimate verification of our hypothesis needs to come from experimental neuroscience. However at the moment we intentionally avoid discussing molecular or physiological aspects, which although essential to understand the problem experimentally, at this point would simply obscure understanding what we propose are the strategic means through which the brain works at the level we aim to describe.
 
 \subsection{Analogy with Darwinian Evolution}
  As stated above, so-called Neural Darwinism does not account for the generation of post-developmental variation repeatedly in circuitry, on which selective mechanisms could act. However, this is still not enough for a complete implementation of adaptive evolution in the functioning of the brain. What is missing is an interpretation of heredity in terms of neurophysiology, so that the selected variants can be expanded and, from them further variants be generated. In this way the interaction between selection, variability and heredity can find the right spiking patterns to solve a problem \cite{Fernando:2012ju}.
 
The mechanisms for generating variability of neural spiking patterns are relatively simple to rationalize, and there are many works in the literature that take this aspect as modelling objective \cite{Ullman:2012wq}. But it is less obvious, of deeper implications and of far-reaching consequences, to realise that a mechanism of `neuronal heredity' between local complexes might exist.

	As explained above, for neuronal heredity to occur it is necessary that several local networks can act in parallel; stochastic variation in spiking is not enough. Heredity occurs when circuits that have reached satisfactory solutions transmit their contents to some other circuits that did not perform as well (Fig. \ref{Fig:Layers}). Although there is no replication of the population of neurons per se (as in a biological population), these repeated rounds of evaluation and replacement implement a mechanism of heredity  \cite{Fernando:2010is,Fernando:2012ju}  that is analogous to genetic inheritance. Admittedly, models of `neuronal replication' thus far have relied too much on accurate topographic mapping between local networks. In general this assumption is unjustified. We hasten to note that a neurobiologically realistic solution to this problem is already in our hands and will be subject of forthcoming publications. We regard our present paper a `formal' one partly because we treat the component process of replication as a black box of which the content will be revealed later. Thus we perform an abstract analysis of evolutionary neurodynamics by linking basic theory in neuroscience and evolutionary biology under the assumption that neuronal heredity is solved. Note that the discussions on two mechanisms of accumulating knowledge (by evolution and by learning) have been largely isolated from each other. These two sides of the discussion are not exclusive. We of course recognize that spiking neurons, Hebbian learning and SSP exist, and are central components of cognition, but we argue that on their own, they do not suffice for explaining how complex tasks are solved.

In this paper we merge these concepts formally and investigate how learning and Darwinian selection can work together to drive the system to optimal solutions. Our approach is akin to models coupling learning and selection by directly rewarding over stimuli (cf. the direct actor; \cite{Dayan:2005fr}, pp. 344-346), which, as in our evolutionary approach, results in maximising the average reward (i.e. fitness). However, our work has a wider scope since it also merges other ideas such as neuronal copying and structural synaptic plasticity (though see Discussion). We take a mathematical and computational approach to understand the principal factors that are relevant and isolate aspects that are in principle falsifiable. We show that for eyes educated in evolutionary biology, the equations that describe the whole process are astonishingly similar to the mutation-selection equations, albeit with a twist. The quantity that is analogous to mutation rates is not constant since it depends on the state of the whole system. The relevance of this difference is that such `mutation rates' derive from the associative (Hebbian) learning mechanism, where the synaptic weights represent the strength of the associations, effectively represented by a graph with weighted edges. (This would be equivalent to mutation rates that depend on the extant population diversity.) When coupled to selection, the learning system is more efficient than simple hill-climbing. This is our first central result. Second, we will consider asymmetric landscapes and show that the learning weights correlate with the fitness gradient. That is, the neuronal complexes learn the local properties of the fitness landscape, resulting in the generation of variability directed towards the direction of fitness increase, as if mutations in a genetic pool were drawn such that they would increase reproductive success. Third, we study how this mechanism is efficient in reaching optimal solutions in rugged landscapes. We show that the system often reaches sub-optimal solutions when we assume random networks of synaptic connections. We identify these as impasse states. That is, the network reaches a sub-optimal state where each possible modification of the synaptic and spiking activity only decreases the quality of the solution. However, when synaptic plasticity is considered, these states are easily overcome. For this dynamics, we characterize the distribution of lifetimes of synaptic connections. This is a central aspect because it is a quantity that can be measured experimentally. Although at this stage we are not concerned with parameter estimation or inference, we do point out that our results are in principle falsifiable.

\section{Models}
We note that on short time scales (milliseconds) spikes take place and the selective dynamics can act by rewarding different sub-networks of the neuronal circuit. Yet, variation in spiking can be produced due to changes in synaptic weights. On a larger time scale, SSP generates novel circuitry. For simplicity we separate these two time scales.
We first describe the joint action of selection in several groups and learning. For now we assume that all groups have the same topology of connections, but each one has a different spiking pattern; we describe SSP below.

\subsection{Learning in parallel groups}
In the spiking models, learning occurs by updating the weights that determine the probability that a neuron fires. This update follows Hebb's rule, verbally stated as `neurons that fire together, wire together'. Hebb's rule has been modelled with fixed connection topology where the weights are allowed to change according to the covariance amongst neurons, as for example \cite{MacKay:2003wc}:
\begin{equation}
\label{Eq:HebbsRule}
\Delta \phi_{ij}=\lambda X_i Y_j
\end{equation}
where $\lambda$ is the learning rate, $X=1$ if the neuron fires (on) and $X= -1$ if it does not (off), \(Y_k= \sum_i \phi_{ki} X_i \) is the output of neuron $k$, and $\phi_{ij}$is the weight between neuron $i$ and $j$. (Note: in the neuroscience literature, weights are denoted by $w$, however this notation is potentially confusing in the context of evolutionary analyses because a similar symbol is employed for fitness; see Table \ref{Table:Analogy}).

\begin{table}[t]
\caption{
{\bf Analogy between the concepts in evolutionary genetics and neurodynamics.}}
\begin{tabular}{lll}
\hline
Evolutionary genetics & Neurodynamics & Symbol \\ \hline 
Loci/genes & Neurons & $i,j,k$  \\
Nr. of loci & Nr. of Neurons in a group & $n$  \\
Alleles & Neuron state (on/off)& $X$  \\
Allele frequency  & Firing probability & $\rho$  \\
Population & Groups of neurons & $N (=\infty) $  \\
Adaptive landscape & Rewarding mechanism, score & $W$  \\
Mutation rate & Switching probability & $A,M$  \\
$-$ & Hebbian weights & $\phi$  \\
$-$ & Learning rate & $\lambda$  \\
$-$ & Synaptic cost & $k$\\ \hline
\end{tabular}
\label{Table:Analogy}
\end{table}

Hebb's rule is problematic because it allows weights to increase unboundedly. Thus, for computational convenience, we employ Oja's rule, which is a version of Hebb's rule with normalised weights:
\begin{equation}
\Delta\phi_{ij} = \lambda Y_j (X_i - \phi_{ij}Y_j)~.
\end{equation}
Oja's rule ensures that the weights are normalised, in this case with Eucledian norm, so that \(\sum_i \phi_{ik}^2 =1\).

Whether any one neuron spikes or not is assumed to be a random event. The probability with which neurons change state (switch on or off) is given by an update rule $A$ that depends on the state of the input neurons and their weights. Thus, the probability that a neuron $i$ is on, \(P[X_i=1] \equiv \rho_i\) is given by the master equation
\begin{equation}
\label{Eq:ActivityChance}
\frac{d\rho_i}{dt} = A_i^{\text{on}}(1-\rho_i)-A_i^{\text{off}}\rho_i
\end{equation} 
where \(A_i^{\text{on}} = \Pr[X_i: 0 \rightarrow 1] \) and \(A_i^{\text{off}} = \Pr[X_i: 1 \rightarrow 0] \) are the probabilities that inactive neurons spike and spiking neurons shut dow, respectively. We assume that the update rule takes into consideration the state of both the focal neuron and the rest of the neurons in the group at the previous evaluation round. We also assume a time scale that is larger than the refractory period, so that that spiking is only affected by the previous state of the network.

Some models, such as Boltzmann machines (see \cite{Dayan:2005fr}), assume that the neuron itself has no memory, and its spiking probability is independent from previous neuronal states. In this case \(A_i^{\text{on}} = \Pr[X_i=1|\bf{X}]=1-\Pr[X=0|\bf{X}] \); the master equation gives
\begin{equation}
\frac{d\rho_i}{dt} = A_i^{\text{on}} -\rho_i
\end{equation} 
and the `on' probability is given by
\begin{equation}
A_i^{\text{on}} = \left[ 1+\exp(-Y_i) \right]^{-1} ~.
\end{equation}

For firing neurons the weights $\phi$ increase, so $A_i^{\text{on}}$ tends to grow to 1. Hence, co-spiking neurons become more likely to fire in each evaluation round. 

	However, we take a different approach by assuming that learning can be modulated more efficiently by allowing $A_i^{\text{on}}$ and  $A_i^{\text{off}}$ to have an effect on the network. Note that this description of learning is coarse-grained: it only tracks how often a neuron tends to be on as learning proceeds. This is a different view than that of machine learning, where neural networks are trained by a set of examples from which the weights are inferred. Then, from this inference the model can be used to predict or classify data that were not included in the training set. Our goal in this paper is different: we consider parallel networks that try to solve a specific problem. 

Solving a particular problem requires some comparison between the target, $T$ and the output or solution $\Omega$. For this purpose, we can employ the square deviation, \(\Delta^2=(\Omega-T)^2\) which we want to minimize. We assume that $T$ is a given parameter and $\Omega$ is the output evaluation of the network. Each network presents an alternative solution, thus having a different deviation $\Delta$ from the target. Hence, we minimize the mean value of the deviation, \( \bar{\Delta^2}=E\left[(\Omega-T)^2 \right] \). Under a proper scheme of neuronal network replication, this minimization amounts to Darwinian selection. This selective scheme is the following: First, each local network is weighted according to its fitness, given by \(W= \exp(-\beta \Delta^2]\). Second, groups that have larger fitness are kept. Third, networks with lower fitness are overwritten with the content (spiking and/or weight states) of the groups with large fitness. (There are several ways in which this copying can be implemented: this is the black box part as explained above.). Since in the present model we assume that there are infinitely many groups, replacement need not be done explicitly: we simply consider the proportions of groups (this in order to have a direct link to classical population genetics models that assume infinite population size). Since we assume that copying is random across different neuronal loci, then the proportion of a group with a specific configuration is simply the product of the probability of the state of each neuronal locus (this is analogous to the Hardy-Weinberg assumption of population genetics; \cite{Crow:1970wa} pp. 34-39). Mathematically, we track the proportions, $\rho$, of active neurons and the distribution of weights across groups. For the former:
\begin{equation}
\label{Eq:FreqChange}
\frac{d\rho_i}{dt} = \rho_i (1-\rho_i) \frac{\partial \log( \bar{W})}{\partial \rho_i}   +M_i (2 \rho_i-1) ~
\end{equation}
where $W_i$ is the fitness, and \(\bar{W}=\sum_i\rho_iW_i\) the mean fitness. The first term is well known to evolutionary biologists and computer scientists: it describes the process of hill-climbing in the direction of fitness increase \cite{Crow:1970wa}. Note that we can approximate \(\log\bar{W} = \bar{\Delta^2}+\text{var}(\Delta)\). The second term represents the variability that is generated throughout learning. For simplicity, we assumed that the switching probability is symmetric (\(M_i=A_i^{\text{off}} = 1-A_i^{\text{on}}\)), and which is given by the activity rule 
\begin{equation}
M_i = \frac{1}{1+e^{Y_i}}
\end{equation}
where \(Y_i= \sum_j\phi_{ij}X_j \) is the activity or current of the the input neurons, and \(\phi_{ij}\)are the weights determining the associations amongst neurons, and which evolve according to Oja's rule. As more spiking neurons are connected the activity of the focal neuron increases and its switching probability decreases asymptotically to zero. Whether a neuron stays on or off however depends non-trivially on the collective success of reaching the target $T$.

Appendix \ref{SI:HebbianWeights} shows that to first-order approximation we can track only the mean weight at every synapse and apply a general learning rule to all the average activities of the ensemble of groups. That is, we approximate that each network has, on average, input activity \(\bar{Y}_i= \sum_j\bar{\phi}_{ij}(2\rho_j-1) \). This still allows each network to have a different spiking pattern from those of other groups. We will see below that even under these simplifying assumptions, evolution  has a dramatic effect by accelerating convergence to maximum fitness (or minimum \(\bar{\Delta^2}\)). We will assume small initial values of the synaptic weights. Moreover, the variance of these becomes increasingly small as the neuronal complexes converge to a solution. Thus, we will make no further distinction between \(\bar{\phi}\) and $\phi$. 

Summarizing, for $n$ neurons and $k$ synapses amongst them we work with $n+k$ ordinary differential equations. Initial conditions are assumed to be random and the topology of the network is assumed to be fixed during each learning round.

\subsection{Information content of a synapse}
How to measure the information content of a synapse is not obvious. First, we point out that it is important to discern between local measures, such information in a particular synapse, or global information in the whole circuitry of a complex, or even across complexes \cite{Tononi:2004ki}. Different choices depend on the specific purpose, and can be made according to distinct epistemological views and/or experimental purposes. For the aim of this article we choose one of the simplest measures, mutual information, $H$, because it describes the interdependency amongst two specific neurons in the context of a specific complex. Mutual information is defined as
\begin{equation}
H_{ij}=\sum_{r,s \in \{0,1\}} \Pr[X_i=r | X_j=s] \rho_j \log \left[ \frac{ \Pr[X_i=r|X_j=s]}{\rho_i} \right] A .
\end{equation}

To calculate the conditional probabilities we first evaluate the input activity $E_{i|j}$ of neuron $i$ by fixing the value of neuron $j$ to $\rho_j=0$ or $1$. This gives a conditioned value of the switching probability, $M_{i|j}$. Then, the solution to Eq. \ref{Eq:FreqChange}, using the conditioned switching probability $M_{i|j}$, gives the desired conditioned probabilities. In this case we keep the weights constant because we are only assessing the information capacity of the specific synapse and not the information capacity of the whole network. The exact expression of $H$ is derived in the Appendix \ref{SI:MutualInformation}, where we also show that for Gaussian selective landscapes information is approximately:
\begin{equation}
\label{Eq:SynapticInfo}
H_{ij}= 6 \phi_{ij}^2 M_i M_j ~.
\end{equation}

Mutual information quantifies how likely it is that, if one neuron spikes, the other one will also do so. If the neurons are not connected, $\phi_{ij}=0$ implying $H_{ij}=0$. If the focal neuron $i$ spikes randomly (large $M$) then the information content is low (in that case $\phi_{ij}$ is expected to be close to zero). Since $0<|\phi_{ij}|<1$, the quadratic term dominates over $M$, which makes $H_{ij}$ proportional to  $\phi_{ij}^2$. Hence, the information of a synapse increases as the weight increases. Note that for a given switching probability, the learning weights are higher in sparse networks than in fully connected ones. Thus, to an extent, the former encode more information than the later.

\subsection{Structural synaptic plasticity}
We implement SSP on a time scale different from that of associative learning. The system described above is in terms of statistical averages, and can be regarded as conditioned on a given network of connections. We assume that synaptic connectivity changes occur in one arbitrary group (below we explain how changes are introduced). If the new topology improves fitness, it spreads across all groups. However, we allow for a random component to avoid trapping in local optima, where the network is stranded in a state of impasse. For simplicity this is implemented through a Metropolis-Hastings algorithm. That is, if fitness increases with the new topology, this spreads to all groups. If fitness remains unchanged or decreases, then the layer might spread with probability \( \exp(W_{new}-W_{old})\). Allowing for this fitness decrease facilitates the escape from states of impasse.

We assume that changes in synaptic structure follow two heuristic rules inspired from neuroscience. First, if two neurons are unconnected but they are highly likely to spike, then a new synapse amongst them can be introduced. There is evidence that synaptic rearrangements result from circuit rewiring upon (e.g. in neocortical pyramidal neurons) stimulation \cite{Caroni:2014ey,Trachtenberg:2002cy,LeBe:2006eq}. Algorithmically, we randomly choose pairs of neurons with probability $\rho_i\rho_j$ amongst the set of unconnected pairs of neurons. Second, we allow existing synapses to be disconnected randomly with probability 
\begin{equation}
\label{Eq:Disbanding}
R=\exp[-\alpha H] ~,
\end{equation}
where $H$ is the synaptic information (Eq. \ref{Eq:SynapticInfo}. That is, if a synapse is informative, then it is unlikely to be disconnected, whereas if it contains no information, it is likely to be disconnected \cite{Fauth:2015jn,Chklovskii:2004eb,Poirazi:2001uy}.

In addition, we perform a Metropolis acceptance/rejection system: if the change in the network structure results in fitness increase, it is accepted. If on the contrary, the change results in fitness decrease, it is accepted with a probability proportional to the fitness ratio between that of the new network to that of the old one.

\section{Results}
\subsection{Selection and learning together speed up finding solutions}
\begin{figure}[t]
\includegraphics[scale=0.35]{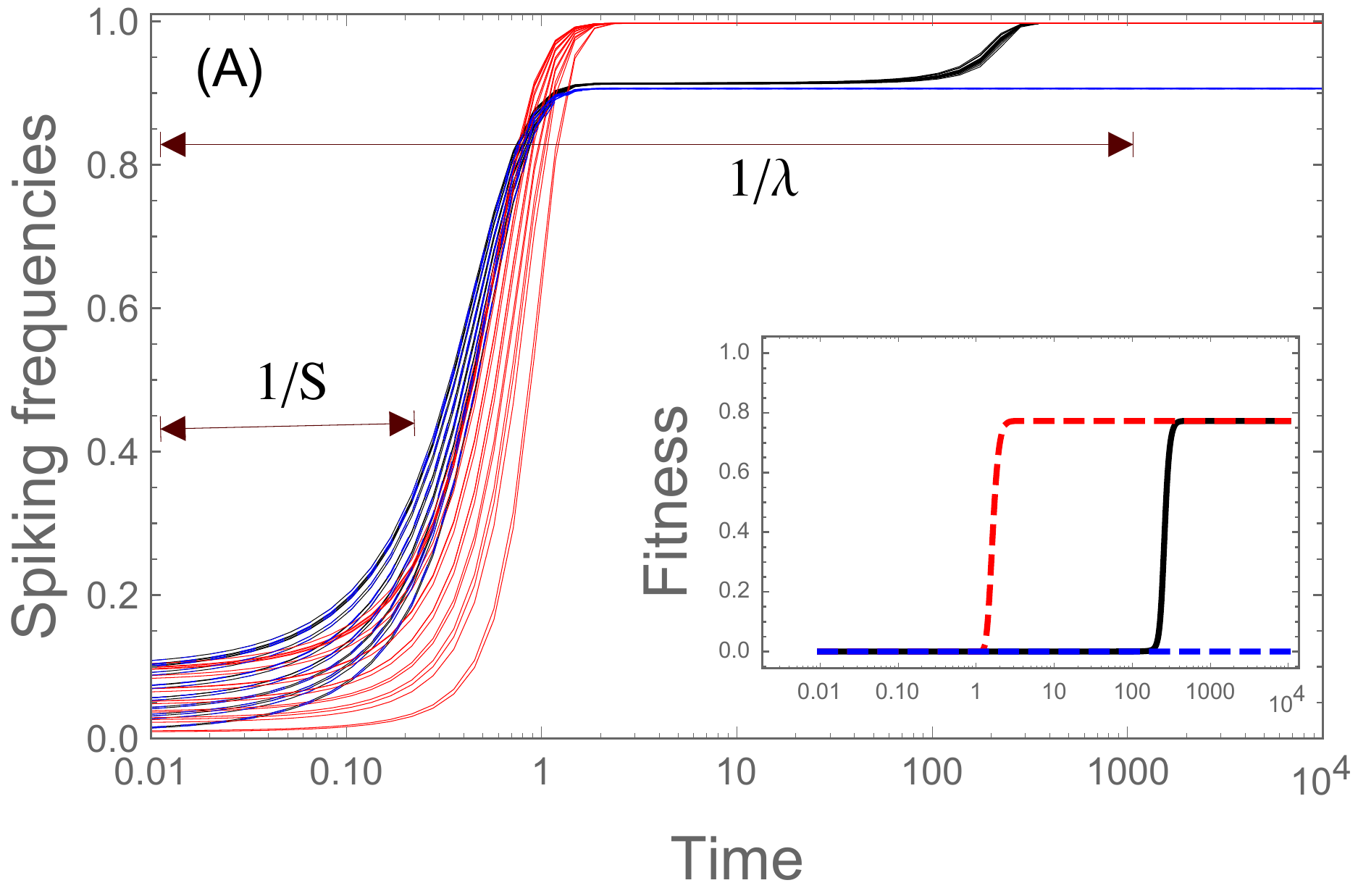}
\includegraphics[scale=0.35]{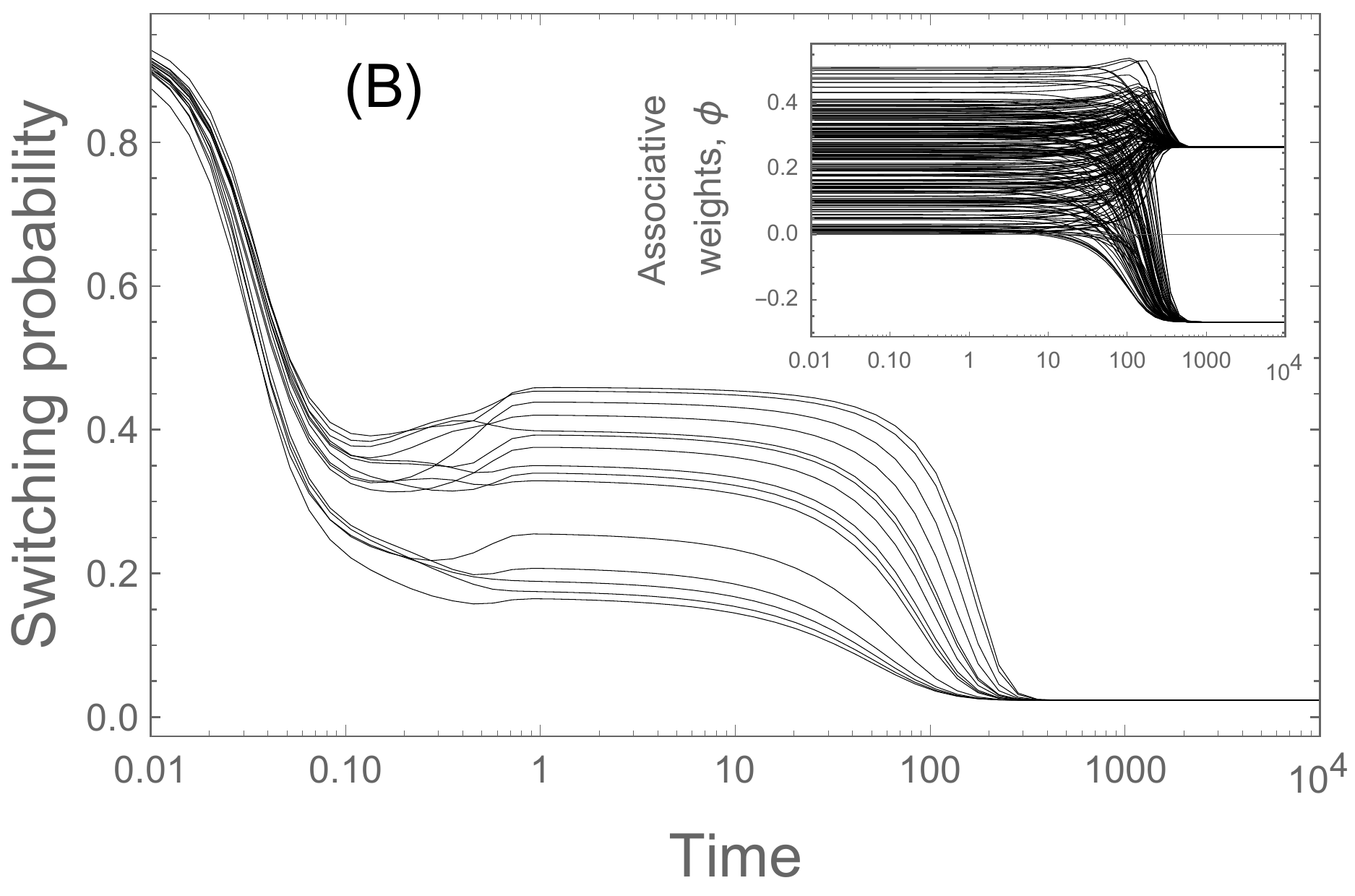}
\caption{\bf Example of selection-learning dynamics.}
(A) Selection-learning dynamics (black lines) compared to standard mutation-selection with na\"{\i}ve transition rates ($M\sim1/2$; blue) and to the run with the transition rates already learnt (red). Inset: evolution of fitness. (B) Evolution of the transition rates. Inset: evolution of Hebbian weights. $n=20, S=5, \lambda=0.001$. Initial conditions for allele frequencies and for initial weights are randomly sampled from a uniform distribution $U[0 , 0.01]$. The learning network is fully connected. Note the log-time scale.
\label{Fig:SelectionLearningDS}
\end{figure}

To understand how learning and selection jointly act we first assume an elementary scenario, i.e. a simple hill climbing process where we target for all neurons to be on. This situation results from a fitness function given by \(W=\exp[S\sum_i\rho_i]\), which has a constant gradient, \(\partial_\rho \log\bar{W} = S\) for every neuron. (This can be seen as a limit where $T$ is far from the current state, thus $S=2\beta T)$. We assume that Hebbian learning is slower than selection; i.e. $\lambda <S$. This regime describes the coupling of rounds of learning with copying across groups. Otherwise, learning would be completed independently in each group, associating random spikes and would not be able to learn the relevant features of the landscape. The associations that the network makes would relate back to previous states, effectively acting against hill climbing. However, in the regime where learning is slower than selection, fitness increases the representation of solutions, and once these are stabilised, learning can create meaningful associations. Figure \ref{Fig:SelectionLearningDS} presents a typical outcome where the process is characterized by three stages. We first observe an initial exponential increase in the proportion of active neurons. Compared with systems that do not learn the dynamics are similar on short time scales. This is because selection increases the representation of groups that provide better solutions, but these are initially in very low proportions. These fitter solutions are simply products of lucky stochastic events. Initially there is hardly any learning, indicated by light weights, and selective expansion simply amplifies those groups that have higher activity. This amplification takes on the order of $1/S$ rounds of evaluation. As groups become selected learning takes over, entering an incubation period where associations are built up because a good proportion of neurons fire correctly. After a learning period, associations are fully strengthened and the solution is finally reached whereby switching probabilities reach a minimum (Fig. \ref{Fig:SelectionLearningDS}B). The width of the plateau has a duration of roughly \(\frac{1}{\lambda}-\frac{1}{S}\). This regime is noticeable on a log-scale. Although in absolute time the selective process is so quick that it might pass unnoticed, this selection stage is crucial to explore configurations that can be fixed through learning. We emphasize that this early stage corresponds to the selective stabilization in the Neural Darwinism theory.

Instead of favouring equally all neurons to spike, the landscape can be set to favour distinct neurons to fire preferentially over others. For instance, making  \(W=\exp[\sum_iS_i\rho_i]\) and allowing $S_i$ to take any arbitrary value introduces asymmetries to the landscape. Crucially, if the dynamics are re-run with the learnt weights, the equilibrium is reached order of magnitudes faster. We stress that this is true even if initial spiking probabilities are randomised (Appendix \ref{SI:InitialConditions}).

\begin{figure}[t]
\includegraphics[scale=0.7]{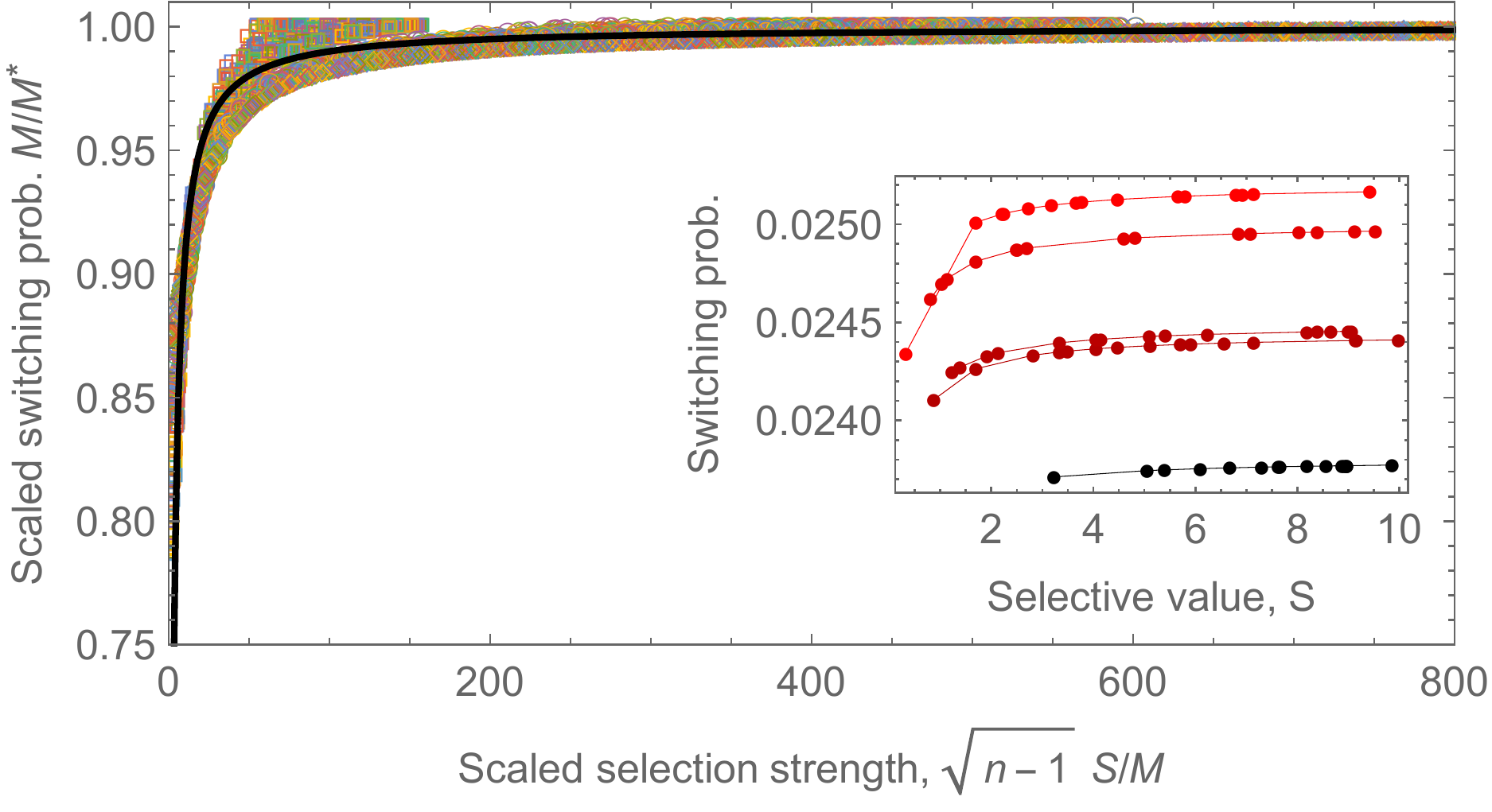}
\caption{\bf Relationship between selection intensity and associative weights.}
By scaling the switching probabilities with the maximum $M$ of a network, and scaling selection intensity as $\hat{S}=\sqrt{n-1} S/M$, we obtain a universal behaviour (black line). The distinct symbols represent runs with (squares) $n=5$, (circles) $n=10$ and (diamonds) $n=15$ neuronal loci. For each $n$ we report 300 independent runs. The selective values at each locus are chosen independently from a $U[0,10]$. $\lambda=0.01, \rho_0~U[0,0.1], \phi_0~U[0,0.01]$.
\label{Fig:LandscapeAssociation}
\end{figure}

Importantly, the switching probabilities become strongly correlated with the fitness gradient, meaning that the associative weights learn the local properties of the landscape. Figure \ref{Fig:LandscapeAssociation} shows that scaling the relative switching probabilities as $\mathcal{M}_i=M_i/M_m$, where $M_m$ is the largest $M$ of all the neuronal loci, then there is a universal behaviour with $\hat{S} = \sqrt{n-1} S / M_i$ of the form:
\begin{equation}
\mathcal{M} = e^{-\frac{1}{\hat{S}}} ~.
\end{equation}
This is an approximate form which breaks down at very low values of $S$ (see Appendix \ref{SI:MSscaling}). Intuitively, we would expect that larger values of $S$ would result in small switching probabilities. What happens is that the coupling of the system is sensitive to the distribution of $S$. For example, if all the selective values are similar (i.e. the variance is low)  then the corresponding $M$'s are all low (e.g. red curves in the inset of Fig. \ref{Fig:LandscapeAssociation}). If the $S$'s are distributed in a wider range, then the $M$'s are also distributed more widely, but have, overall larger values (e.g. black curves in the inset of Fig. \ref{Fig:LandscapeAssociation}). In other words, for a given mean value of $S$, the average $M$ increases with var($S$). In fact, there is a positive correlation between the variance of $S$ and the mean switching probability (data not shown). In turn, for a given variance of $S$, $M$ decrease with the mean of $S$. This means that most efficient complexes are composed by neurons that are required to fire more specifically.

For a given system, the associative weights increase (asymptotically) with the strenght of selection (data not shown). We performed performed Spearman's ranked correlation test to measure the strength of the dependency. (Because of the non-linearity,  `standard' Pearson's correlation is not a good measure for the dependency between $S$ and $\phi$.) In absolutely all cases the $p$-values were numerically zero, indicating strong dependence amongst $S$ and $\phi$ (data not shown). This strong statistical support indicates that the synapses encode the fitness gradient, directing variant spiking patterns accordingly: strong selection results in strong weights, which in turn decrease the switching probability. This leads in minimal variability of spiking, which maximises speed of fitness increase. Conversely, weak selection leads to poor associations resulting in large spiking variability, which allows exploration of the landscape.

We note the learnt equilibrium point is independent of the learning rate $\lambda$. This turns out to be generally true, regardless of the fitness landscape. We also note that under these `directional landscapes' the initial conditions (of both weights and allele frequencies) do not affect the equilibrium state of the system. (However, later we will see that under more complex fitness landscapes this is not true.)

\subsection{Formal analogy between evolutionary dynamics and neurodynamics}
At this point we formalise further the analogy with evolutionary biology, and more specifically with population genetics. First we realise that the bimodal neuron model is analogous to a biallelic genetic system. We start by clarifying a small but crucial difference in the notation. While in the models considered in this paper neurons take states $\{-1,+1\}$, in population genetics alleles are typically denoted as $\{0,1\}$. The +/- notation is convenient mathematically in order to describe Hebb's rules, thus in our evolutionary analogy we also require this property. Hence if $G$ is the value of a gene or allele, then we define $X =2G-1$. In this way we can readily apply the machinery from evolution to neuronal networks.

	Second, we consider the spiking probability of a neuronal locus (node) across all groups (Fig. \ref{Fig:Layers}). This average, which is the probability $\Pr(X_i)$ that a neuronal locus i fires in some of the groups, is thus analogous to the average $\mathbb{E}[2G_i-1]=2\rho_i-1$, where $\rho_i$ are allele frequencies at locus $i$. Note that allele frequencies are interpreted as the probability of sampling a particular allele in the population. Thus, for the analogy to be consistent, population size needs to be analogous to the number of groups involved in the learning. Although in both populations of individuals and of neuronal groups numbers are in fact finite, in this work we consider, as a first approximation, an infinite number. In this way we do not need to worry about stochastic effects that complicate the analyses. However, we recognise that randomness due to finite population size (a.k.a. genetic drift) can play a crucial role in both evolution and in learning. This is because randomness facilitates escaping local peaks and exploring the landscape in a less constrained manner. But before taking stochastic factors into consideration we want to focus on the interaction between selection and learning in the infinite population model.

Third, upon reproduction a population generates a new set of individuals, which sooner or later replaces the parental population. However, in neurodynamics, reproduction has to be interpreted in a particular way, because there is no generation of a new set of neuronal groups. However, the selective copying into groups with inferior performance effectively corresponds to a new population of groups (Fig. \ref{Fig:Layers}D). 
	
Given the analogies above, we can ask the converse question: what is the interpretation of the learning process in evolutionary dynamics?

Equation \ref{Eq:ActivityChance} describes the activity changes of neural networks across iterations, leading to an update rule of the spiking frequency of each neuron. In population genetics, this transition probability corresponds to a mutation rate. In molecular evolution mutation rates are normally state-independent, dictated by, for example, copying errors of the polymerases that replicate DNA, repair mechanisms, or other molecular processes that do not depend on the genetic states of the individual or population. (Although there are genetic models that consider evolvable mutation rates; see Discussion.) The switching probability \(M_i=1/(1+\exp[Y_i]) \) is dependent on the state of the system, and follows directly from the update rule. Apart from this dependency, the equations (Eq. \ref{Eq:FreqChange}) are analogous to a selection-mutation equation. The resemblance is a natural outcome from the analogy laid out above.

But beyond the cosmetic similarity between the replicator-mutator equation and neural dynamics, the crucial difference is that the update rule is able to learn the local properties of the fitness landscape. By doing so, hill climbing towards a fitness peak is facilitated by generating variation directed towards the fitness increase. 
	
\subsection{Learning in rugged landscapes}
We now consider the more complex adaptive landscape, given by \(W=\exp\left[-\beta \Delta^2\right]\). In evolution this kind of landscapes are known as `stabilizing selection'. The complexity of this landscape results from the non-linear effects (known as epistasis in genetics and evolution). These are hard landscapes to explore because there are many local peaks or solutions, some equally optimal, some sub-optimal, and simple hill-climbing algorithms often fail to converge to an absolute maximum of fitness. 

\begin{figure}[t]
\includegraphics[scale=0.35]{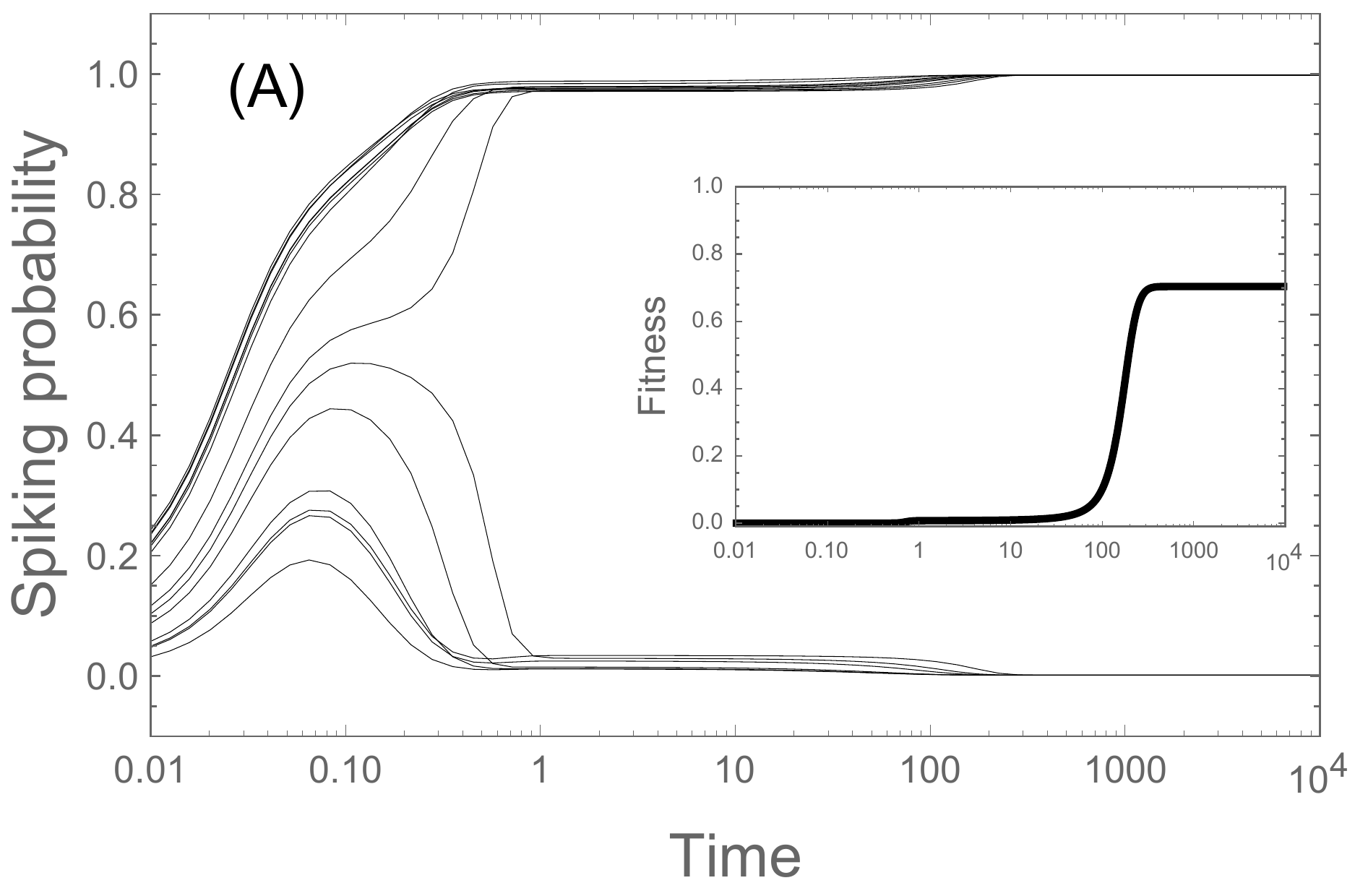}
\includegraphics[scale=0.35]{SwitchingProbSS}
\caption{\bf Neurodynamics in a stabilizing rugged landscape.}
Model parameters: $T=15$, otherwise as in Fig. \ref{Fig:SelectionLearningDS}.
\label{Fig:SelectionLearningSS}
\end{figure}

Figure \ref{Fig:SelectionLearningSS}  shows the neurodynamics. We find that exactly 15 neurons fire (with probability $\rho=0.995$) and the remaining five remain off. In this case the uninformative neurons are shut down. Which neurons spike and which do not is contingent on the initial conditions, but in this landscape the identity of each neuronal locus is meaningless. Different initial conditions can lead to different but equivalent solutions (data not shown).

\subsection{Random and sparse topologies of the neuronal connections impair learning}
So far we assumed that there are synapses amongst all pairs of neurons. Relaxing that assumption corresponds mathematically to fixing certain weights $\phi_{ij}$ to zero, indicating that no synapse exists among neurons $i$ and $j$. Under these circumstances the equilibrium switching and spiking probabilities are more variable, with the spread determined by the connectivity of the underlying learning network. Appendix \ref{SI:NetworkTopology} presents some neurodynamic outcomes using different random topologies under directional and stabilising landscapes. These topologies are drawn from different random graph models with various degrees (see Methods). We tried Erd\H{o}s-R\'enyi, Barab\'asi-Albert (scale free) and Watts-Strogatz small world topologies. Each of these models has different statistical properties. Irrespective of these, there are two central conclusions. First, random networks lead to unfit solutions, where the systems cannot reach the target. This is true regardless the target value, number of neurons and type of topology. The systems typically converge to a suboptimal solution where no further learning can happen and cannot escape local optima. We regard this as a situation where a network that was previously functional for another task is repurposed for a new task, and the initial topology is, regarding to the new task, arbitrary. Thus, the initial circuit is not expected to be adapted to the new task. Consequently, what the system can learn is only limited, and in the vast majority of cases, suboptimal. We identify these solutions as states of impasse, which means there is no further progress possible, because any small modification to the synaptic weights or spiking patterns leads to a state with lower fitness score.

The second central conclusion is that poorly connected neurons have very low input activity, leading to high switching probabilities. Highly connected nodes have small switching probabilities with spiking frequencies close to unity. Hence, only highly connected nodes (the less frequent) can learn efficiently. Since random topologies give suboptimal results, we consider that details regarding specific network distributions are secondary and discuss them only in the Appendix \ref{SI:NetworkTopology}.

Our choice of network distributions is arbitrary, motivated by mathematical convenience and scientific hype. Hence, the results above do not necessarily imply that brains are suboptimal unless fully connected. However, they reveal that in sparse networks, as in real neuronal complexes, Hebbian learning does not suffice to solve complex problems because it too often leads to impasses.

\subsection{Structural Synaptic Plasticity}
Structural synaptic plasticity is a mechanism that goes beyond the update of existing synaptic weights (i.e. Hebbian learning) by allowing new synapses to be established and old ones eliminated. This dynamical restructuring of the topology of the neuronal networks as the system learns has been shown to be important for the transfer of short-term to long-term memory \cite{Kilgard:2012ci}. However, we test the role of SSP in the more general scenario of problem and impasse solving.

Above we found that network topology impairs problem solving on complex learning landscapes. This is paradoxical because there is clear evidence that the brain is not fully connected, even though the type of connectivity is disputed and tissue-dependent. But our negative results do not rule out that there might be specific topologies that facilitate or optimize learning. We now show that under synaptic plasticity, the neuronal complexes form particular structures, which are unlikely to be recovered randomly, and thus accounts for the negative results above.

A mechanistic description of SSP rooted in neurophysiological processes is beyond the scope of this article, in part because much is unknown. Instead, we propose a simple phenomenological model to show that SSP can not only affect the outcome of the learning process fundamentally, but also that this mechanism can resolve impasses. Further specific aspects keep the essential features unchanged, even though some of these might be crucial for the actual implementation of the cellular mechanisms that we explain using a simple model.

A central feature of SSP in the neurodynamics is that, by modifying the distribution of synapses, it provides new pathways to explore the space of solutions. This is why SPP can be an efficient mechanism to escape impasses.

\subsection{Dependency on the number of neurons}
\begin{figure}[t]
\includegraphics[scale=0.65]{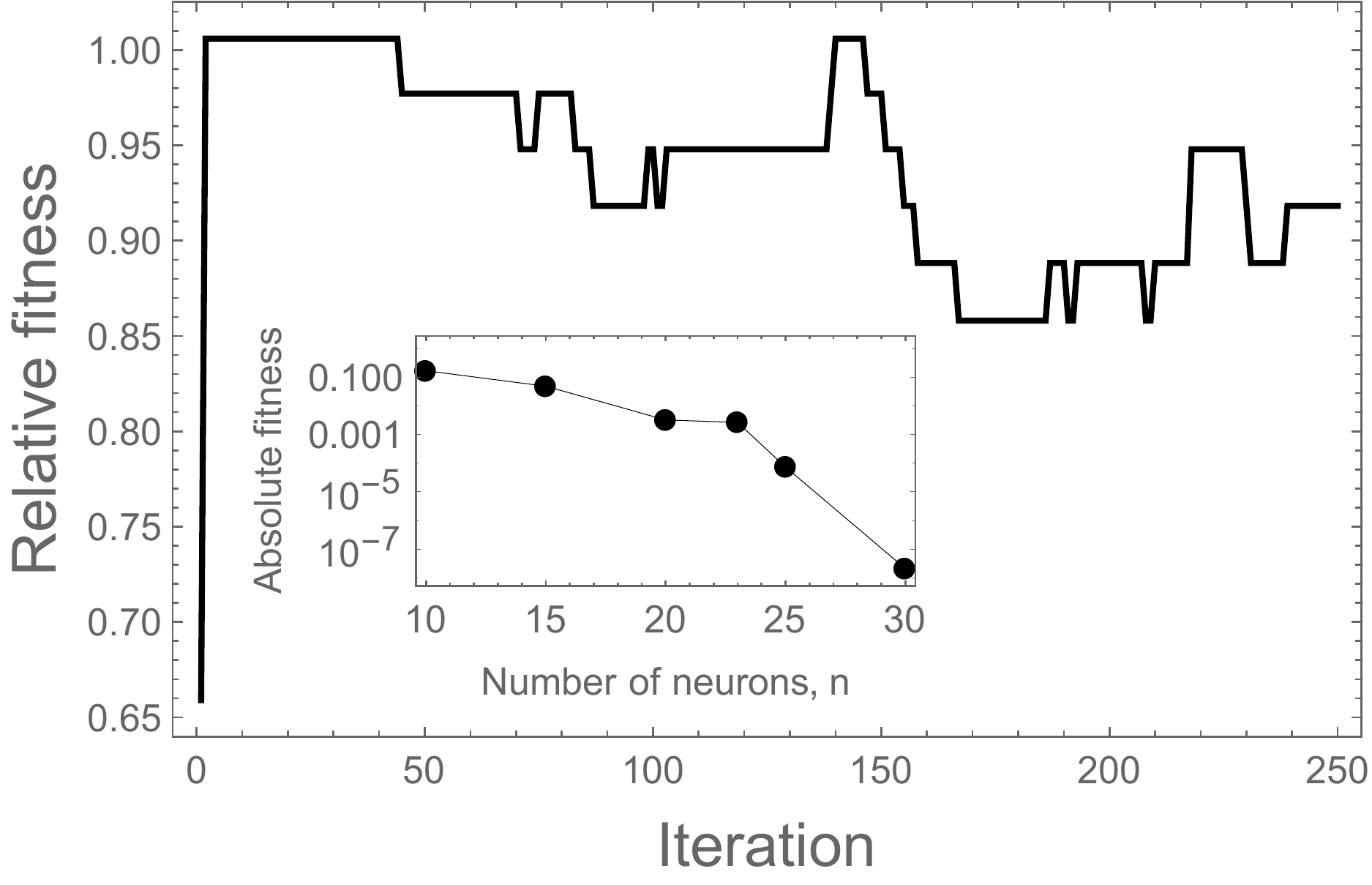}
\caption{\bf Dynamics of structural synaptic plasticity.}
Example of a random realization with $n=30$ neuronal loci (fitness is scaled to the maximum value). Inset: Absolute fitness as a function of neuronal loci. Parameters: $T=7$; otherwise as in Fig. \ref{Fig:SelectionLearningDS}. Synapses are assumed to have no cost.
\label{Fig:SPPDynamics}
\end{figure}

This random mechanism allows the neuronal networks to explore the space of configurations, leading, on average, to an increase of fitness. Figure \ref{Fig:SPPDynamics} shows that systems with few neurons evolve good solutions more easily than larger systems. This is clear: finding an optimal configuration with a few neurons requires fewer evaluations than larger networks simply because the search space of the former is much smaller than that of the latter. The number of possible configurations increases with $n^2$, thus on the basis of trying one modification at a time the convergence time increases non-linearly with the number of neurons. However, although this holds true for our model, there is no reason to think that there cannot be parallel evaluations of different topologies in different complexes, dramatically alleviating this inefficiency. However, in this paper we restrict ourselves to evaluating one modification per iteration. Note that one step in the iteration does not correspond to a physiological time unit because the Metropolis algorithm only ensures convergence to the equilibrium distribution as dictated by detailed balance and considers no information regarding the diffusion leading to said equilibrium.

In each round of learning (i.e. after the system converged to equilibrium with a newly tested network) the current weights are being kept, and new connections are assigned to new random initial values. Alternatively, we can simply reset all weights to random values. The second strategy proves to be more efficient than the first. Whether spiking probabilities are reset or not, proved irrelevant (data not shown).

\subsection{Structural plasticity leads to maximal connectedness}
\begin{figure}[t]
\includegraphics[scale=0.3]{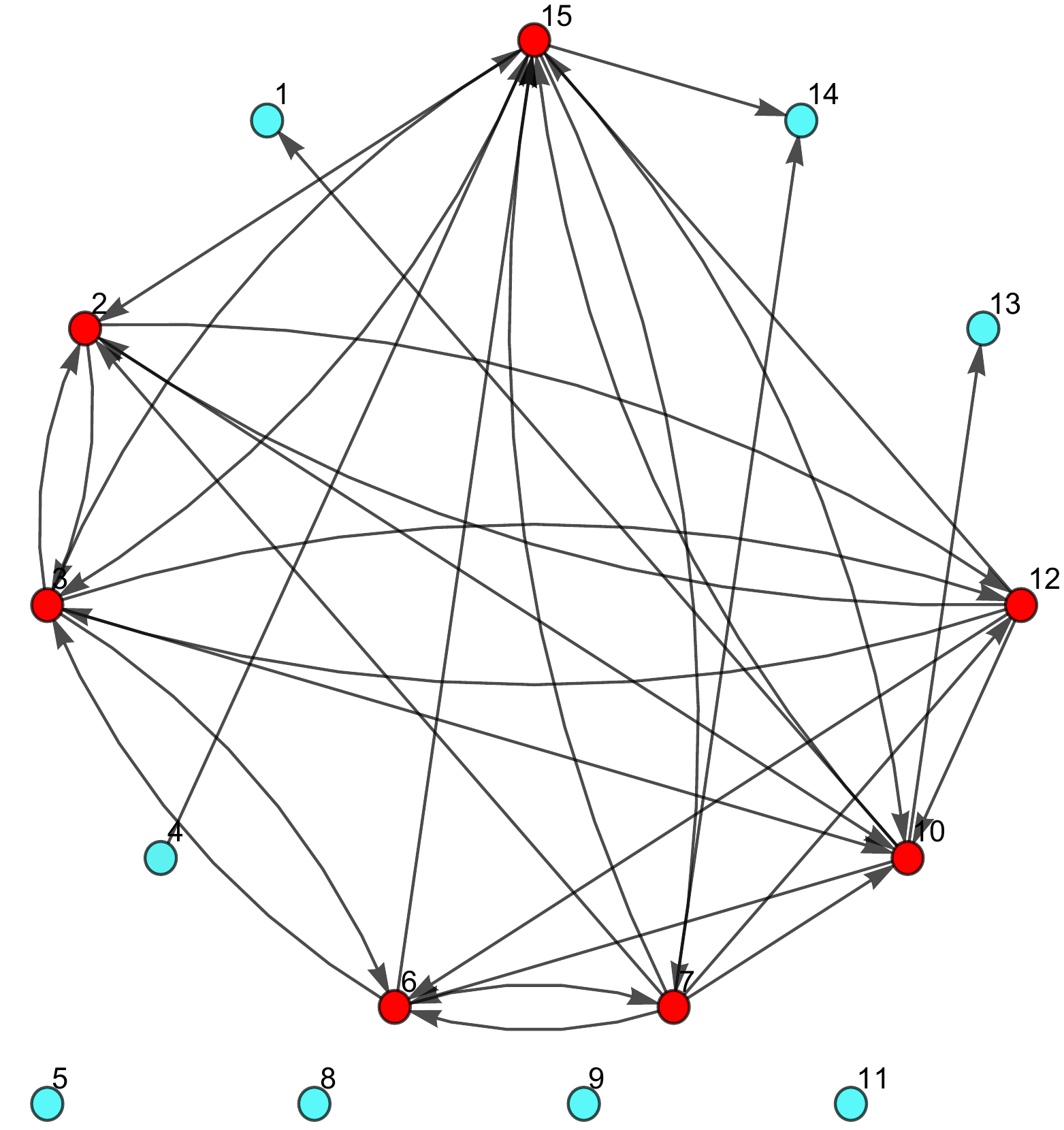}
\includegraphics[scale=0.35]{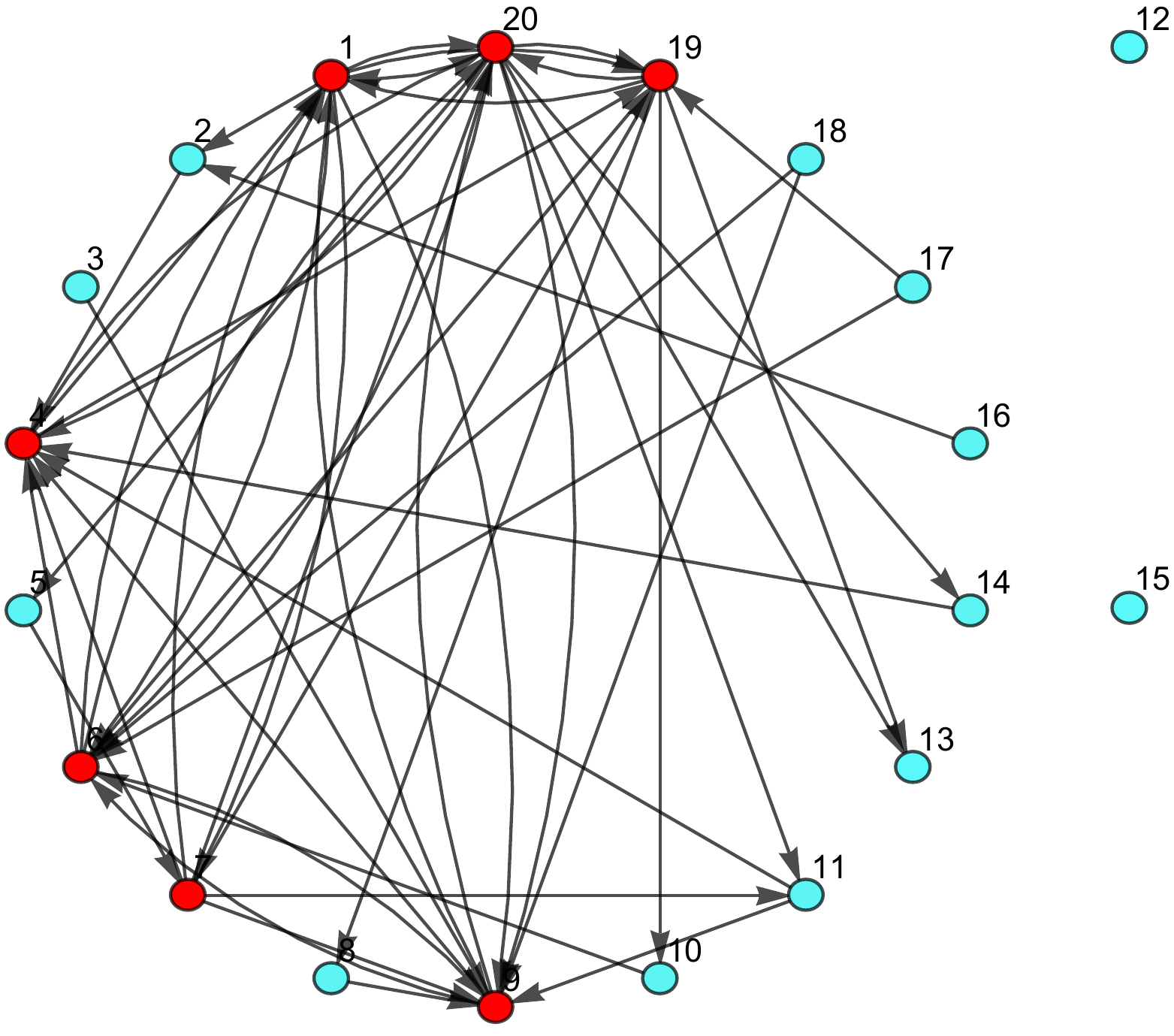}
\includegraphics[scale=0.25]{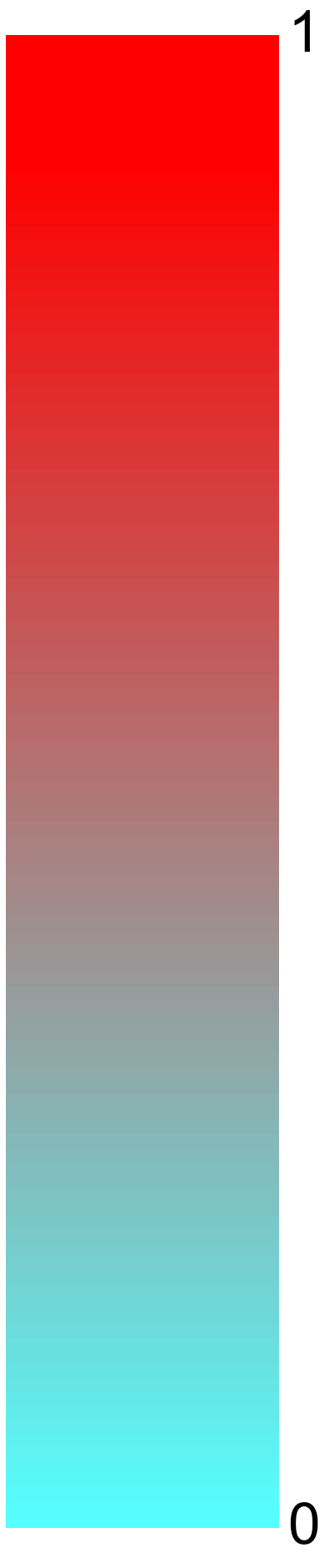}
\caption{\bf Evolved learning networks.}
Example of evolved networks with $n=15$ (left) and $n=20$ (right) neuronal loci. The colors indicate the frequency of active neurons. In both cases the particular node labels are irrelevant, and the proportion of active neurons depend on the initial conditions and the history of the process. Note that irrespective of the number of neuronal loci, the number of active components is correct. Parameters as in Fig. \ref{Fig:SPPDynamics}
\label{Fig:EvolvedNetworks}
\end{figure}

The resulting synaptic networks are straightforward. Recall that our test problem chooses for a target number $T$ of spiking neurons. Thus, the optimal state has exactly $T$ neurons on and the rest are off. Ideally, these $T$ neurons are fully connected amongst them. We find that the complexes correctly converge to solve the problems, and, thanks to SSP, the networks that evolve fully connect these active neurons (Fig. \ref{Fig:EvolvedNetworks}). In other words, the systems converge to networks that fully connect the required components to solve the problem.

The convergence to fully connected networks is due to two factors. The first is the need to switch on the right number of neurons, which requires strong synapses amongst them. The second is to switch off the unneeded components; this also requires connected components because negative weights between active and inactive decrease the probability of firing. If negative weights are not allowed, the system can only maximise fitness by ensuring the right neurons are on, and the networks converge to fully connect these components (Fig. \ref{Fig:EvolvedNetworks}). 

\subsection{Neuronal networks are robust to small costs of synaptic connections}
\begin{figure}[t]
\includegraphics[scale=0.42]{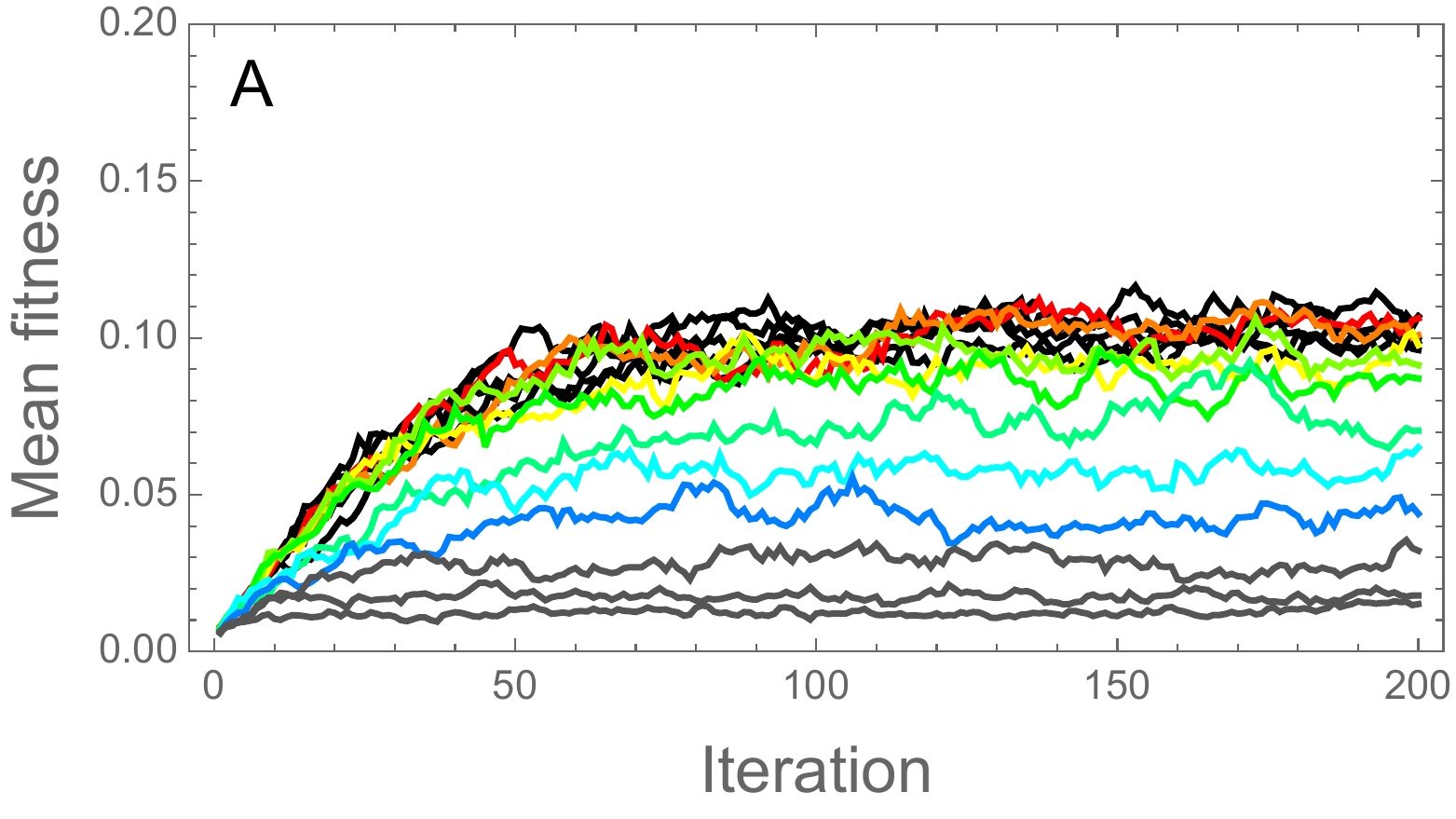}
\includegraphics[scale=0.42]{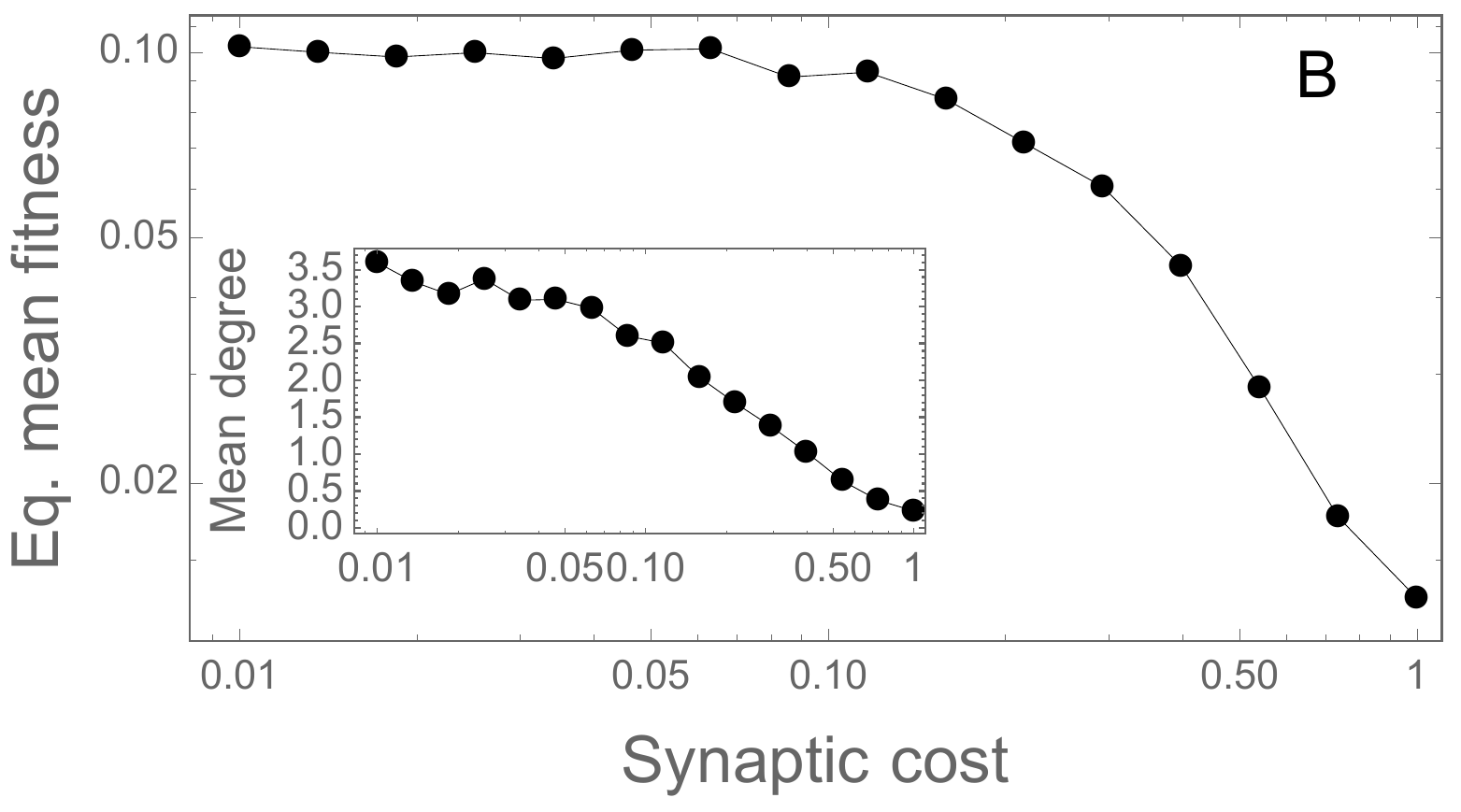}
\begin{center}
\includegraphics[scale=0.42]{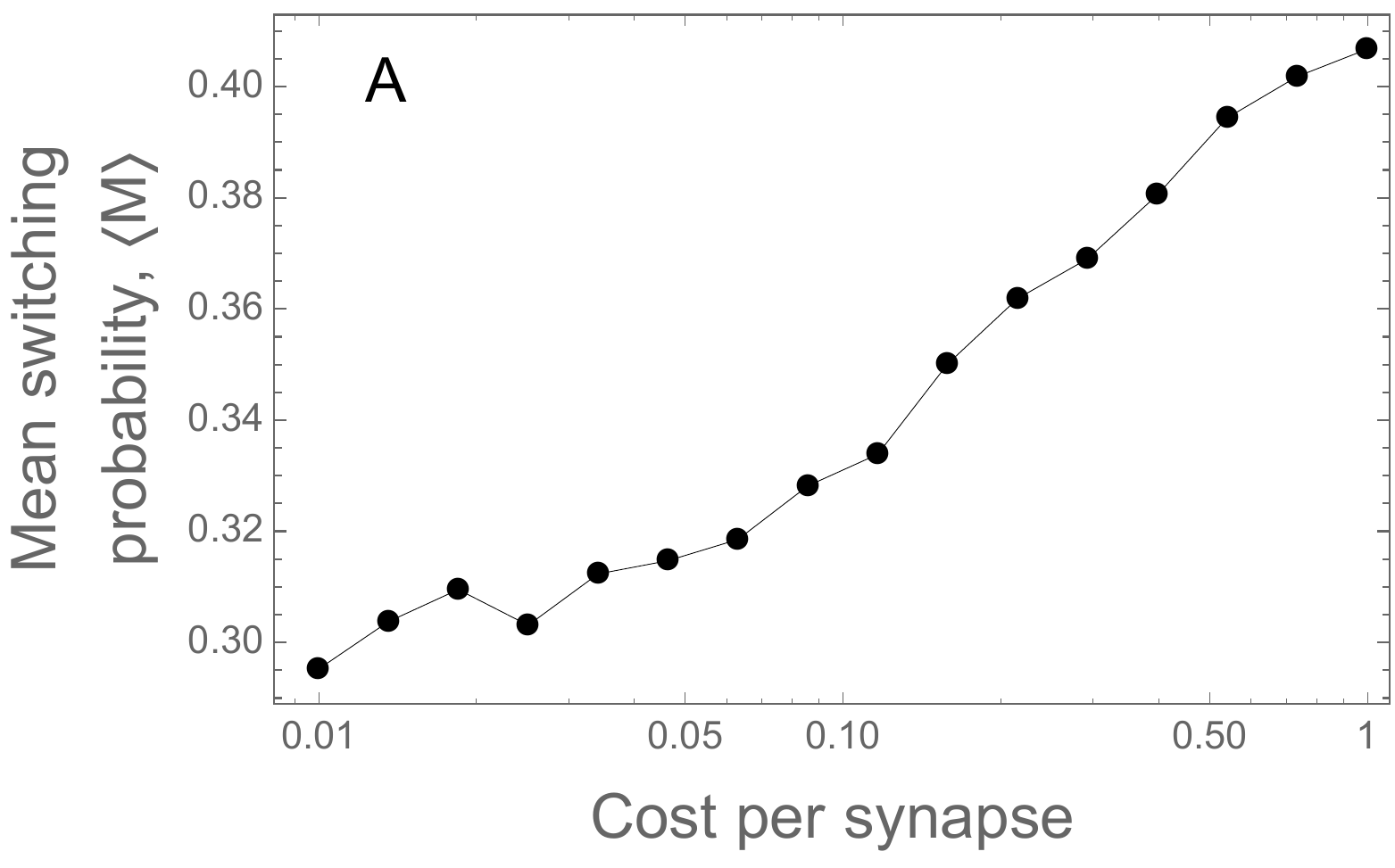}
\end{center}
\caption{\bf Neuronal systems under costly synapses}
(A) Dynamics of the fitness (relative to the maximum) of neuronal systems under different cost per connection. Black curves on top: low costs ($k < 0.05$). Colour curves: intermediate costs increasing from $k=0.05$ (red) to $k=0.5$ (blue). Grey curves at the bottom: high costs ($k>0.5$).  Each curve is a replica of 77 independent simulations. (B) Equilibrium fitness as a function of the synaptic costs. Inset: Average number of synapses of each neuron (network degree) against the cost per synapse. (C) Mean switching probabilities of the connected nodes against synaptic cost. Each point is an average at the stationary values (last 50 time points) and over the 77 simulations. Parameters: $n=10$, $T=7$, $S=10$, $\lambda=0.01$.
\label{Fig:SPPDynamicswCost}
\end{figure}

Now we penalize for the amount of connections that the networks have (Fig. \ref{Fig:SPPDynamicswCost}). There are various reasons to assume this constraint. First, there are costs associated to synaptogenesis. Second, there are higher metabolic costs due to the transmission of action potentials, which increases at least proportionally (if not allometrically) with wiring. Third, there are major spatial constraints in the brain, limiting the amount of white matter that can be packed. In order to take into account these and other reasons for limiting the amount of neurons, we include a fitness cost to the system, \(\exp[-kd]\), where k is the cost per synapse, and $d$ is the number of synapses (number of edges in the learning network).

In Fig. \ref{Fig:SPPDynamicswCost} we observe that finding the solution to a problem is impaired as the cost of establishing synapses is increased. (In these examples we target $T=7$, but the particular choice of the target value is unimportant; in the Appendix \ref{SI:StructuralPlasticity} we present results for other target values.) Clearly, this is because the number of connections decreases with increasing cost, which in turn compromises spiking specificity. 

It might be unsurprising that the number of synaptic connections decreases with their cost, and naturally, the networks become less discriminative as they lose connections. However, it is also true that they show a notable level of resilience (graceful degradation), in the sense that even if performance is somewhat impaired as synapses are eliminated the required number of connected neurons is robust to the cost. In other words as the cost increases the networks still converge to structures that connect (even if sparsely) the necessary neurons (see Figs. \ref{Fig:SPPDynamicswCost}B). The networks lose performance as they lose synapses because neurons receive less input and therefore their switching probability becomes higher. Nevertheless, they tend to remain connected with as many components as possible.

In the stationary state the distribution of networks is broad. As indicated in Fig. \ref{Fig:SPPDynamicswCost}B, the average number of synapses decreases with the cost; this is also true for its variance (in Appendix \ref{SI:EvolvedNetworksDist} we present the degree distributions). The meaning is that as the cost increases, each neuron establishes fewer synapses with other ones. Also note that some of the components tend to have only few connections. This is indicated by the notion of connectivity (Fig. \ref{Fig:SPPDynamicswCost}B), which is the number of synapses that we need to remove to separate the network into two unconnected subsets. Typically in the evolved networks this is due to a single poorly connected network, rather than to connected sub-complexes interconnected by a few synapses. As costs are very high ($k\sim1$), the networks are sparse and have several unconnected components.

We point out two important differences between random networks and the evolved distribution of networks. First, taking the random network as a null model (the Erd\H{o}s-R\'enyi, is the one that best matches the evolved distributions,  Appendix \ref{SI:EvolvedNetworksDist}) we expect a binomial distribution $B[n-1,p]$. The observed distributions are reminiscent of the binomial using the empirical $p$'s. However in all cases we rejected the null hypothesis ($\chi^2$ tests, all $p$-values numerically zero); the expected variances are too low.

Second, despite the variability in the distribution of the evolved networks, these solutions are not in states of impasse. With fewer synapses, the input activities of the neurons are lower, translating into larger switching probabilities (Fig. \ref{Fig:SPPDynamicswCost}C). This does not reduce specificity of firing: still the correct neurons are more likely to fire in an idiosyncratic manner. However, there is more `background' noise due to fluctuations. We could say that for larger costs, neurons are still accurate, albeit less precise.

\subsection{Distribution of synaptic lifetimes}
\begin{figure}[t]
\includegraphics[width=\columnwidth]{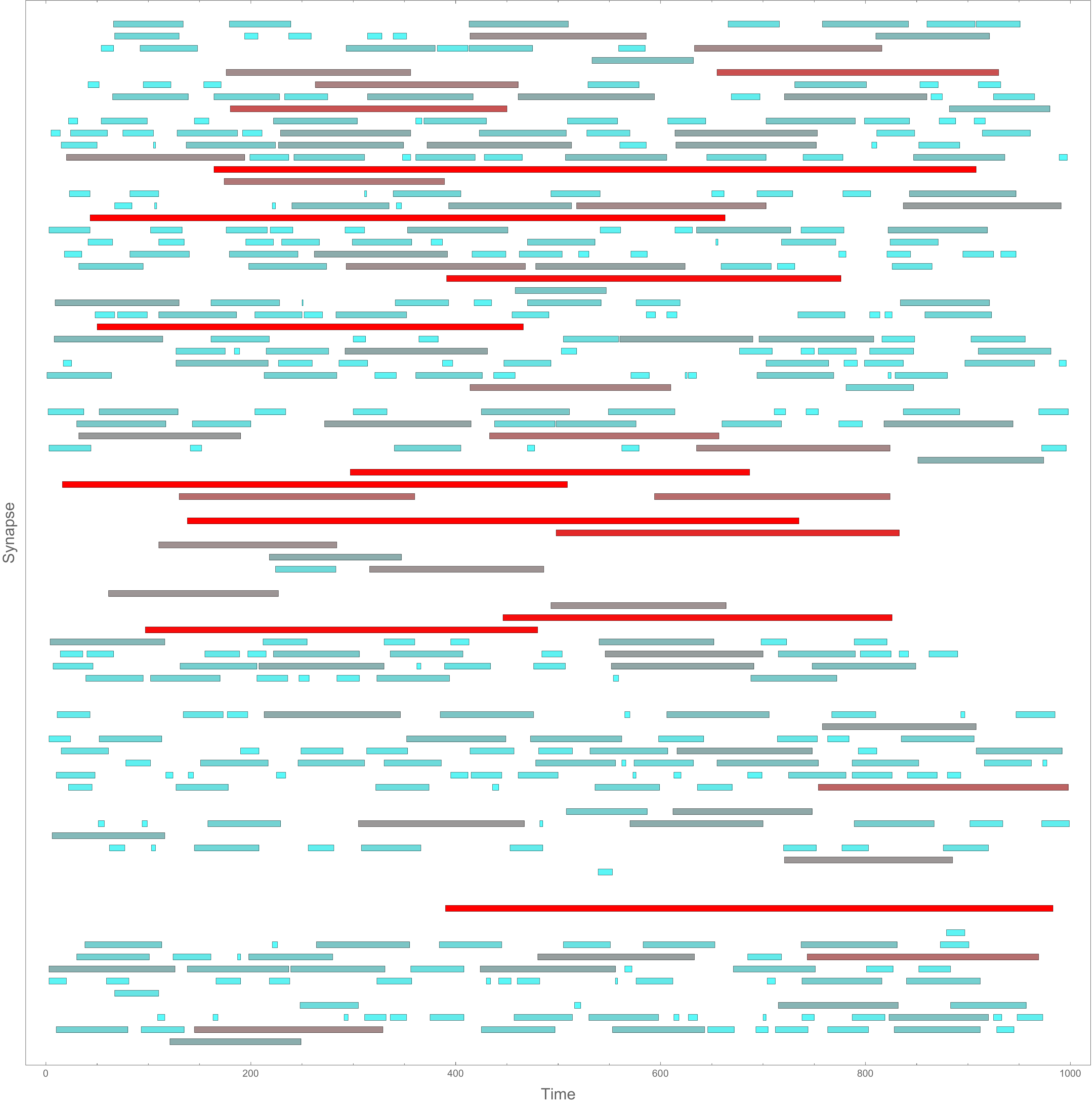}
\caption{\bf Timeline of synapses. }
Each row corresponds to a particular synapse between a given pair of neurons. Colors reflect the synapse length (only for visualization): blue: short lasting synapses; red: short lasting synapses. In this case the cost of synapses is $k=0$; otherwise as in Fig. \ref{Fig:SPPDynamicswCost}. 
\label{Fig:SynapticLifetimes}
\end{figure}

Part of the reason why our model shows a large variance in synaptic network topologies is that we allow for synapses to continuously be established and disbanded. We still ignore the quantitative and dynamical extent of these processes, reason why we can only make arbitrary assumptions. Precisely this, however, is an important point of falsification of our ideas. Irrespective on the details regarding the \emph{assumed} synaptic distribution, the \emph{observable} distribution can lead us to refine our model. Yet, the relevant aspect here is not the details of the predicted distribution (since, as argued above we present a simplified model), but the fact that we predict that there is a distribution of synaptic lifetimes at all. For instance, Fig. \ref{Fig:SynapticLifetimes} presents the timelines of synapses. We note that many are short lived and a few live long. This result is expected to hold regardless of the specific problem being solved, in as long as we do not trace the identity of the neurons that the synapses are connecting. Clearly, the number of long-lived synapses must increase with the number of neurons in a complex. However, the lifetime distribution should hold relatively robust. This is expected because what determines the lifetime of a synapse is essentially its local information, and this is independent of other properties of the network. For example, increasing the parameter $\alpha$, which controls how weaker connections are penalised, modifies the probability $R$ of synaptic disbanding so that the average synaptic lifetime become shorter. If however the target or intensity of selection is changed, the effect on the distribution of synaptic lifetimes is minimal (Appendix \ref{SI:SynaptilLifetimes}).

\begin{figure}[t]
\includegraphics[scale=0.35]{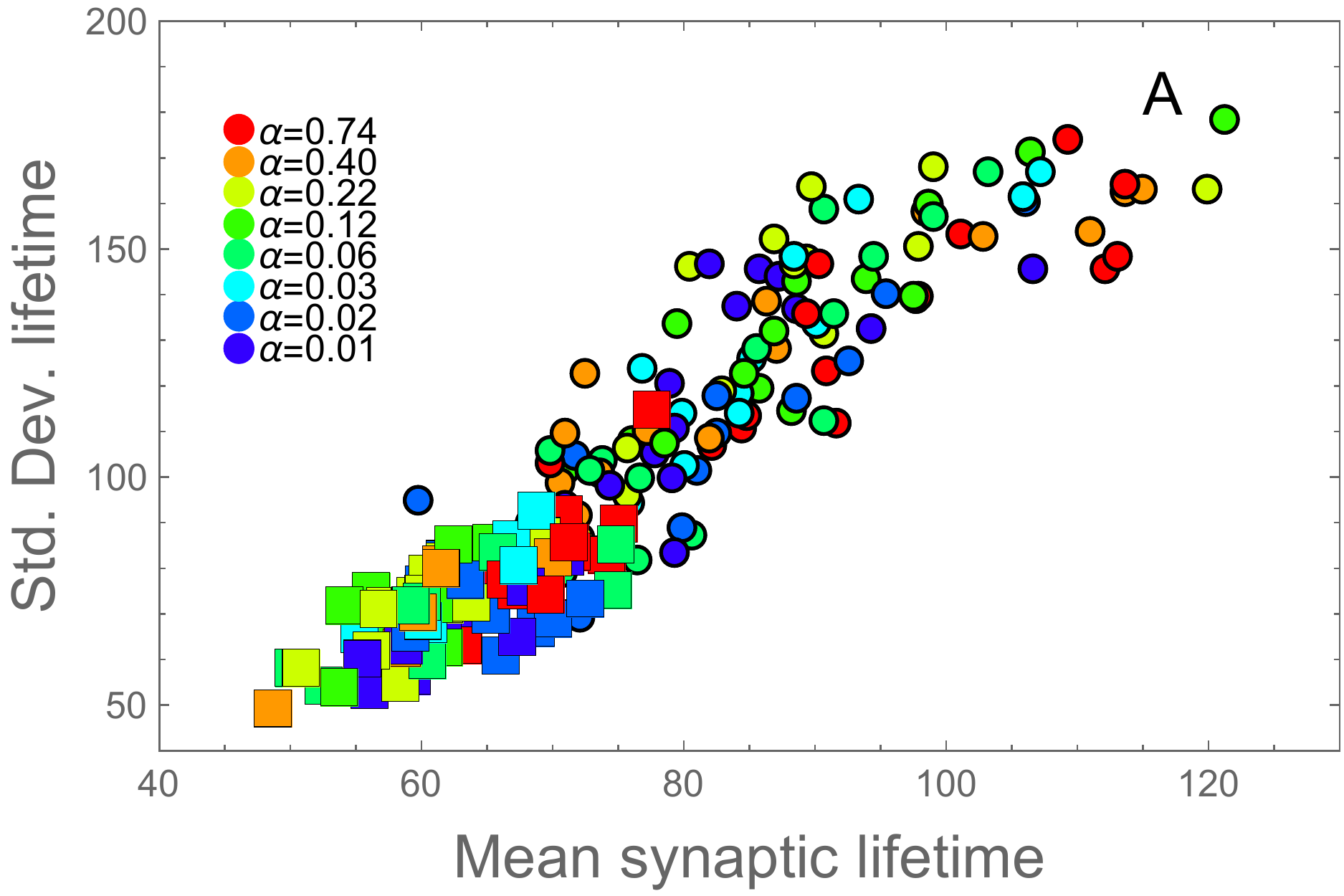}
\includegraphics[scale=0.35]{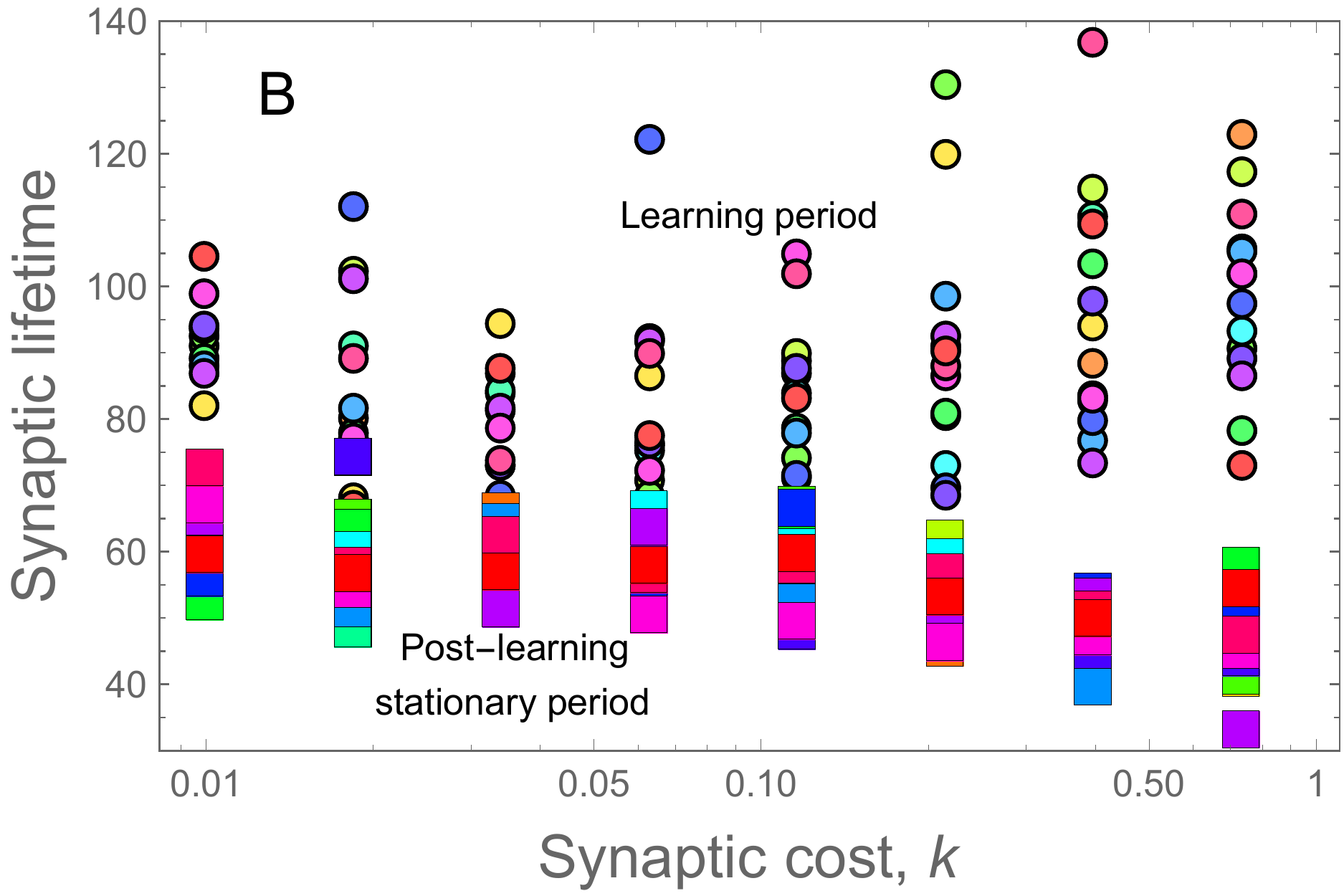}
\caption{Average lifetime of synapses, measured over one history.}
(A) Relationship between average and standard deviation of the synaptic lifetimes for 14 independent realisations. Colors indicate the synaptic costs $k$ as in the inset legend. (B) Average lifetime as a function of the cost per synapse $k$. Different colors represent 14 different (independent) realisations. In both panels: squares indicate synapses established at early stages (before  200 iterations) and the bullets: synapses established after reaching stationary states (later than 200 generations). Parameters as in Fig. \ref{Fig:SPPDynamicswCost}.
\label{Fig:SynapticLifetimesStats}
\end{figure}

	Our results show that synapses that are established early tend to last longer than synapses established in later developmental stages (Fig. \ref{Fig:SynapticLifetimesStats}A). This is a natural outcome of the model; we have not imposed any mechanism that presupposes this behaviour purposely. This happens due to historical contingencies. Some of the synapses that are initially established lead the network close to a local peak. Afterwards other synapses fine-tune the network, increasing fitness in smaller amounts. Only after the network has been populated, new connections can substitute the original ones. Figure \ref{Fig:SynapticLifetimesStats}B reveals how synaptic lifetime increases with the cost. For newly formed synapses, low costs do not show a central trend, implying a certain degree of resilience or robustness. In the long run, in the small cost range, synapses established later live shorter as the cost increases. For larger cost range, synaptic lifetimes tend to increase with cost. However, recall that for costs \(k \gtrsim 0.5\) the networks have low or no connectivity. As the cost increases, the network starts to dispense with inhibitory connections. Recall that shutting unwanted neurons down only decreases the background noise, but does not affect the result of the `good' neurons. Thus, synapses amongst these `good' neurons tend to live longer, as they are essential for the system's performance.

	Due to the redundancy of optimal solutions occasional major restructuring of the networks are expected to occur during lifetime for two reasons. First, once a peak has been found, disbanding the principal connections leads to major function impairment (fitness decrease). Second, establishing new synapses that are potentially as good as the existing central ones is unlikely due to the costs of synaptic connections. However, we do find occasional major network restructuration. Because of the strong coupling amongst several neurons and synapses, once a major connection is destroyed this can cause further impairments by subsequent changes that result in even worse fitness. At some point there is a restitution of the system when a new fitness peak is approached. This has the qualities of self-organised criticality. However, it might also be that the stochastic behaviour allows a few groups to shift from sub-optimal solutions to better ones, effectively jumping across fitness peaks. The subsequent replication of these successful solutions to other groups can result on a full escape from impasse states. (In evolution this process is known as \emph{shifting balance};  \cite{Wright:1931,Wright:1982ey}.

\section{Discussion}
\subsection{Relationship to previous models}
Using a Bayesian framework, \cite{Ullman:2012wq} proposed a model that explains aspects of cognitive learning in children. In their model, the brain implements a Bayesian update algorithm to form theories based on observed data. They stress that in contrast to a more reductionist `connectionist' view, Bayesian learners explain neuropsychological aspects of the dynamics of learning. In their framework, theories map to a multidimensional landscape where well-formed theories lie at peaks. The dynamics include learning, but only at as a local process. They argue that learning cannot account for the invention of new theories, but rather, only modify the degree to which we believe in any given theory. That is, learning acts to fine tune the theory around a peak. The proposal of new theories does not happen through learning, but through stochastic modification of the existing theories (in a data-independent manner, that is, there is random variation). If a new theory scores better than the previous ones, it is adopted with certain probability \cite{Kemp:tn}.

This model is similar to ours, particularly in the implementation. However, there is nothing mystical about this coincidence. What Ullman \emph{et al.}. and we describe belongs to the Markov Chain Monte Carlo class of models, which is a popular methodological toolbox in stochastic processes. One important common aspect is that learning alone does not produce any new configurations (networks in our case, theories in theirs), but only improves local adaptation given the current configuration. Whilst their model describes processes occurring at high level of cognition, we describe simpler processes at the neurophysiological scale. However, we reach similar conclusions regarding the need and limitations of learning in relation to other processes that can generate variant configurations that lead to a better performance. This coincidence and its consequences (see discussion in Ullman's paper) are preliminary evidence supporting our proposed physiological mechanisms.

Although Ullman \emph{et al.}. do not discuss the states of impasse these are implicit in their models. Also, the explanations they give in terms of `theory formation' are to a large extent equivalent to those of our model's. That is, learning a local peak restricted to a given configuration results in weight values that always lead to lower scores if any modification is introduced. They resort to stochasticity (see below) as a mean to jump across peaks. In our case this stochasticity is introduced via synaptic plasticity. There can be other sources of stochasticity, which we address below. However, synaptic plasticity is a component accounting not only for the necessary randomness to escape peaks, but is also known to be an important component of learning. The question of what generates circuit diversity remains so far open: we propose that SSP can account for this diversity.

Note also the crucial difference in the search mechanisms in the two models. Kemp and Tenenbum \cite{Kemp:tn} use a greedy search algorithm on stochastically generated variation, whereas we adhere to the view of the Darwinian neurodynamics, also known as the neuronal replicator hypothesis \cite{Fernando:2010dn,Fernando:2010ul}. The point is that on vast combinatorial landscapes, evolutionary search is known to produce impressive results. The greedy algorithm works for relatively small spaces but for larger spaces more efficient search is needed, as Kemp and Tenenbaum \cite{Kemp:tn} acknowledge. It is remarkable that although Ullmann \emph{et al.} \cite{Ullman:2012wq} explicitly draw a rugged conceptual landscape, the possibility of evolutionary search is not mentioned. 
Interestingly, this can lead to predictions. That is, we expect an inverse correlation between the rate of synaptic turnover and the capacity to overcome impasses. For instance, the rate of synaptic turnover tends to be slower in adults, who are often less able (or do it at a slower rate) to overcome impasses in certain neuropsychological test problems than children. It has not escaped our attention that this can also be addressed through experimentation in animal models from several points of view, from behavioural to neurophysiological.

Besides stochasticity in circuitry, as addressed in this paper, we identify at least two more stochastic sources. One has been addressed already in the first part of or work. That is, the stochastic nature of neuron firing. Particularly in our infinite layer model, the solution to any posed problem exists \emph{a priori}, because possible spiking patterns combinations are represented. Clearly, any arbitrary configuration will occur in an infinitesimally low proportion. However, any random spiking pattern that is not causally supported by a synaptic network is extremely unlikely to be repeated in the following rounds of evaluation. Therefore even though it might be selected for one round, this layer will be overwritten at a later stage. Thus the stochasticity from neural spiking is not a source of heritable variation across groups. Since neurons spike with probabilities given by their input activity the only meaningful variation that can be produced is that which follows from modification of synaptic weights. However, as we have extensively argued, only this provides the substrate for selection to act on a local basis.

We call attention to a partly related model by Seung \cite{Seung:2003we} invoking `hedonistic synapses' that would release neurotransmitters stochastically, and an immediate reward would either strengthen or weaken them according to whether vesicle release of failure preceded reward, respectively. It was noted that such randomness in synaptic transmission would play the role of mutations in a Darwinian analogy. Seung also notes that stochasticity in action potentials could play a similar role, and that mechanism would be faster \cite{Werfel:2004vl}. But since ultimately these mechanisms operate on a fixed topology, the limitations without structural plasticity remain. Note that `copying' in our mechanism is a fast component, intermediate between spikes and Hebbian synaptic plasticity. This is a valid assumption if we assume that copying is aided by dedicated adaptations (cf. Adams, 1998). 

Another source of stochasticity can result from the competition amongst not an infinite but a relatively small number of groups. In this case the details of how neuronal copying occurs can be important. Some of the $N$ groups that have a solution with superior fitness score have to overwrite some groups that have lower score. This cannot be done in a proportional way, as in the infinite layer model, partly due to the topographic, rather than merely topological structure of neuronal networks. Therefore, there is some randomness of which groups are overwritten and which ones not. Naturally, the ones with higher fitness are more likely to be transcribed to other groups, but ultimately the process introduces fluctuations due to finite size. The nature of these fluctuations is simply binomial on each neuronal locus (assuming that each one is copied independently), introducing a variance of the order of $\rho(1-\rho)/N$. Thus if there are relatively few groups, the stochasticity is strong, allowing for a shift in configurations, avoiding impasses and facilitating the convergence to the optimum. This might lead us to think that fewer groups would be advantageous for this purpose. However, with very few groups, this sampling effect overrides selection, no hill climbing can occur, and thus no learning can happen. Hence, it becomes clear that there must be a compromise between these factors, or even an optimal layer number that ensures learning but facilitate impasses. (However we are of the opinion that what limits number of groups are physiological costs, not this trade-off.)

Summarising, there are three central sources of stochasticity: in spiking, in circuitry and random sampling. The first one is important for short-term learning, whilst the second and third are important for long-term learning.

\subsection{Gating can act as a selective mechanism}
It could be argued that we do not require invoking layer copying in order to solve problems or even impasses. For instance, gating mechanisms could account for the selective amplification simply by overweighting the outcome of solutions with larger score. In other words, gating can implement selection as efficiently as a more complex network to network transmission mechanism. This is because instead of increasing the number of groups that produce good solutions, a single group with such a solution is rewarded preferentially.

To visualise the difference we make an analogy. Suppose we are mixing several paint colours to reach a given tonality. For this purpose we have a stock of basic colours and a set of pipes to a recipient where the colours are mixed. Thus we need to pour a given amount $a\%$ of base paint A, $b\%$ of base paint B, etc. Suppose that $a>b$. We can increase the flow of A over B (i) through a single pipe, by adjusting the tap or (ii) we can use identical pipes all of which have the same flow, just that for A we use more pipes than for B. The former case is analogous to a gating mechanism, whereas the second case is analogous to increasing the representation of the groups.

The direct actor models \cite{Dayan:2005fr} describe a hill-climbing situation that is most consistent with this gating mechanism. This is because the rewards, being directed to neuronal activity, do not require a mechanism of neuronal copying. Rather, it is enough for two (or more) competing agents to adjust output voltage. This effectively implements hill climbing. In the case of the direct actor, gating mechanisms are conspicuously clear: learning acts directly on the strength of selection, and the closer a complex is to a solution, the stronger selection for it becomes, and the weaker the competing complexes become.
Although invoking gating seemingly renders the assumption of neuronal replication obsolete, it also poses some issues. First of all, selection through gating remains a local learning mechanism that can lead to local fitness peaks resulting on impasses. In fact, if we assume that during copying the whole content (not just one or few neuronal loci) is transferred to a new layer, the equations that describe gating or copying are the same. Second, if we want to consider SSP for long-term learning, then we require invoking a mechanism that creates and tests novel circuitry, rather than acting on the standing variation of the available strategies. In essence this mechanism has to include copying of the group in order to evaluate which circuit is better, and then discard the worse. Thus in any case, processes of neuronal copying must exist.

However this criticism does not mean we are rejecting gating as a selective mean; we just point out that it is not enough to implement the complete procedure of evolution-enhanced learning. In fact we embrace the possibility that both mechanisms, namely gating and copying, can act together. This is not only plausible for finite number of groups, but also an efficient way to modulate the signal between selection and random sampling \cite{Fernando:2008mw,Fernando:2010is}, potentially facilitating overcoming impasses.

\subsection{Size of the neuronal complexes}
The complexity of the brain is reflected by the dimension of its constituent cells (billions of neurons) and by the intricate number of synapses (on the order of trillions). In some way this accounts for the cognitive capacities of humans, although how, it is not fully clear. We have presented a hypothesis that serves as an organising principle for this complexity. However, we have considered systems that employ as few as 10 neuronal loci on each layer. Although this small number is partly motivated by computational easiness, there are reasons to think that each layer might not require excessive number of neurons. First of all, it is well known that the brain is highly modular, with different neuronal complexes allocated to specific functions. We believe that some of these modules might be specialized for processing information in the way we propose, and thus, are expected to be sub-structured into smaller functional complexes, each of which is constituted by a system of interconnected groups acting in parallel and competing to solve tasks. Thus, the actual number of neurons dedicated to any given task depends on the number of groups that are recruited for processing a given input, not just on the number of neuronal loci. Second, modular networks also facilitate rewiring because finding the right network configuration becomes increasingly harder for larger numbers of neurons. Hence, for an efficient implementation of SSP, brains might work on a modular way to facilitate rewiring of small complexes. Third, most complex tasks are likely to be split into subtasks, each of lower complexity employing relatively small circuits. In this divide \& conquer strategy, these smaller circuits can in turn be included on larger complexes to accomplish more elaborate tasks.

We may additionally argue that neuronal systems can show invariant properties, so that larger networks behave in a way similar to smaller networks. For instance, distribution of synaptic lifetimes does not depend on the size of the system, and is therefore invariant. Although we expect the time required to reach a given solution increases with the size of the system, we expect this time to scale in a particular way. This scaling may also imply that more complex problems need to recruit more groups, although this is not a straightforward requirement.

\subsection{The expansion-renormalisation model}
Kilgard proposed a verbal theory based on Darwinian dynamics that accounts also for circuitry variation and stabilization, which he termed the expansion-renormalisation model \cite{Kilgard:2012ci}. As mentioned in the introduction, the ERM assumes the generation of variant circuitry, which results in accelerated learning. He correctly points out that previous Darwinian frameworks do not take into consideration mechanisms of variation. Kilgard accounts for such variation in a verbal model; we have a similar stance regarding these ideas, but our approach is a formal one. More specifically, unlike Kilgard's work, we assume specific neuronal rules that account for circuit variability. SSP constitutes the basic process that allows circuits to be modified. However, resorting to SSP also demands understanding or assuming factors that drive SSP with the purpose of circuitry modification. We have assumed two principal means for variation of the network structures, namely establishment of new synapses and disbanding of old synapses. The former follows a higher instance of Hebb's rule, which simply means that unconnected neurons that co-spike can become connected. The specific neurophysiological processes that facilitate rewiring amongst two arbitrary neurons are unknown. Regarding the second factor of SSP, the disbanding of existing synapses, we have introduced a novel idea. That is, we assume that the disbanding probability decreases with the amount of local information. Moreover, we have shown that synaptic information is proportional to the square of the synaptic weights. This is an interesting result because it relates the mechanistic aspects to the intuitive notion of neuronal function. Together, these two mechanisms determine the dynamics of variation after layer replication.

On this line, Fauth \emph{et al.} \cite{Fauth:2015jn}  propose and analyse a model for the distribution of synapses between two neurons, by studying the interplay between Hebbian learning and SSP in a similar way we have done. They assume a constant rate of synaptogenesis for unconnected neurons, unlike the Hebbian-like mechanism we employ. Synaptic disbanding occurs with probability  \(Pr=p_o\exp(-\alpha \phi_{ij}^a ) \), where \(p_o,a\) and $\alpha$ are positive constants, which is of similar form to our's, $R$ (Eqns. \ref{Eq:SynapticInfo} and \ref{Eq:Disbanding}). Although in their case this form is not motivated by information content, they do point out that the topology of the network might constitute the basis of information storage, and that the role of this storage in memory. The topology, in turn, is determined by the balance between synaptogenesis and disbanding. Thus, although they do not explicitly assume that disbanding decreases with information capacity of a synapse, in the context of our model their model comes down to precisely that.

Kilgard's ERM hypothesizes that there must be a transient increase of circuitry variability (expansion) with a subsequent pruning of sub-optimal synapses, reducing the variation (renormalization). We have not seen evidence for this. Rather, we find an increase of circuitry variability, with an eventual stabilization. The Monte Carlo implementation allows for persistent fluctuations and occasional  `avalanches', which afterwards recover and re-establish the network functionality. Although the behaviour of Kilgard's model and ours is different, we think that the disparity is superficial. The expansion that creates a standing variation that is later reduced might occur under specific fitness landscapes and might thus be problem-dependent. In our case, since the fitness landscape has many equivalent maxima allowing for equally good solutions, there is no force that generates excess variability. However, certain types of non-linarites in the fitness landscape can certainly lead to that behaviour. It maybe that the `expansion' phase of the ERM is not a pervasive feature of neurodynamics, but rather, a context and stimulus-dependent attribute. However, we must also point out that in our implementation we only allow for circuit modifications in one layer at a time (coincidentally, as  Fauth \emph{et al.} \cite{Fauth:2015jn} do). This circuit might be either copied to all groups, or discarded altogether. By assuming that many groups can develop different circuits in the same evaluation round we might nevertheless find the expansion phase. (In fact in evolutionary models that allow for high mutation rates there can be a transient increase of genetic variability, which is equivalent to the expansion phase of the ERM; e.g. \cite{Burger:2001wy}). Finally, it may well be true that new synapse formation is adaptive in that its rate is increased by the appearance of novel tasks, for which there is some evidence \cite{Levy:2004sy,Butz:2009qq,Butz:2014dz,Butz:2014pi}. This provocation-based mechanism would easily lead to the expansion phase. 

\subsection{Towards replicative neurodynamics}
In this article we theoretically test a novel mechanism for neural function. We have found a crucial synergy between learning and fitness climbing, strengthening previous, related findings \cite{Fernando:2010is}. We have used simple models to show that the combination of selection and learning is an extremely efficient one. Moreover, we also showed the relevance of SSP in this context: modifications of the learning topology of networks. However, there is another open possibility that can result from a combination of learning and plasticity. We showed that certain networks are in general terms more efficient learners. Thus, the recruitment of an existing, efficient network could in principle lead through synaptic plasticity to the copying of its structure in the current network. In an analogous way in which DNA is copied, an existing network could replicate and such a structure would spread, allowing for problem solving. This has been previously proposed as the \emph{neuronal replicator hypothesis} \cite{Fernando:2008mw,Fernando:2010is,Fernando:2010dn,Fernando:2010ul}. The idea is very recent, reason why there has still been no experimental verification. But in this article we advanced the mechanisms that justify the neural replicators. It remains open to study how the copying of the network topologies can occur.

Related to the issue of exponential strengthening versus exponential replication, the path evolution algorithm by Fernando \emph{et al.}. \cite{Fernando:2011ey} is a remarkable suggestion. In that model, neurons along a path are assumed to code for some behaviour. Whilst neuronal activity is fixed, paths grow collaterals and thus recruit new nodes. Neuronal activity can spread along different paths probabilistically that can be evaluated and compared according to some performance (fitness) measure. Good paths become strengthened by reward, whereas bad ones are weakened. Various paths can have few or many common neurons. This algorithm explicitly incorporates structural plasticity and selection and, despite the differences, is thus the closest precedent to our model. We improve on that model in two respects: we present a mathematical framework (as opposed to mere simulations) that takes the first steps to unite theory of learning with that of natural selection, and we consider recurrent networks that posed a special problem for path evolution.

\subsection{Mutations and recombination as creative sources}
We should call attention to two possible usage of the term `mutation' in the neuronal context. One we have seen before: stochasticity in firing or transmitter release. Another one structural plasticity itself: the term `synaptic mutation' was coined by Adams\cite{Adams:1998ft} in this latter sense, who by the way foresaw the potential importance of the phenomenon for the performance of the nervous system. Note that \cite{Fernando:2011ey} in their path evolution model use structural plasticity to implement `crossover' between different paths.

Although visually reminiscent to genetic recombination, the synaptic mutations and the path crossovers are not formally equivalent to DNA crossover. In genetics, recombination does not create new allelic variability (on/off probabilities). Instead, it reshuffles the existing variants at any given locus. This certainly results in a `macromutation' at the phenotypic level, but the genetic variability of the population remains intact. Thus, recombination does not increase allelic variation, but it does increase variation across groups. The distinction is important in the context of our work: we assume the equivalent to high recombination. Namely, at any given neuronal locus, the copying can occur from any other group, irrespective of the state of the other neuronal loci. This provides the highest rate of reshuffling, and is thus `creative'. The contrary limit is when only the complete content of a selected group can overwrite an out-selected group. This is an `asexual' limit in that there is no recombination. The latter provides the fastest selective response, but is less creative because it has no combinatorial power.

\subsection{Relationship to evolvability}
Adam's synaptic mutations and Fernando's path-crossover are analogous to modifications in the architecture of traits. This is a more powerful type of macro-mutation because it truly modifies the decoding of the information stored in neuronal states, in an equivalent way as development decodes the genetic information into a trait, which is one of the fundamental aspects of evolvability.

Evolvability is understood as the potential of a population to respond to selection. How fast the response to selection is depends on the amount of genetic (or heritable) variation that can be produced. This can be given by standing variation, cryptic variation (due to epistasis, for example), or due to mutational variance. Although high mutation rates will provide source `material' to respond to selection, these will also create load that keeps the population maladapted. However, the optimal scenario is achieved if mutation rates can be increased as selection is started, and tuned down once the population approaches adaptation.

	As we saw above, this is precisely what happens with the neurodynamics we have described. Of course, genetic systems do not have a learning mechanism as the brain does. Nevertheless, these are analogous. We want to bring the analogy further and interpret that the input current $E$ as a quantitative trait, with the weights $\phi$ taking the role of additive effects.

	A previous model in quantitative genetics has taken an approach reminiscent of ours \cite{Jones:2007ub}. They did not apply a learning mechanism, but considered modifier alleles for the mutational effects. These are selected indirectly, increasing the transition rate in the direction of the highest increase in fitness, in a way that is analogous to our switching probability, which generates variability in order to improve fitness increase.

\subsection{Information storage in neuronal circuits}
Understanding the relationship between information capacity and synaptic changes is central in order to understand learning, memory and other aspects of cognition \cite{Chklovskii:2004eb,Caroni:2014ey}. Under Hebbian learning, information storage relies solely on the modification of synaptic weights and is contingent on the existing connections. In the SSP scenario, the information capacity of neuronal complexes is adjusted through modification of the connections \cite{Poirazi:2001uy}. Despite the advantages of this mechanism in learning, it poses a challenge because of all possible synapses that can exist between every pair of neurons, only a small fraction is realised, even within a complex.

Even granting that SPP is a mechanism for the exploration of alternative circuits, discovering the right synaptic configurations in this vast combinatorial space is a major problem. Merely grasping the efficiency of the brain in managing multiple tasks of combinatorial complexity demands an understanding of the mechanisms behind such capabilities. In other words, how is the brain implementing algorithms for searching the combinatorial space of solutions?

Our model introduces means to explore the combinatorial space, implementing mechanisms that are analogous to biological evolution. From this perspective, understanding the algorithmic means used by the brain comes down to understanding and identifying what constitutes the units of selection and, crucially, what are the units of variability. This has been one of the central questions of our research, which we have clarified by employing our analogy with genetic systems.

Understanding the physiological basis of learning and cognition requires the identification of the modifications that occur during learning, and which result on specific circuitries directed by different stimuli and experiences. However, we point out that there is a fundamental relationship between units of learning and units of variability, which we must consider in order to understand and identify which are the units of learning.

In our model information content is stored not in the activity of neurons, but in the synaptic weights and switching probabilities. This has two important implications. First, this suggests that the loci of memory are circuits (not neurons), even though the mapping between memory loci and cognitive functioning might be mediated through the coordinated spiking if individual neurons (cf. \cite{Rolls:2012ix}).

However, most importantly, we have shown that SSP is the process that directly mediates establishment and disbanding of synapses resulting in circuits that represent solutions to specific problems. Previous findings also support the notion that information is stored in circuits, not neurons \cite{Poirazi:2001uy,Stepanyants:2002tr}. Furthermore, at a higher cognitive level, it has been proposed that consciousness can be gauged through information integration measures between neuronal complexes \cite{Tononi:2004ki}. At a phenomenological level it is well-known that the SSP at the level of both spine growth and modifications of the synaptic networks is directly inflected by sensory experience \cite{Trachtenberg:2002cy,Caroni:2014ey}, or by manipulating neuromodulators \cite{LeBe:2006eq}. Despite these lines of evidence, which are compelling for our theory, we still require and lack direct experimental verification regarding the minimal complexity in the circuit distribution that results from solving particular tasks. 

The second important implication is the procedural relationship between learning and variability. Even when these two are not the same and they constitute fundamentally two different processes, we have shown how learning fine-tunes the generation of variability. At the synaptic level, Hebbian learning modifies switching probabilities, which are the mechanism for generating variability in spiking. At the level of circuitry, SSP dictates longer-term changes, where informative synapses persist and uninformative synapses are disbanded. Altogether these two processes mediate the exploration of the complex combinatorial space by generating the required variability, guided by learning. These are the `fuel' for the motor that results in effective changes, which are, ultimately, selection mechanisms.

Although the question of neuronal and circuit information storage has been extensively discussed, to our knowledge, we are the first to consider informational aspects as mechanisms that direct synaptic `survival' and lifetimes (see below). By showing that synaptic information between two neurons is proportional to the learning weight and to the switching probabilities, we identified a plausible and verifiable mechanism through which SPP can be directed. This is crucial because we still ignore the causes and consequences of structural plasticity at a cognitive level. Still our hypothesis is supported by known facts, chiefly the experiences-driven SSP, where dendritic spine growth is observed \emph{in vivo} in the neocortex of adult rats \cite{Trachtenberg:2002cy,Poirazi:2001uy,Stepanyants:2002tr}. 

\subsection{The lifetime of synapses}
Despite the known potential of neuronal complexes to undergo experience-driven synaptic network restructuration \cite{Buonomano:1998oe,Knott:2002xp}, we still ignore how frequently synaptic connections are modified in adult brains \cite{Caroni:2014ey}. On the basis that the number of synapses remains practically constant during adulthood, it has been long argued that the rates of synapse establishment and disbanding balance each other. Our result indicates that the length of synaptic lifetimes follows a statistical distribution, and the question of synaptic stability is a quantitative one rather than a yes/no statement. Once a stationary state has been reached, the number of connections remains more or less constant with an unchanging distribution of synaptic lifetimes. Even then, circuits are by no mean fixed but, rather, show quite a degree of dynamism.

The SPP model of Fauth \emph{et al.} \cite{Fauth:2015jn} studies the equilibrium distribution of synaptic connections between two neurons. In their work, as in ours, the general shape of the distribution of synapses is robust to model parameters. However, they analyse synaptic stability between two neurons, not the general distribution of synapses of a network (whereas they allow multiple synapses between neuron pairs, we do not take into account this possibility).

Through three-dimensional reconstruction of cultured neocortical cells it was determined that 75\% of the connections can be explained by assuming that pairs of neurons connect randomly. However, the remaining 25\% of the connections requires invoking function, anatomic differences and positioning, amongst others factors that facilitate chemical mechanisms for attraction or repulsion in order to complete or avoid the establishment of new synapses \cite{Hill:2012fl}.

Direct experimental information regarding the rates of structural plasticity comes from murine models. Although dendritic spines in the cortex of adult brains tend to be rather stable, a proportion of them show considerable variation, and experience can induce not only their growth or shrinkage \cite{Boothe:1979nr} but also \emph{de novo} formation or elimination of spines, which, importantly, occur at balanced rates. Importantly, these structural spine dynamics are known to be similar in distinct parts of the brain \cite{Boothe:1979nr,Bourgeois:1994fp,Rakic:zw}. Consequently, it is safe to assume that the physio-anatomical principles, if not the rates, that maintain the distribution of spine size and numbers are similar in distinct cortical regions .

Although direct observation of synaptogenesis is more elusive (mostly due to technical limitations), it is known that dendritic spines sometimes result in new synapses \cite{Toni:1999kr}. Changes in axon terminals occur over several weeks in young brains \cite{Purves:1985xw}. A compelling example is that of the visual cortex at early ages, where synapses established during he developmental critical period are stable for more than 13 months \cite{Grutzendler:2002jf}.

\section{Numerical and Simulation Methods}
\subsection{Neuronal dynamics}
To simulate the time evolution of a complex of multiple groups (effectively infinite in number), where each has a fixed number of $n$ of neurons, we solve a system of coupled ordinary differential equations where n of them represent the change in spiking probability, and, along with it, a set of differential equations that describe the change in learning weight (i.e. Oja's rule for Hebbian learning). The number of learning equations depends on the connectivity of the learning network, which we assume undirected. The initial conditions for the spiking equations are random deviates from a uniform distribution between zero and one (unless otherwise stated). The initial conditions for the learning equations are random deviates from a uniform distribution between $(0 , 0.01]$, unless otherwise stated. The system of equations are solved numerically for $t=10000$ time units. (This number ensures convergence of the system for all parameters used, and overall is considered to be on the order of $\sim 10$ ms. Although we could choose $S$ and other parameters such that the system is measured in the relevant units, numerically it is more convenient to employ the mentioned scale). All simulations were implemented and solved in \emph{Mathematica} 9.0 and/or 10.0.

\subsection{Random networks}
To explore learning network topologies, we generated random graphs from three classes of distributions. The first one is the Erd\H{o}s-R\'enyi model (ER), where nodes are connected randomly. This model assumes a fixed number of nodes n and certain probability r that each node is connected to any other node. The second model is the Barab\'asi-Albert (BA), famous for its scale-free properties. The BA model also employs two parameters that control the network topology: the fixed number of nodes n and number k of vertices that are preferentially attached to each node. The third model is the Watts-Strogratz (WS), or so called small-world networks, which takes as parameters n nodes and a probability r of rewiring a vertex amongst two nodes in such a way as to avoid loops. In all cases we forbid multiple edges and self-connections. These network models are built-in \emph{Mathematica}, and employed as indicated in the software's Documentation Centre.

\subsection{Structural synaptic plasticity}
We assume that Hebbian learning happens faster than SSP, leading to a separation of time scales. Consequently, we allow the learning equations to reach equilibrium, and only then modify the learning network. We consider the three modifications of a network indicated in the Model section. First, we allow synapses to be formed by adding an edge amongst two unconnected nodes i and j with a probability \(q_{ij} \propto \rho_i \rho_j\). In this way, neurons that co-fire tend to be wired together. Second, we allow synapses to be eliminated by randomly removing edges between two connected neurons $i$ and $j$ with probability $R_{ij} = \exp[-\alpha H_{ij}]\), so that neurons that do not co-fire tend to be disconnected. Third, we also allow random rewiring (irrespective of firing probabilities) with a small probability $u=0.01$: we randomly and uniformly choose a connected pair $i,j$ and eliminate the edge, and at the same time choose an unconnected pair l,m and establish an edge. In each time step any (including all) of the above events are allowed to happen. Once the networks have been rewired, a new round of learning is performed. Initial conditions may or may not be modified (see Results section). After a new equilibrium is reached, the new fitness is compared to the fitness before the rewiring. This is implemented through a Metropolis algorithm: if the fitness is increased, the change is accepted, but if the fitness decreases, the change is accepted with probability proportional to the ratio of new to old fitness. We additionally impose a multiplicative fitness cost per synapse of $\exp[-kd]$, where $k$ is the penalty of each edge in the network, and d is the number of edges of a given network. We typically run the simulations for at least 250 steps, to ensure convergence. However, in different experiments larger step numbers are used, as indicated in the figures.

\section{Acknowledgments}
This work was financially supported by the EU FET-open project \href{http://www.insightproject.eu}{`INSIGHT'}, agreement number no 308943, by the European Research Council project `EvoEvo', grant agreement no 294332 and by the Templeton Foundation project `Darwin and Hebb', FQEB Grant \#RFP-12-19.

\clearpage

\appendix
\section{Hebb's and Oja's rule on the group ensemble}
\label{SI:HebbianWeights}

In this appendix we want to show that we can approximate the ensemble of weights by their mean. We will show that under certain assumptions, Hebb's and Oja's rules apply on the average, plus some variance terms. We argue that the variance terms are small and thus the average learning is enough. This approximation allows a tractable analysis of the neurodynamics without needing to follow the full distribution of associative weights in the ensemble.

We first work out the simpler case of Hebb's rule (Eq. 1 in the main text) and then Oja's rule (Eq. 2 in the main text). We interpret the change in the average weight as the average weight change, i.e.
\begin{equation}
\frac{d \langle \phi_{ij} \rangle}{dt} \simeq \left\langle \frac{d \phi_{ij}}{dt} \right\rangle ~,
\end{equation}
where $\langle\ldots \rangle$ denotes average on the ensemble population (for simplicity we will also use the `bar' notation, e.g. $\overline{\phi}$ to denote the same quantity. The average is in principle taken on a joint distribution \(\mathcal{P}[\bm{\psi},\bm{X}]\), where the bolds indicate the vector of values (weights and neuronal states). However, we will make the simplifying approximation that all these quantities are independent. In population genetics this corresponds to the `Hardy-Weinberg' equilibrium, which we take throughout this article.

\paragraph{Mean-field Hebb's rule}
From Eq. 1 in the main text we have that
\begin{equation}
\frac{d \overline{\phi}_{ji}}{dt} = \lambda \left\langle X_i Y_j \right\rangle.
\end{equation}
Using the definition of the activity, \(Y_j = \sum_{k \neq j}^n \phi_{jk} X_k\) we have that 
\begin{equation}
\frac{d \overline{\phi}_{ji}}{dt} = \lambda \left(\sum_{k \neq j}^n \left\langle X_i \phi_{jk} X_k  \right\rangle + \left\langle\phi_{ji} X_i^2  \right\rangle  \right).
\end{equation}

We now use the independency assumption to  further develop the two averages. The first term is
\begin{equation}
\left\langle X_i \phi_{jk} X_k  \right\rangle = \left\langle X_i  \right\rangle  \left\langle\phi_{jk} \right\rangle  \left\langle X_k  \right\rangle = \overline{\phi}_{jk} (2\rho_k-1)(2\rho_i-1)
\end{equation}
(in as long as $i\neq j$). The second term is
\begin{equation}
 \left\langle\phi_{ji} X_i^2  \right\rangle = \left\langle \phi_{ji}\right\rangle  \left\langle X_i^2  \right\rangle =  \overline{\phi}_{ij}.
 \end{equation}

Now we put together the two expressions, complete the sum to include terms $i$ and some algebra gives that
\begin{equation}
\frac{d \overline{\phi}_{ji}}{dt} = \lambda \left( (2\rho_i-1) \sum_{k \neq j}^n \overline{\phi}_{jk} (2\rho_k-1) + \overline{\phi}_{ji}4\rho_i(1-\rho_i) \right)
\end{equation}
Note that $(2\rho-1)$ is the average input activity and that $4\rho_i(1-\rho_i)$ is its variance. Then 
\begin{equation}
\label{Eq:AverageHebbsRule}
\frac{d \overline{\phi}_{ji}}{dt} = \lambda \bar{X}_i \bar{Y}_i + \lambda \overline{\phi}_{ji}\text{var}(X_i),
\end{equation}
where the notation $\bar{Y}$ is a shorthand for the output of the average activity (and not the average output, which is a correlation between $X$ and $\phi$).

The last expression shows that, to a first order approximation, Hebb's rule can be applied to the average. In most cases  var($X_i$) is small. Moreover, once the system has substantially or fully learned, then $\rho=0,1$ making the second term vanish.

$\blacksquare$

\paragraph{Mean-field Oja's rule}
We proceed in a similar way as above, averaging over Oja's rule (Eq. 2 in the main text). First note that there are two terms:
\begin{equation}
\frac{d \overline{\phi}_{ji}}{dt} = \lambda \left(  \left\langle X_i Y_j \right\rangle  - \left\langle \phi_{ji} Y_j^2 \right\rangle \right); 
\end{equation}
the first terms is given by the right-hand side of Eq. \ref{Eq:AverageHebbsRule}. Thus we concentrate in working out the second term. We begin by expanding the $Y^2$:
\begin{equation}
 \left\langle \phi_{ji} Y_j^2 \right\rangle = 2 \sum_{k\neq l\neq j} \left\langle \phi_{ji}  \phi_{jk} \phi_{jl} X_k X_l\right\rangle + \sum_{k\neq j} \left\langle  \phi_{ji}  \phi_{jk}^2 X_k^2  \right\rangle.
\end{equation}
Now we use the assumption of independency. However, we must note that not all the terms on the first sum are independent, bacause some symbols can repeat. We first write
\begin{equation}
 \left\langle \phi_{ji} Y_j^2 \right\rangle = 2 \sum_{k\neq l\neq j} \left\langle \phi_{ji}  \phi_{jk} \phi_{jl} \right\rangle (2\rho_k-1)(2\rho_i-1) + \sum_{k\neq j} \left\langle  \phi_{ji}  \phi_{jk}^2 \right\rangle
\end{equation}
Now consider that
\begin{equation}
\left\langle \phi_{ji}  \phi_{jk} \phi_{jl} \right\rangle = \left\{ 
\begin{array}{cc}
 \overline{\phi}_{ji}   \overline{\phi}_{jk}  \overline{\phi}_{jl}  & \text{if }  k\neq l \neq i \\
  \left\langle \phi_{ji}^2\right\rangle  \overline{\phi}_{jl} &  \text{if } l\neq k=i   \\
  \left\langle \phi_{ji}^2\right\rangle  \overline{\phi}_{jk} &  \text{if } k\neq l=i 
\end{array}
\right.
\end{equation}
and in the second sum
\begin{equation}
 \left\langle  \phi_{ji}  \phi_{jk}^2 \right\rangle = \left\{
\begin{array}{cc}
   \overline{\phi}_{ji}  \left\langle \phi_{jk}^2 \right\rangle & \text{if } k\neq i \\
 \left\langle  \phi_{ji}^3  \right\rangle  &  \text{if } k = i     
\end{array}
\right.
\end{equation}

This separates the two sums into five sums. The rest of the derivation is somewhat lengthy but straightforward. It continues by completing the missing terms of sums and then simply rearranging, and it leads to
\begin{multline}
 \left\langle \phi_{ji} Y_j^2 \right\rangle = \overline{\phi}_{ji} \overline{Y}_j^2 + \text{var}(\phi_{ji}) \overline{X}_i\overline{Y}_j  +\overline{\phi}_{ji}\sum_{k\neq j}  \left\langle \phi_{jk}^2 \right\rangle(\text{var}(X_k)-1) \\ +\overline{\phi}_{ji} \text{var}(\phi_{ji})(\text{var}(X_i)-1) +  \left\langle \Delta \phi_{ji}^3 \right\rangle
\end{multline}
where \( \left\langle \Delta \phi_{ji}^3 \right\rangle=  \left\langle \phi_{ji}^3 \right\rangle - \overline{\phi}_{ji}\left\langle \phi_{ji}^2\right\rangle \).

Finally, we note that altogether we can write the mean field on Oja's rule as the rule of the average activity:
\begin{equation}
\frac{d \overline{\phi}_{ji}}{dt} = \lambda \overline{Y}_j(\overline{X}_i - \overline{\phi}_{ji} \overline{Y}_j) + \lambda \times \text{variance terms}.
\end{equation}

$\blacksquare$

\section{Mutual information contained in a synapse}
\label{SI:MutualInformation}

Mutual Information is defined as

\begin{equation}
\label{Eq:GenMutualInfo}
H[I,J]=\sum _{r,s\in \{0,1\}}  \Pr \left[X_i=r\left|X_j\right.=s\right]\Pr \left[X_j=s\right]\log\left[\frac{ \Pr \left[X_i=r\left|X_j\right.=s\right]}{\Pr
\left[X_i=r\right]}\right]
\end{equation}
The unconditioned probabilities are simply the `allele' frequencies. Namely, 
\begin{eqnarray}
\Pr \left[X_j=1\right]=\rho _j\\
\Pr \left[X_j=0\right]=1-\rho _j
\end{eqnarray}

\paragraph{Conditional spiking and switching probabilities.} To compute the conditional probabilities we evaluate the activity of the neuron $i$ with the conditioned value for the neuron $j$, namely
\begin{equation}
Y_{i|j}=\sum _{k\neq i,j} \phi_{ki}\left(2\rho _k-1\right)+\phi_{ji}\left(2X_j-1\right).
\end{equation}
This leads to a conditioned switching probability \(M_{i|j}\) 
\begin{equation}
M_{i|j}=\left(1+\exp\left[Y_{i|j}\right]\right)^{-1}
\end{equation}
This conditional switching probability leads to a dynamical equation with the new activity, whose equilibrium frequencies are \(\Pr \left[X_i=1\left|X_j\right.=s\right]=\rho _{i|j}\) and \(\Pr \left[X_i=0\left|X_j\right.=s\right]=1-\rho _{i|j}\). The conditional frequency is  by the equilibrium of
\begin{equation}
\frac{d\, \rho _{i|j}}{d\, t} = \rho _{i|j}\left(1-\rho _{i|j}\right)\partial _{\rho _i}\log\left[\bar{W}\right]+M_{i|j}\left(1-2\rho _{i|j}\right)
\end{equation}
For the directional selection case the solution is
\begin{equation}
\label{Eq:DSeqFreq}
\rho_{i|j}=\frac{S-2M_{i|j}+\sqrt{4M_{i|j}^2+S^2}}{2S},
\end{equation}
and for stabilising selection
\begin{equation}
\label{Eq:SSeqFreq}
\rho _{i|j}=\frac{1}{2}\left(1+\sqrt{1-\frac{4M_{i|j}}{S}}\right).
\end{equation}

We evaluate on the  evolved weights because we want to measure how one neuron is dependend on one other. Clearly another possible mutual information
measure is to allow the whole network to relax with the constraint (including weights). However, this global measure asseses the capacity of the
whole neuronal complex. By evaluating only the local mutual information we measure directly the information content contained in it.

Note that \(Y_{i|j}\) can be written as
\begin{multline}
Y_{i|j}=\sum _{k\neq i,j} \phi_{ki}\left(2\rho _k-1\right)+\phi_{ji}\left(2X_j-1\right)= \\\sum _{k\neq i} \phi_{ki}\left(2\rho
_k-1\right)+\phi_{ji}\left(2X_j-1\right)-\phi_{ji}\left(2\rho _j-1\right)\\
=\sum _{k\neq i} \phi_{ki}\left(2\rho _k-1\right)+2\phi_{ji}\delta \rho _{\left(X_k\right)}
\end{multline}
with \(\delta \rho _{\left(X_k\right)}=X_k-\rho _k\).  Consequently, we can express the conditional switching probability as
\begin{equation}
M_{i|j}=\left(1+\exp\left[Y_i\right]\exp\left[2\phi_{ji}\delta \rho _j\right]\right){}^{-1}.
\end{equation}
We assume that both \(\phi_{ji}\) and \(\delta \rho _{\left(X_k\right)}\) are small
and expand \(M_{i|j}\) in series of $\delta \rho$, to get
\begin{equation}
\label{Eq:CondSwitchProb}
M_{i|j}=M_i-2M_i\left(1-M_i\right)\phi_{ij}  \delta \rho _j+2M_i\left(1-M_i\right)\left(1-2M_i\right)\phi_{ij}^2\delta \rho _j^2
\end{equation}
Since this last expression does not depend explicitly on \(\rho_i\), the equilibrium value of \(\rho _{i|j}\) is algebraically the same as that of \(\rho _i\) but
with \(M_i\to  M_{i|j}\) (Eqns. \ref{Eq:DSeqFreq} and \ref{Eq:SSeqFreq}).

\paragraph{Approximating mutual information.} The sum in Eq. \ref{Eq:GenMutualInfo} resumes four terms that we can write explicitly:
\begin{multline}
H[I,J]=(1-\rho _{i|0})\left(1-\rho _j\right)\log\left[\frac{ (1-\rho _{i|0}) }{\left(1-\rho _i\right)}\right] \\
+\rho _{i|0}\left(1-\rho _j\right)\log\left[\frac{ \rho _{i|0} }{\rho _i}\right]\\
+(1-\rho _{i|1})\rho _j\log\left[\frac{ (1-\rho _{i|1}) }{\left(1-\rho _i\right)}\right]\\
+\rho _{i|1}\rho _j\log\left[\frac{\rho _{i|1}}{\rho _i}\right]
\end{multline}

Substituting Eqns. \ref{Eq:DSeqFreq}/\ref{Eq:SSeqFreq} and \ref{Eq:CondSwitchProb} into the last formula gives the exact expression for $H[I,J]$, which is lengthy and complicated. Using automated algebra software (\emph{Mathematica} 10.0) we perform a series expansion on $\phi_{ij} $ to second
order. Using the fact that $S>>M$ we get that, in both selective scenarios:
\begin{equation}
H[I,J]=\frac{2 M_i M_j \phi_{ij}^2}{S^2}+\mathcal{O}\left(M^2\right)
\end{equation}
$\blacksquare$

\section{Dependency of the dynamics on the initial conditions of neuronal spiking}
\label{SI:InitialConditions}

In this appendix we present results that show that once a system learns the landscape, the retrieval of the information is much quicker than learning, and it is independent of the initial conditions of the spiking probabilities.

The way we assess this quick retrieval effect is by first solving the full dynamics. Once in equilibrium, we fix the Hebbian weights and compute the corresponding switching probabilities $M_i$ at each neuronal locus.

After the learning phase has been completed, we run the dynamics for the neuronal probabilities $\hat{\rho}_i$ with this fixed $M_i$'s but by choosing random initial spiking frequencies  (uniformly in the interval $(0,1)$). We measure the state of the system relative to the fixed point of the learning neurodynamics, $\rho^*$, and calculate the euclidian distance from this state: \[D_{fp} = \sqrt{\sum_i^n \left[\rho^*_i - \hat{\rho}_i(\tau)\right]^2} .\] Because by time $1/S$ the $\hat{\rho}_i$ have increased in representation, by time $\tau=S$  ($1/S< S <<1/\lambda)$ they are very close to the fixed point, and we expect $D_{fp}$ to be very close to zero.

\begin{figure}
\includegraphics[width=\textwidth]{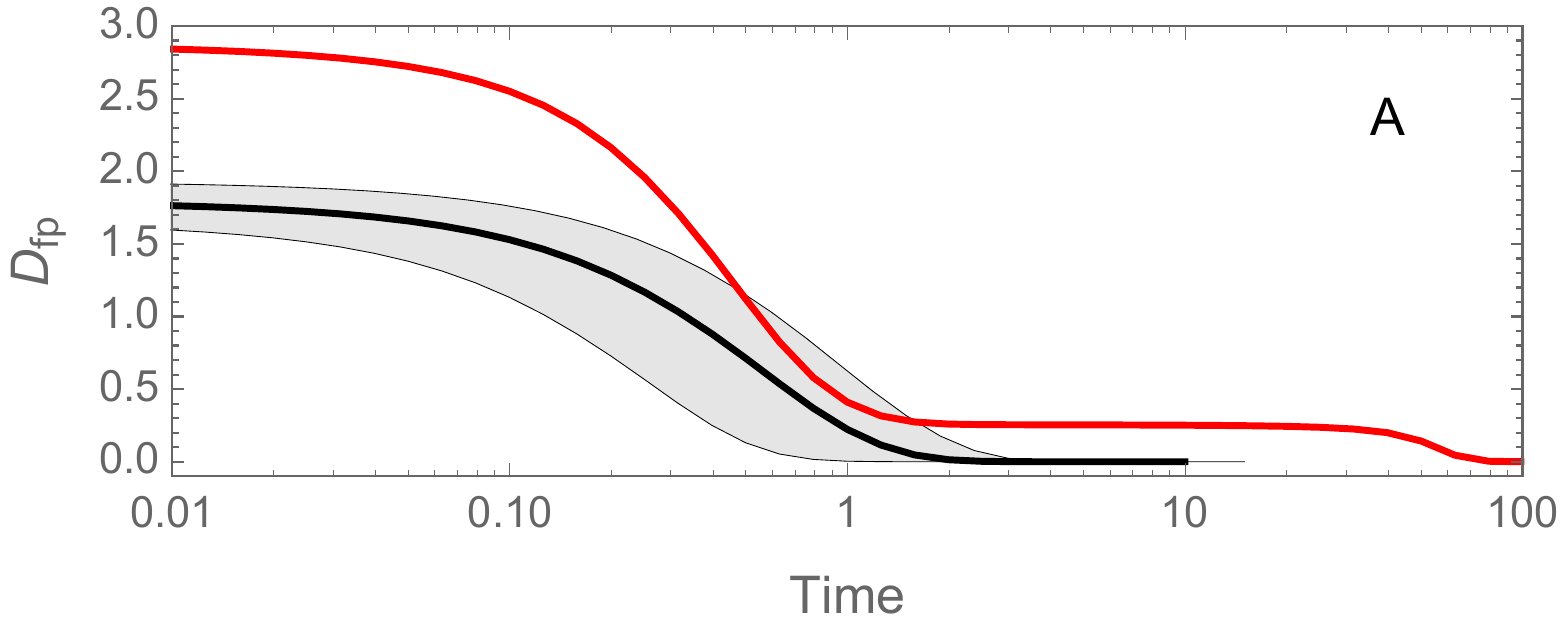}
\includegraphics[width=\textwidth]{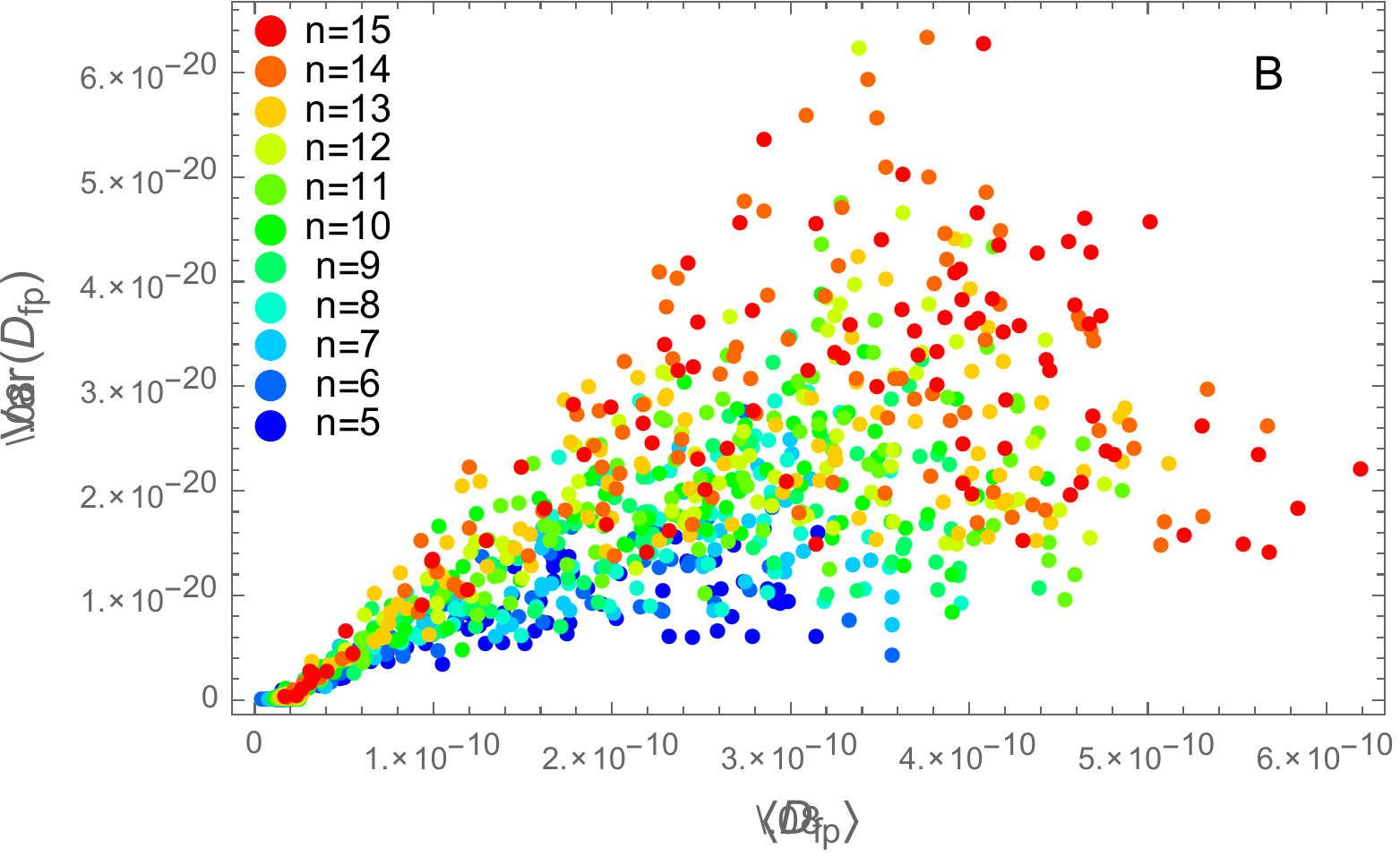}
\caption{\bf Effect of initial conditions on the dynamics of learning.}
(A) Eucledian distance to the fixed point $D_{fp}$ of the learning dynamics (red) and the average on an ensemble (100 samples) of dynamics with learnt weights (black) with random initial conditions for the spiking probabilities. The shadowed region in grey covers the standard deviation. In this example $n=10$ ; the landscape is asymmetric with gradients $S_i$ given by $\{3.1, 3.2, 4.6, 4.8, 4.8, 4.9, 6.2, 7.8, 8.1, 9.9\}$.
(B) Mean and variance of the distance to the fixed point $D_{fp}$ of an ensemble of simulations under different numbers of neuronal loci $n$ (see inset legend) at time $\tau=S=10$. Each point is an average over 30 runs with learnt weights and randomised initial conditions for the spiking probabilities with distribution $U(0,1)$. For each of these ensembles the landscape is a random vector with each component distributed as $U(0,S]$. Other parameters as in Fig. 3 in the main text.
\label{Fig:ICdependency}
\end{figure}

Figure \ref{Fig:ICdependency}A presents the results form simulations where we randomise initial conditions after learning. The figure shows how the average and the variance of the distance (over multiple realisations) to the fixed point quickly decays to zero, and by time $\tau=S$ it is already negligible. Figure \ref{Fig:ICdependency}B shows the mean and variance of a set of ensembles under number of different neuronal loci $n$ and different fitness landscapes.

\section{Scaling of selection intensity and associative weights}
\label{SI:MSscaling}

We show here the scaling of Fig. 3 in the main text.

First, consider that of all neuronal loci there is one that has the maximum switching probability, which we denote with the subscript $m$. We denote the relative switching probability as $\mathcal{M}_i$, which by the definition of $M$ is
\begin{equation}
\mathcal{M}_i = \frac{M_i}{M_m} = \frac{1+e^{E_m}}{1+e^{E_i}}~.
\end{equation}
As a first approximation we note that for large activity values
\begin{equation}
\mathcal{M}_i \simeq \exp(E_m-E_i)~.
\end{equation}
The activity difference $\Delta_i=E_m-E_i$ is:
\begin{equation}
\Delta_i=E_m-E_i = \sum_{j \neq m} \phi_{mj}(2\rho_j-1) - \sum_{k\neq i}\phi_{ik}(2\rho_k-1)~.
\end{equation}
The two sums overlap at all but two indices: $j=i$ and $k=m$, respectively. Hence
\begin{equation}
\Delta_i=  \phi_{mi}(2\rho_i-1) -\phi_{im}(2\rho_m-1)  + \sum_{j \neq m,i} (\phi_{mj}-\phi_{ij})(2\rho_j-1)~.
\end{equation}

As a second approximation we assume that weights are of the order $\phi \sim (n-1)^{-\frac{1}{2}}$. Consequently the sum is negligible and
\begin{equation}
\Delta_i \simeq 2 \frac{\rho_i - \rho_m }{\sqrt{n-1}} ~.
\end{equation}
The equilibrium condition of the neurodynamical equation gives
\begin{equation}
\rho = \frac{1}{2} \left(\sqrt{4 \mu^2+1}-2 \mu+1\right)
\end{equation}
where $\mu = M/S$. Assuming that $S>>M$ we have that to first order in $\mu$, $\rho \simeq 1-\mu$. Thus,
\begin{equation}
\Delta_i \simeq 2 \frac{\mu_m-\mu_i}{\sqrt{n-1}} ~.
\end{equation}

A third approximation follows: since $\mu_m$ is the term with the largest $S$, then, compared to most $\mu_i$, it is negligible. Therefore \(\Delta_i \simeq - 2 \mu_i/ \sqrt{n-1}\).

Summarising, the relative switching probability is:
\begin{equation}
\mathcal{M}_i \simeq \exp\left[ - \frac{2 \mu_i}{ \sqrt{n-1}}  \right]~.
\end{equation}
Now, simply call \(\hat{S} = \sqrt{n-1} S / M\), so we get the scaled relationship
\begin{equation}
\mathcal{M}_i \simeq \exp\left[ - 1/\hat{S} \right]~.
\end{equation}
$\blacksquare$

\section{Effect of the topology of the network}
\label{SI:NetworkTopology}

The number of neurons in the brain is vast; yet it is very sparsely connected, with a modular nature. Empirical measurements suggest small world topologies. However, what lies behind this distribution, and how truly `random' it is, remains unclear. From the point of view of our framework, we can evaluate what effect different topologies have on the neurodynamics.

We can interpret the random topologies in two ways. First, under the Changeux selective (but not evolvable) scenario where neuronal circuits are fixed, we can think that from the point of view of an arbitrary cognitive task,  any pre-established circuitry is essentially random. Second, under the `neuronal replicator hypothesis' , we can think that once some neuronal groups allocated for a new task, has a current state of the circuitry optimised for a  previous task. However from the point of view of the new task, the current configuration is random and potentially uninformative.

In this supporting text, we study how random topologies affect the neurodynamics. We consider learning networks that have topologies drawn from random network distributions, but which remain constant during the learning process. For any given random topology we run the selection-learning system on a stabilising landscape with a randomly placed optimum value (which also remains constant during a run). We consider systems with between 5 to 30 neuronal loci. 

\begin{figure}
\includegraphics[scale=0.35]{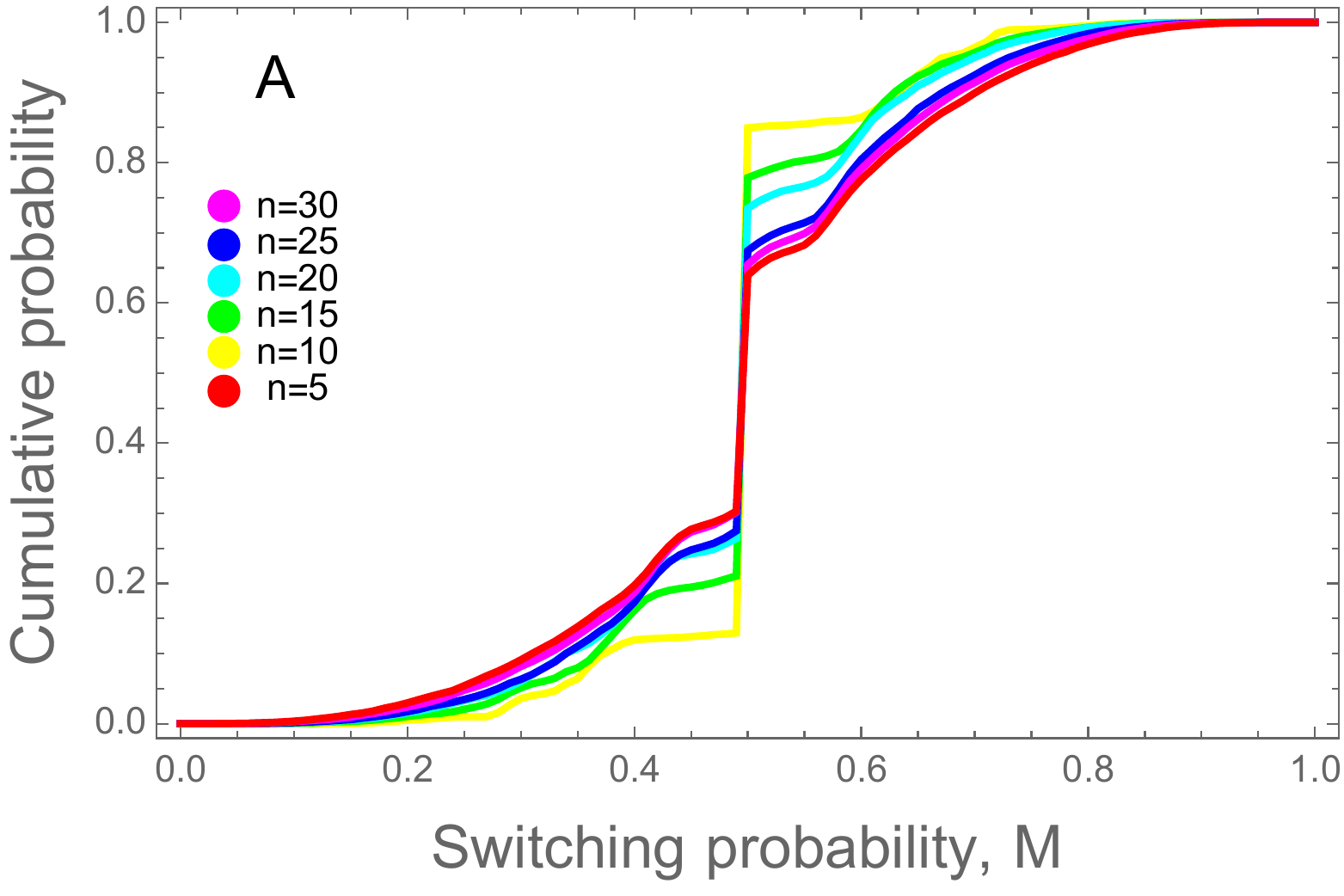}
\includegraphics[scale=0.35]{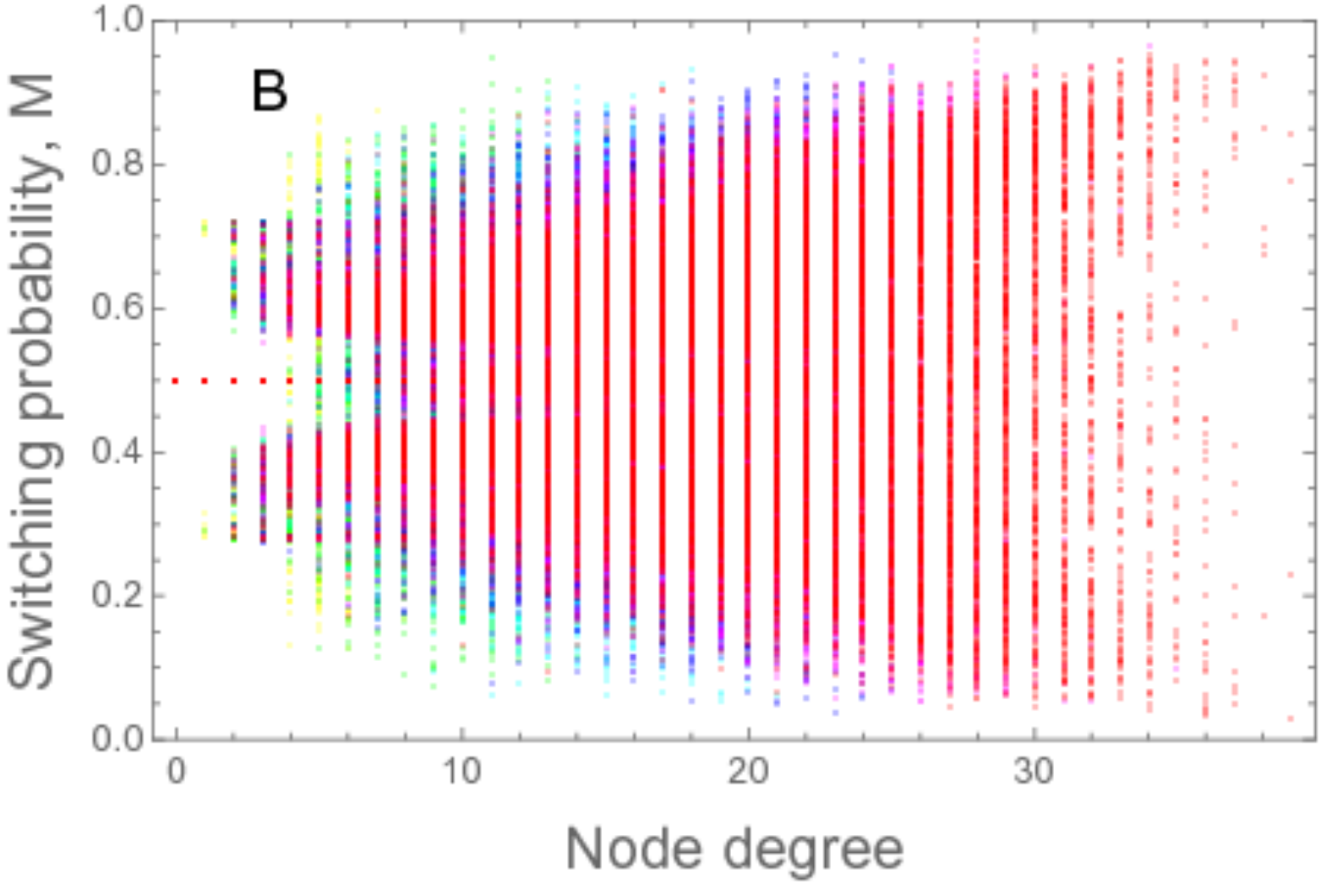}
\caption{\bf Distribution of switching probabilities of neuronal circuits with  Erd\H{o}s-Renyi topologies.}
(A) Empirical probability function on the outcome of the selection-learning process. (B) Relationship between switching probability and the degree of each neuronal locus. In each simulation we randomise the initial conditions (as in the figures in the main text), the position of the optimum (uniformly in the real interval  [1, $n-1]$) and the learning network (using a random $p$). For each $n$ we perform 1000 simulations. $\lambda=0.01$.
\label{Fig:ERTopo}
\end{figure}

We first evaluate the effect of Erd\H{o}s-Renyi (ER) topologies. These are random networks built simply by placing random connections amongst two nodes with a probability $p$. Although we do not have any reason to believe that this ER network distribution is realised in neuronal circuits, it is the most basic pattern of random networks, providing a baseline of uncertainty.  Figure \ref{Fig:ERTopo}A shows the distribution of the equilibrium switching probabilities $M$ under ER distributions for different number of neuronal loci. The pattern is clear: most neurons spike in a random, unspecified way. This is reflecting the fact that the systems are unable to find solutions. Figure \ref{Fig:ERTopo}B shows the dependency of the switching probabilities against the degree of each neutron (numbers of synaptic connections with other neurons). Only as the  degree increases some neurons can be more specific. However, generally speaking the spread the spread of the values is high.

\begin{figure}
\begin{center}
\includegraphics[scale=0.35]{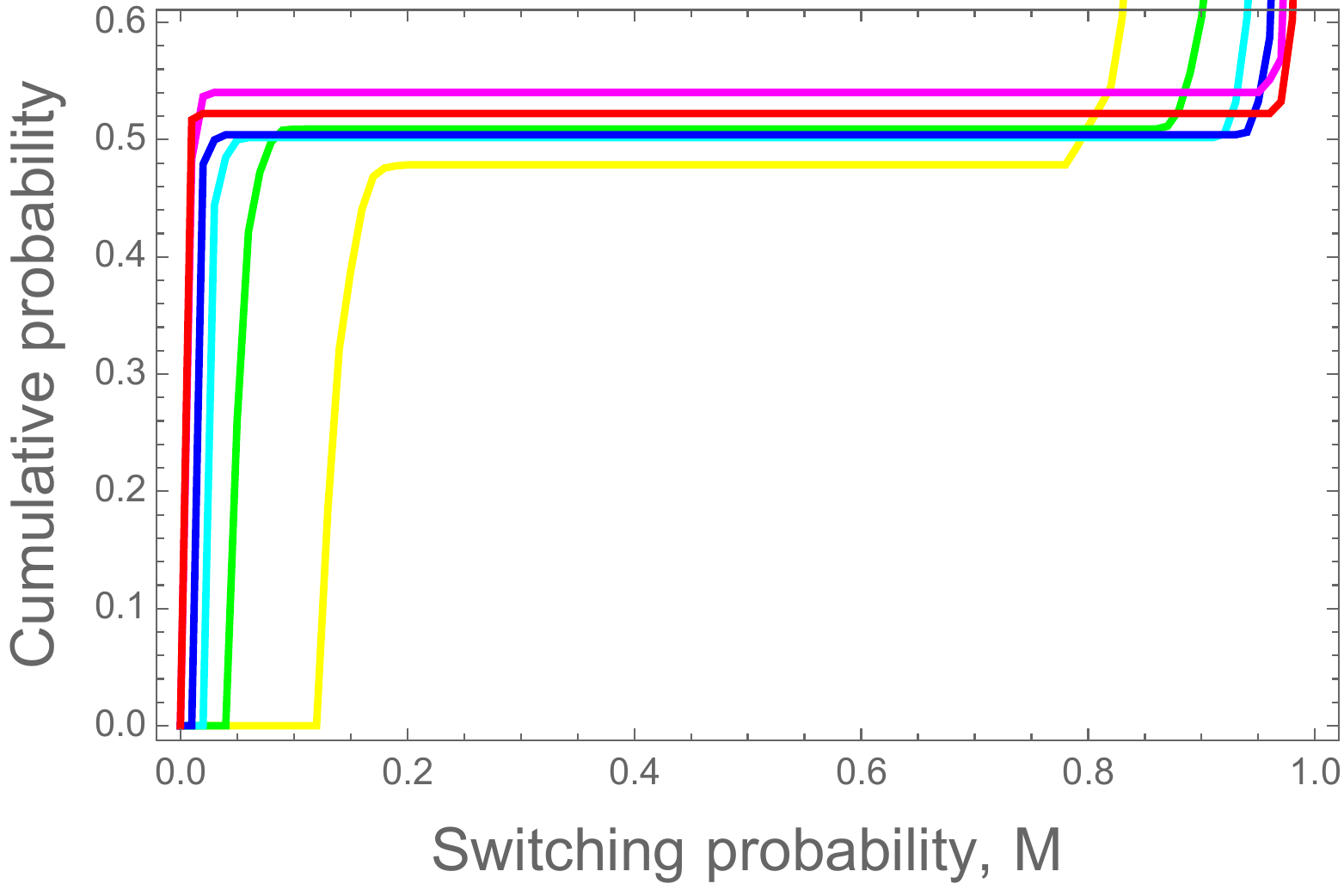}
\end{center}
\caption{\bf Distribution of switching probabilities for fully connected neuronal circuits.}
As in figure \ref{Fig:ERTopo}, but with using fully connected networks. $n$ as in the inset of Fig. \ref{Fig:ERTopo}A.
\label{Fig:CGTopo}
\end{figure}

We can compare these patterns with that of fully connected neuronal circuits (Fig. \ref{Fig:CGTopo}). In this case the switching probabilities are bimodal with peaks close to $M=0$ and M$=1$, indicating highly specific firing, as argued in the main text.

We also perform a similar analysis with Watts-Strogats (WS) `small world' topologies. These kinds of networks have become popularised through  the ``six degrees of separation'' motto. The key property of small world networks is that any node can be reached from any other by connecting through a small number of intermediary nodes. Unlike the ER networks which often have disconnected nodes, the WS networks  are very well connected. Some works have reported that neuronal networks have this kind of properties \cite{}.

\begin{figure}
\includegraphics[scale=0.35]{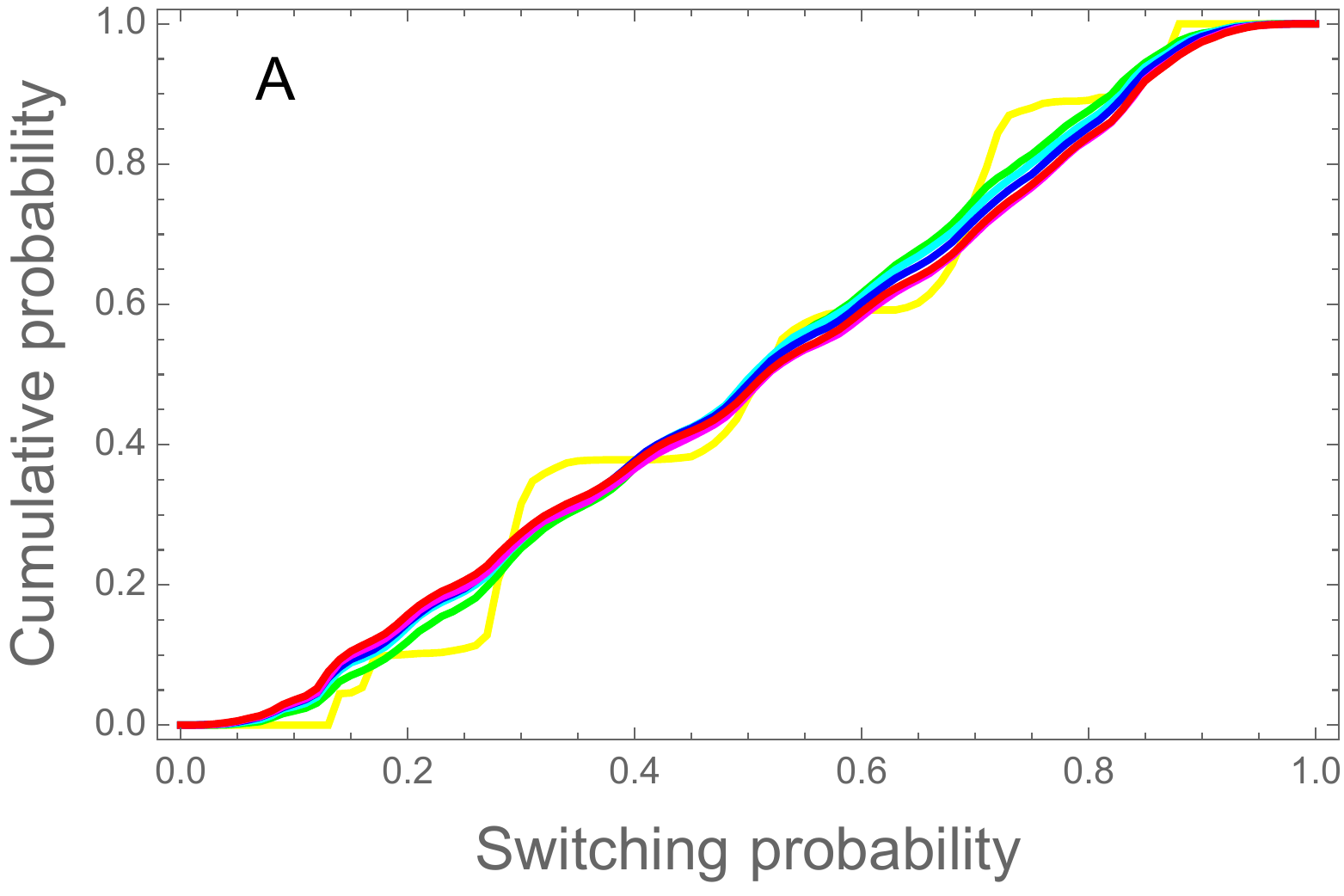}
\includegraphics[scale=0.35]{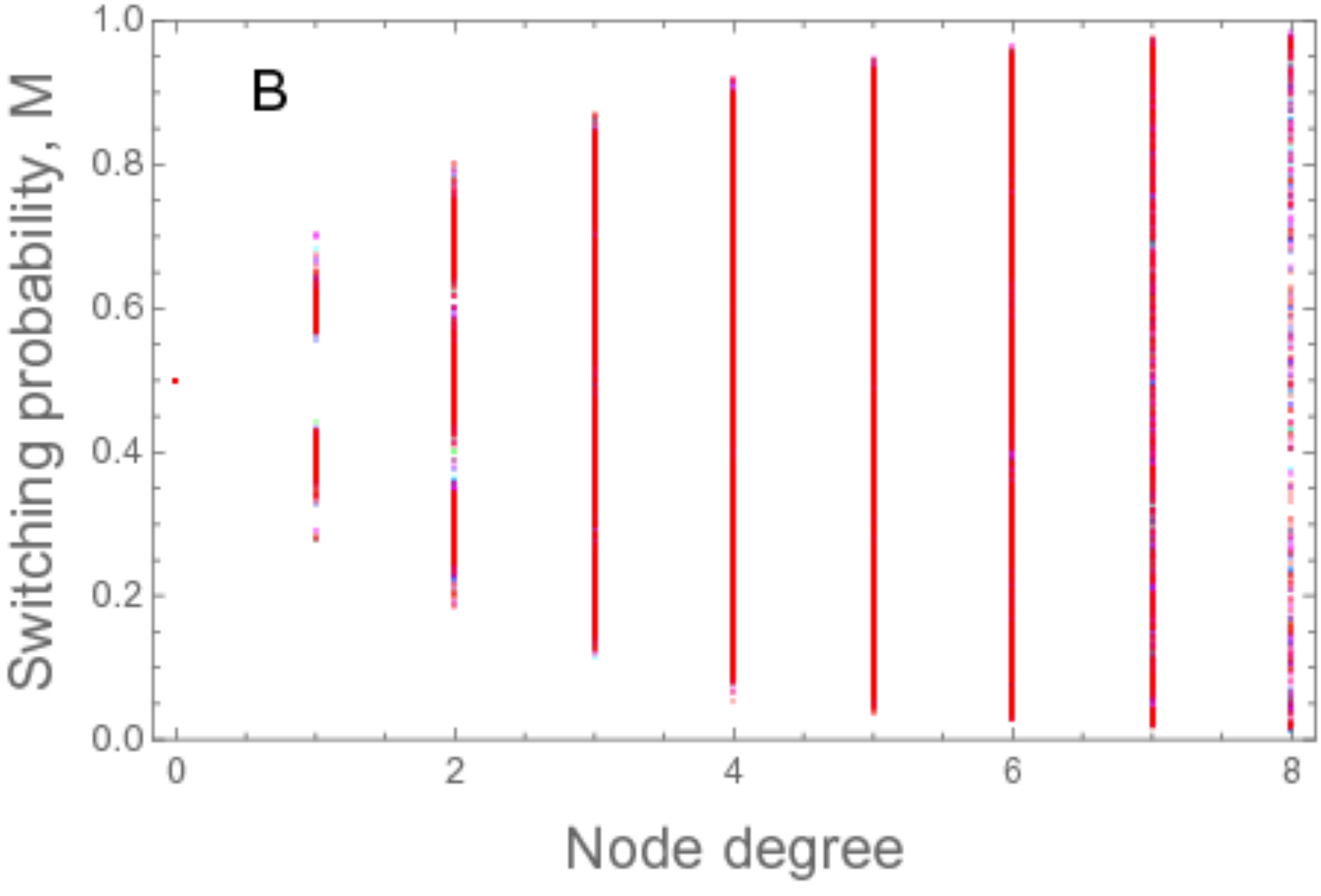}
\caption{\bf  Distribution of switching probabilities of neuronal circuits with Watts-Strogatz small world topologies.}
As in figure \ref{Fig:ERTopo}, but with Watts-Strogatz deviates with a random rewiring probability. $n$ as in the inset of Fig. \ref{Fig:ERTopo}A.
\label{Fig:WSTopo}
\end{figure}

Figure \ref{Fig:WSTopo}A shows that under this random network models, there is an almost uniform distribution of switching probabilities $M$, although there are almost no circuits with neurons that fire specifically. This occurs because although the circuits as a whole are well connected, each neuron has only very few connections (Fig. \ref{Fig:WSTopo}B), which in turn, results in low activities at most neurons.

\begin{figure}
\includegraphics[scale=0.35]{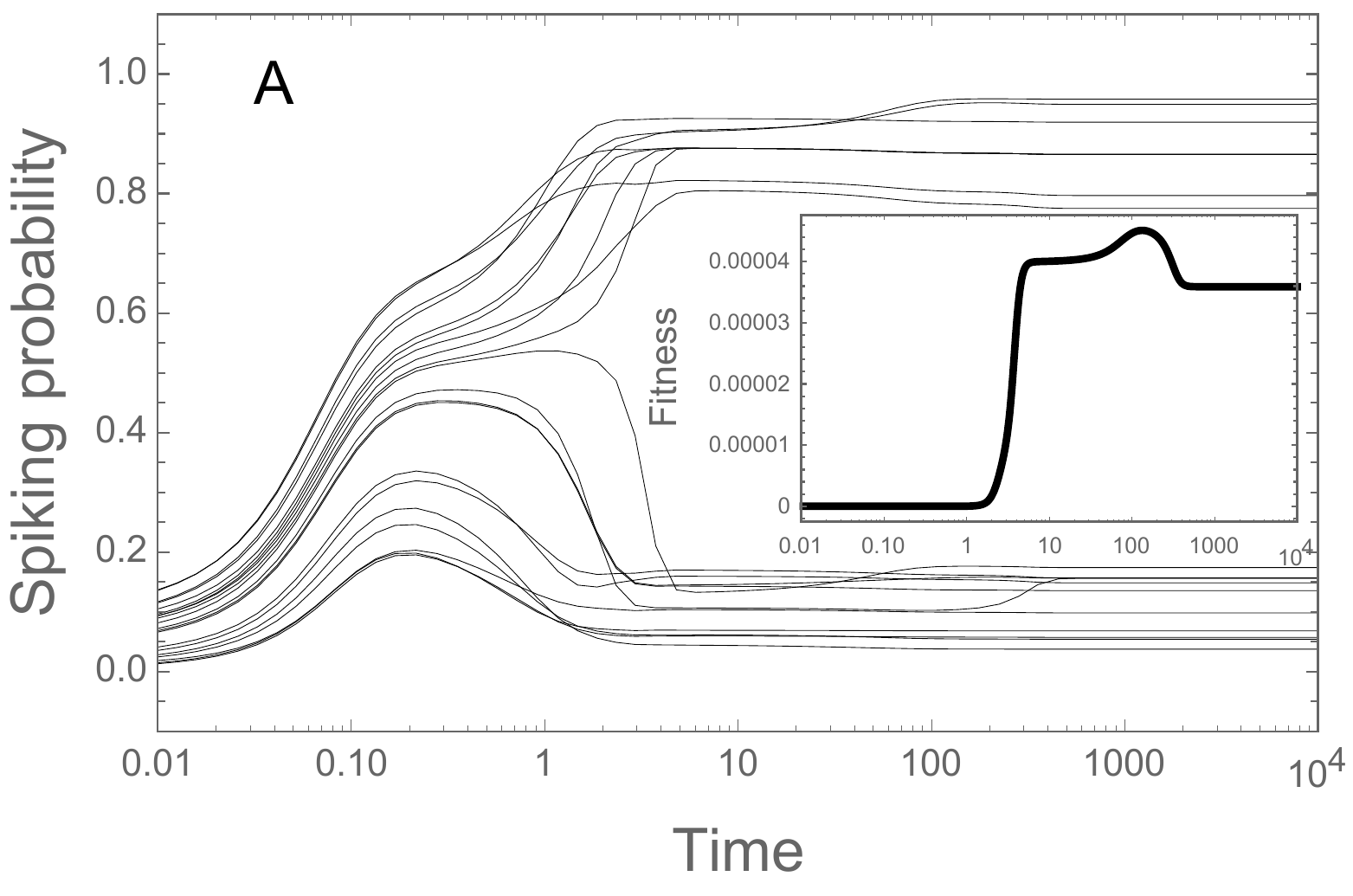}
\includegraphics[scale=0.35]{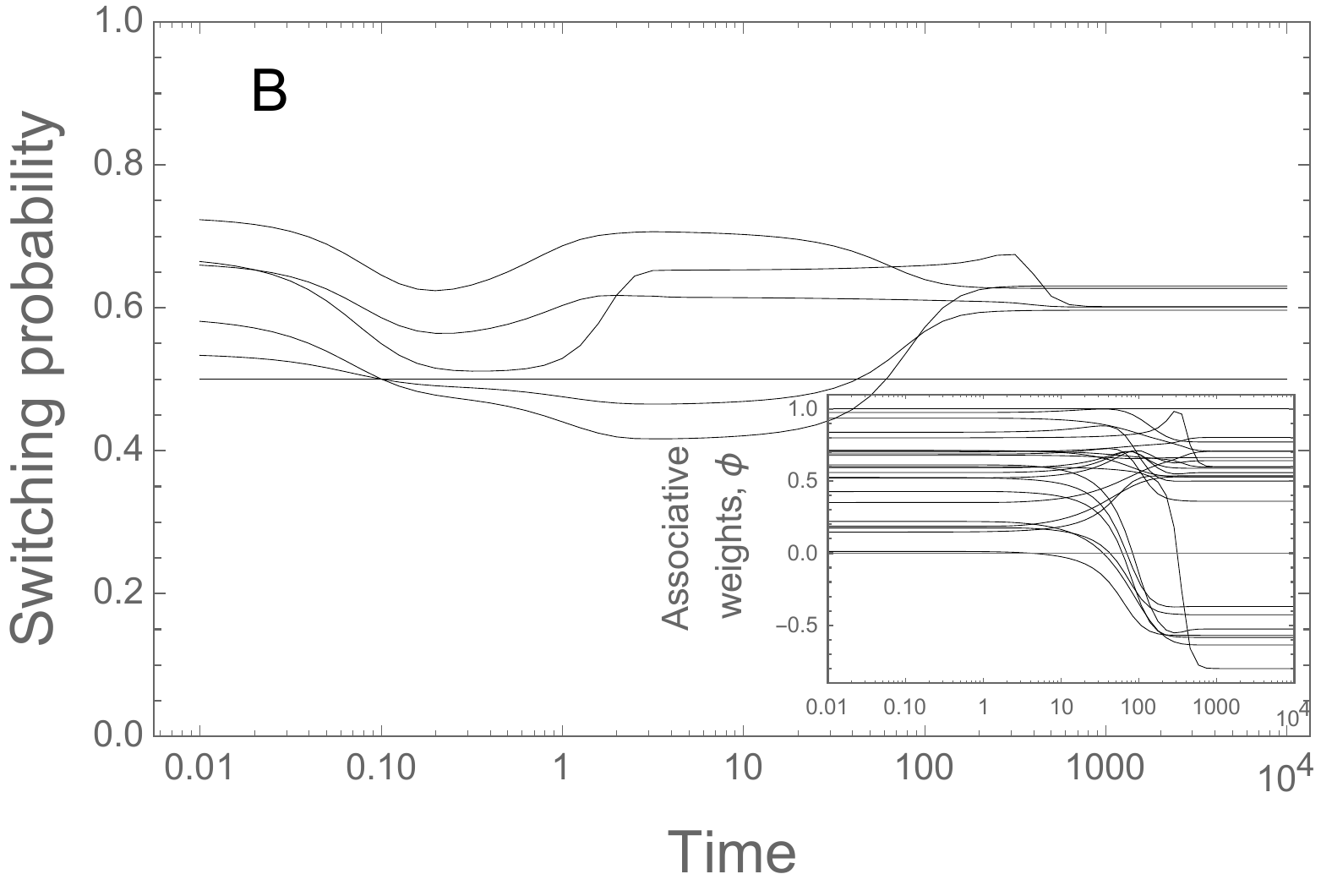}
\includegraphics[scale=0.35]{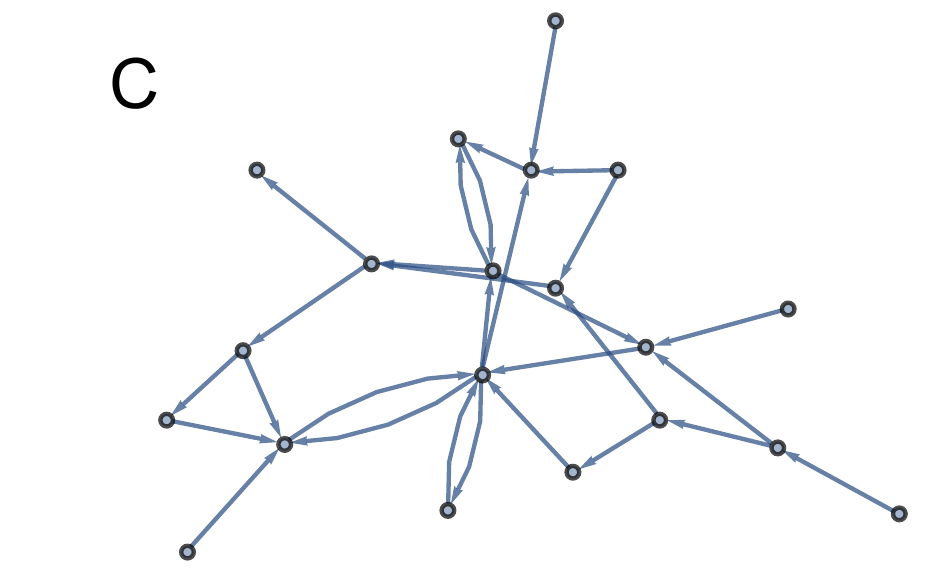}
\caption{\bf Neurodynamics under random (Erd\H{o}s-Renyi) circuit topologies.}
(A) Neurodynamics (inset: fitness) (B) Switching probability (inset: associative weights).  (C) Topology of the neuronal circuit. Parameters as in Figs. 2,4 of the main text.
\label{Fig:ERneurodyn}
\end{figure}

\begin{figure}
\includegraphics[scale=0.35]{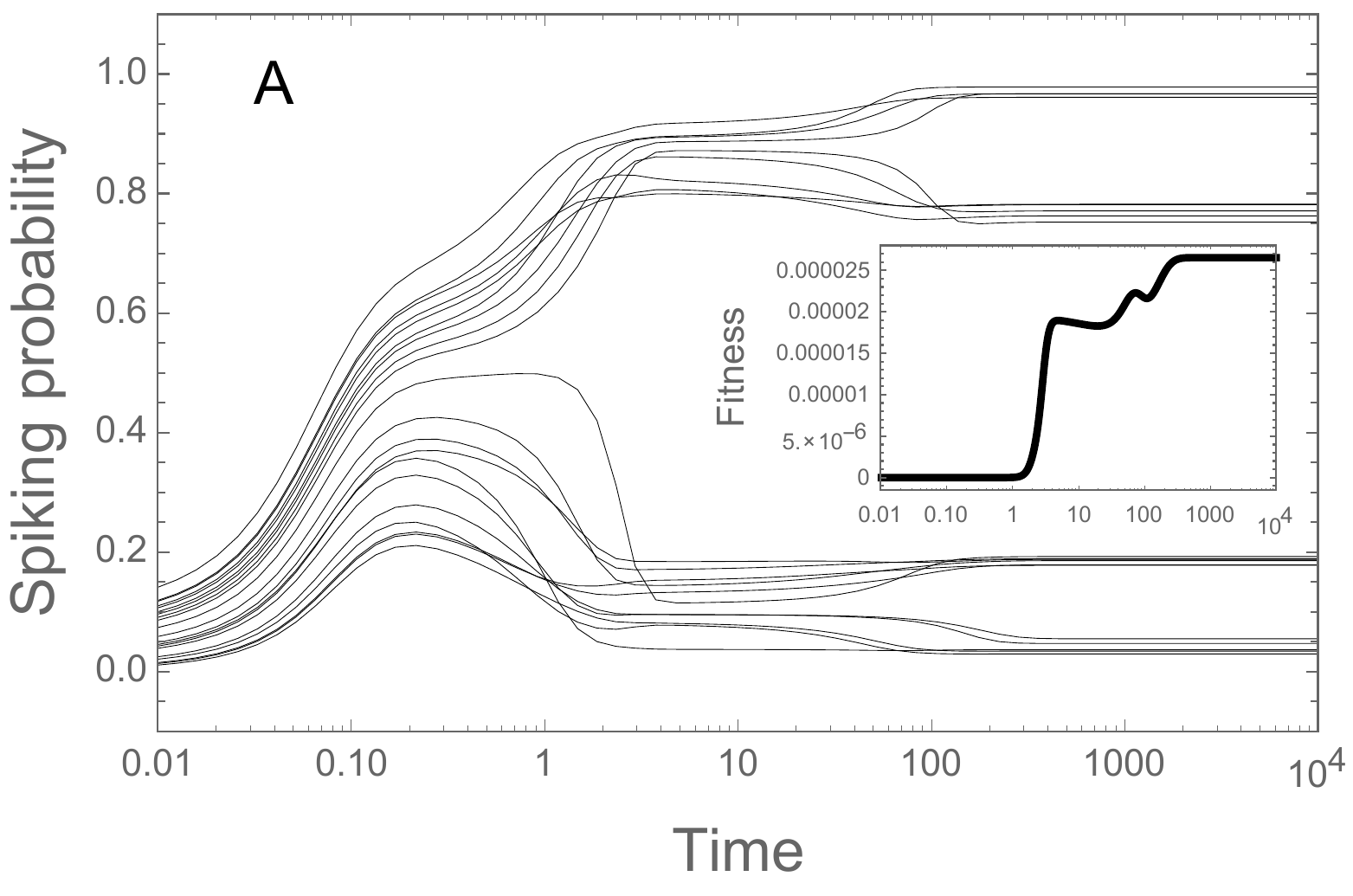}
\includegraphics[scale=0.35]{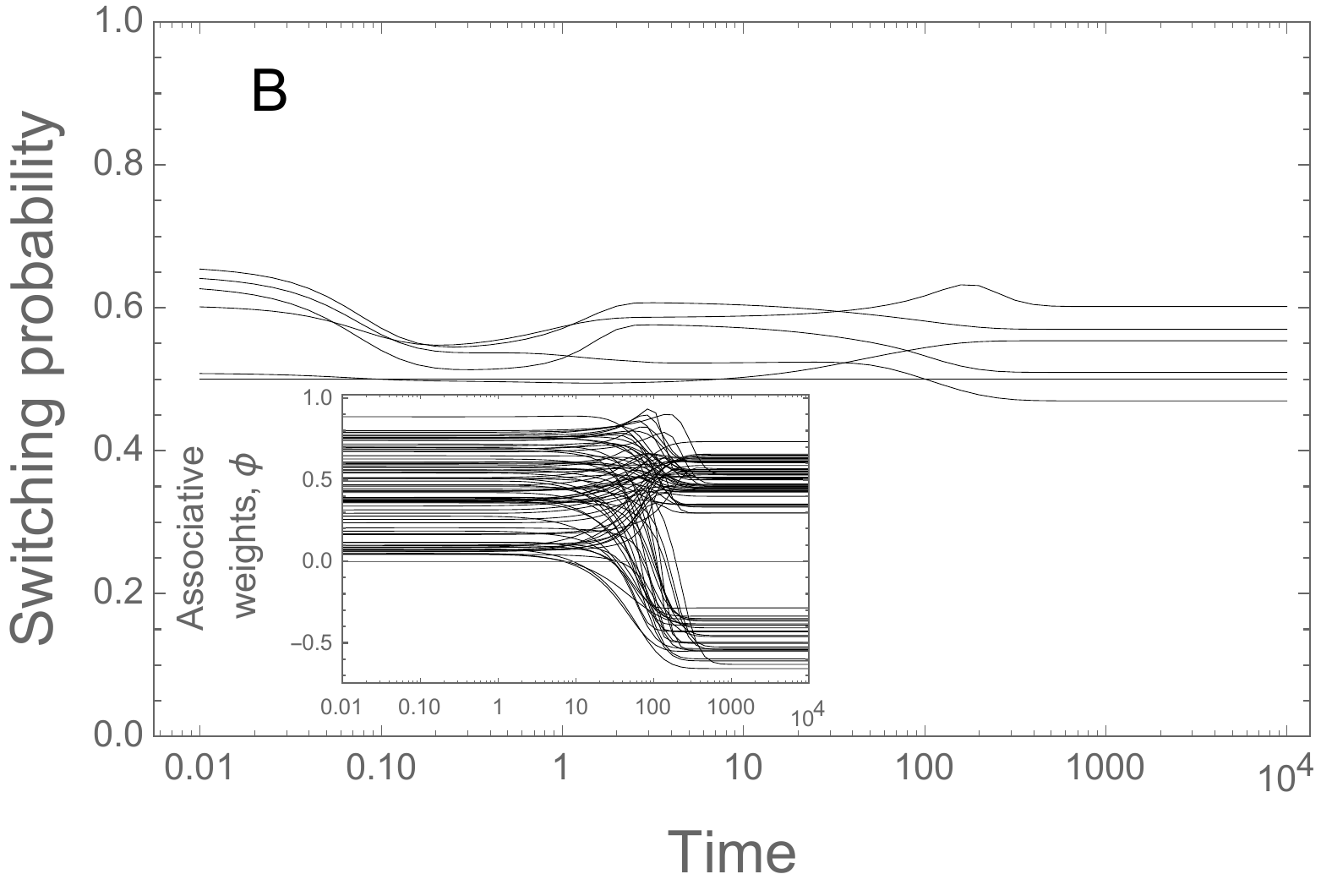}
\includegraphics[scale=0.35]{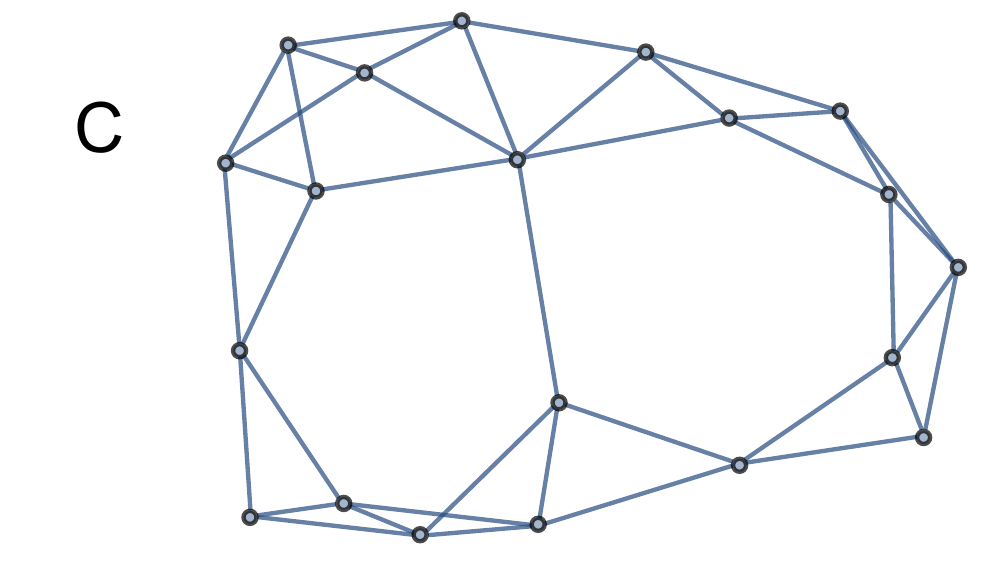}
\caption{\bf Neurodynamics under random (Watts-Strogatz) circuit topologies.}
(A) Neurodynamics (inset: fitness) (B) Switching probability (inset: associative weights).  (C) Topology of the neuronal circuit. Parameters as in Figs. 2,4 of the main text.
\label{Fig:WSneurodyn}
\end{figure}

Figures \ref{Fig:ERneurodyn}-\ref{Fig:WSneurodyn} and shows the neurodynamics assuming a random networks for the circuit topology. These examples show that in both cases the systems are unable to learn do to the constraints imposed  by the network topology. We see that most neurons reach a state where they do not fire specifically. We see this learning process that is unable to learn on an existing circuitry as a state of impasse as defined in cognitive psychology: any alternative that can be reached under the current configuration or representation leads to an even worse `solution'. Algorithmically, this corresponds to a local optimum on the fitness landscape.1

\section{Dynamics of structural plasticity under synaptic costs}
\label{SI:StructuralPlasticity}

\begin{figure}
\includegraphics[scale=0.35]{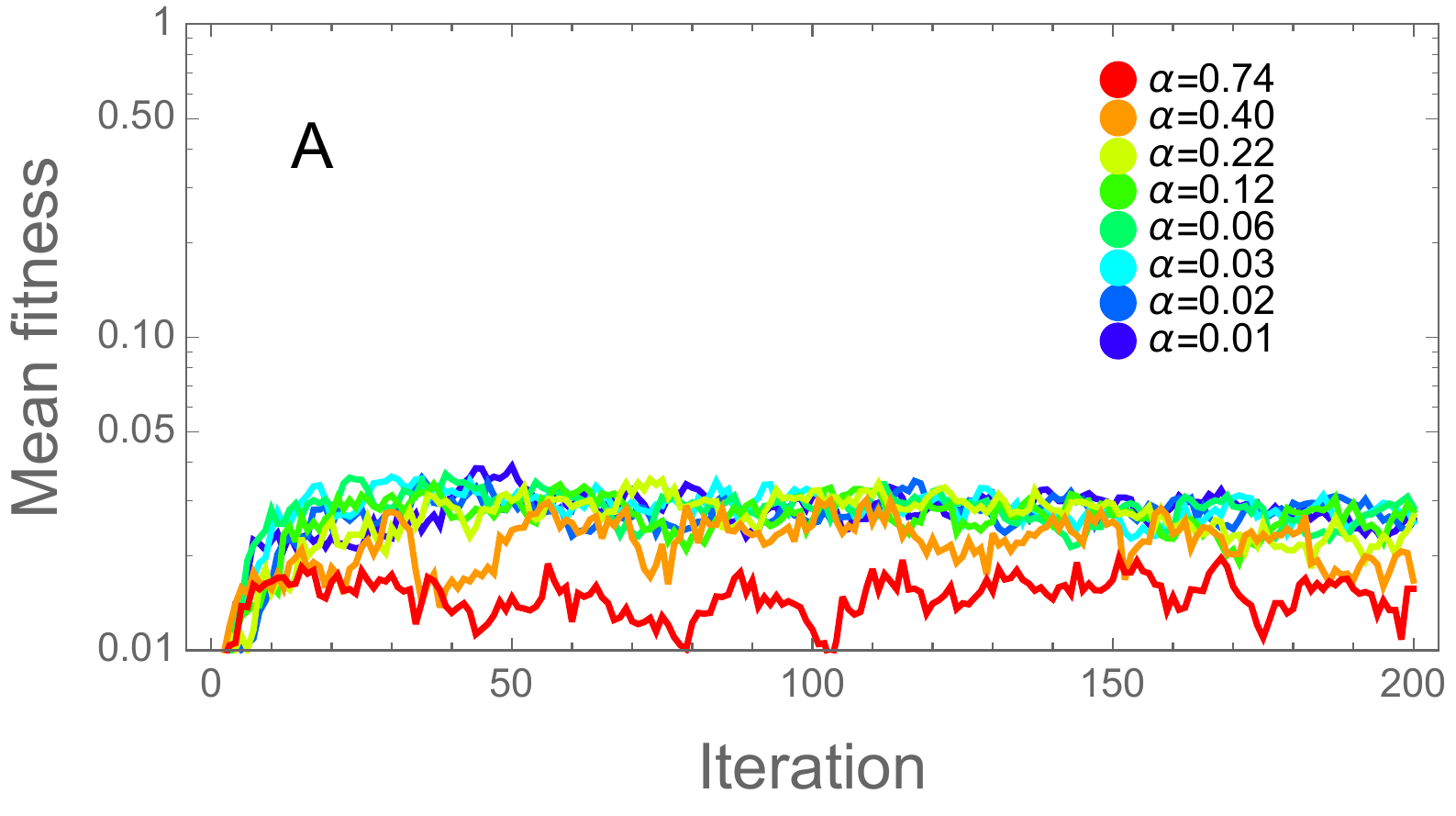}
\includegraphics[scale=0.35]{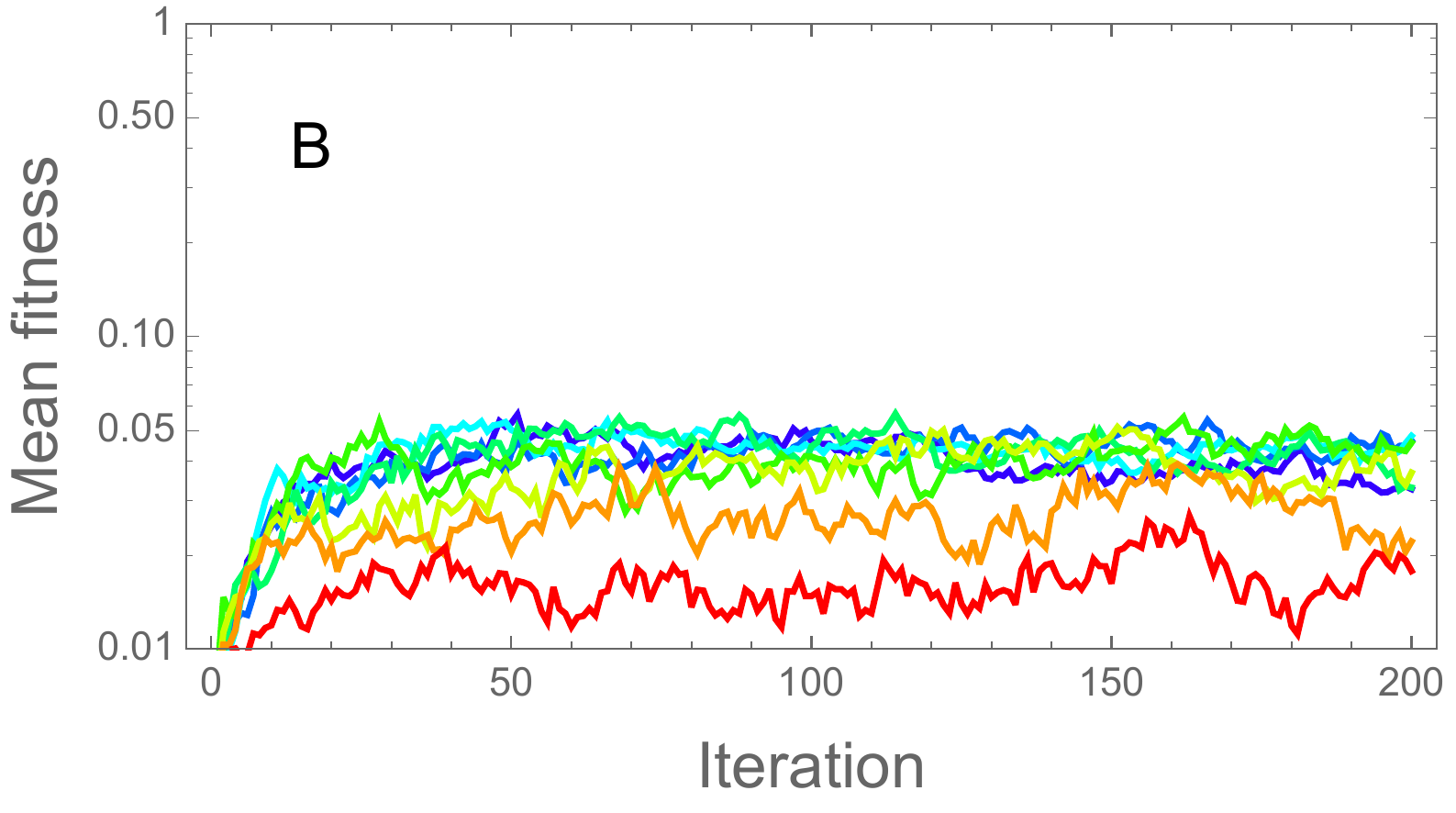}
\includegraphics[scale=0.35]{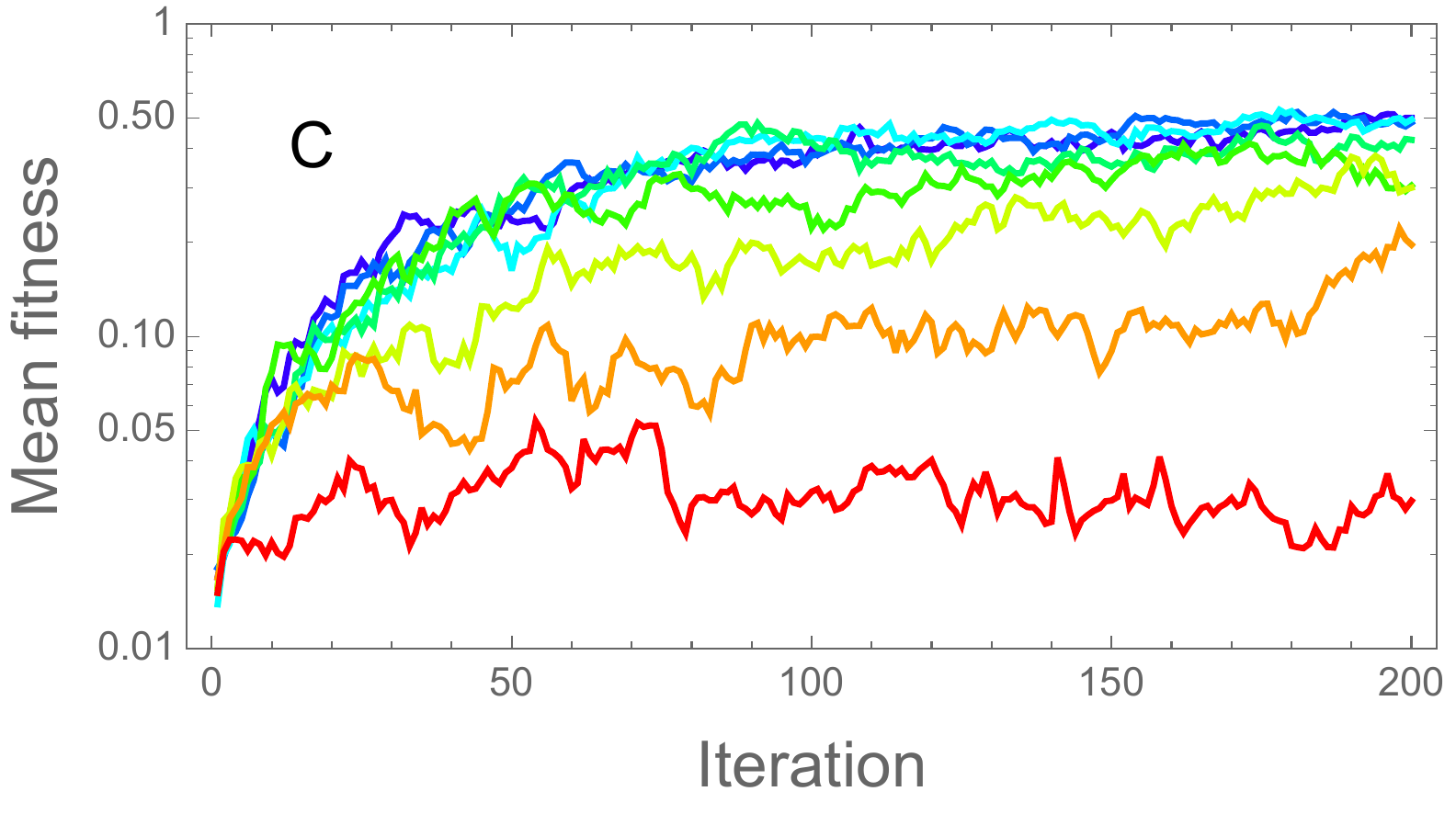}
\includegraphics[scale=0.35]{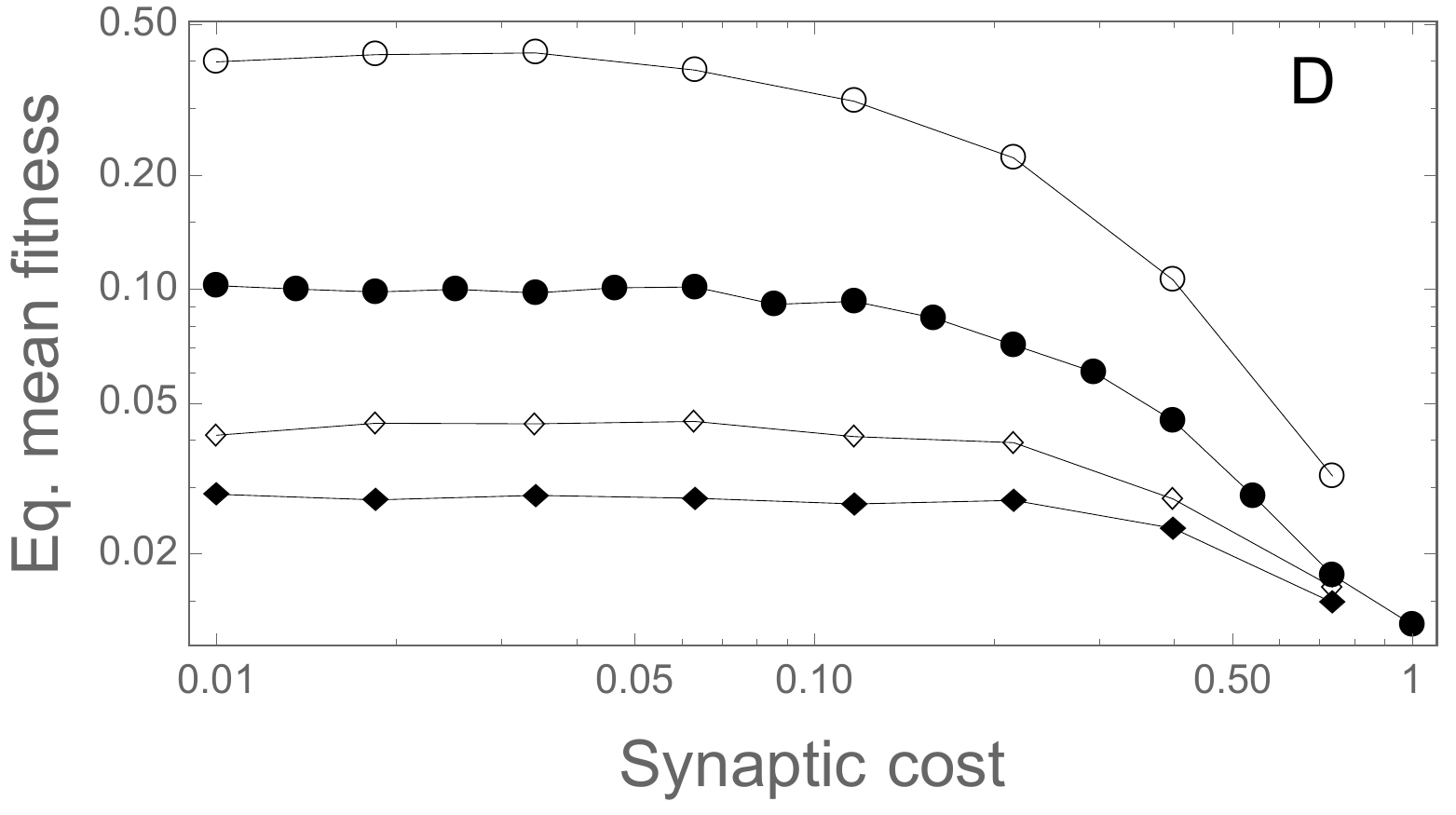}
\caption{\bf Neuronal systems under costly synapses}
(A-D) Dynamics of the fitness (relative to the maximum) of neuronal systems under different cost per connection (see legend in A) and different target values (A: 4; B: 5; C:10). Each curve is a replica of 16 independent simulations. (D) Equilibrium fitness as a function of the synaptic costs. Each point is an average at the stationary values (last 50 time points) and over 16 simulations (except for $T=7$ which is the data in Fig 7 int he main text). Filled diamonds $T=4$; open diamonds $T=5$; filled bullets $T=7$; open bullets $T=10$. Parameters as in Fig. 7 in the main text.
\label{Fig:CostlySSPdynamics}
\end{figure}

Figure \ref{Fig:CostlySSPdynamics} shows the evolution of fitness for neuronal systems targeting different values $T$. Although the mean fitness decreases with $T$, this has no consequences on the outcome of the system. This is because the success of neuronal group only depends on the relative advantage to other neuronal groups with different configuration. The relevant pattern is the decrease of performance as the synaptic costs increase.

\section{Distribution of the evolved networks}
\label{SI:EvolvedNetworksDist}

\begin{figure}
\includegraphics[width=\textwidth]{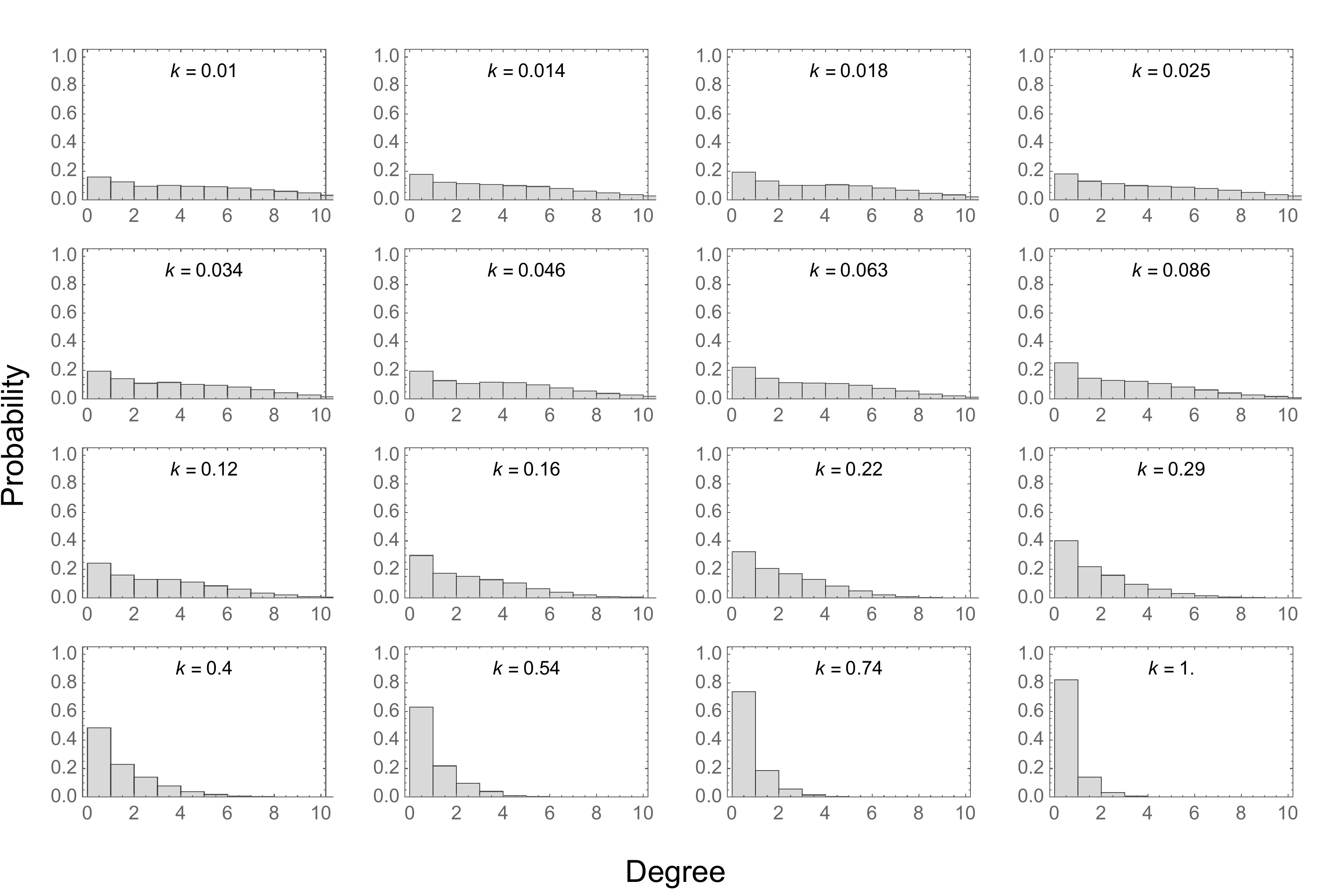}
\caption{\bf Degree distribution of the evolved neuronal circuits.}
\label{Fig:DegreeDist}
The data corresponds to the simulations of Fig. in the main text. The histograms are constructed by taking the last 50 points of 112 simulations and pooling the degrees of all networks. Since each circuit consists of $n=10$ neuronal loci, there are in total 55000 data points in each histogram.
\end{figure}

As reported in the main text, the  neuronal circuits become less connected as the cost per synapse $k$ increases. Figure \ref{Fig:DegreeDist} shows that the distribution of degrees (number of synapses per neuron) becomes more skewed as $k$ becomes larger.

\begin{figure}
\includegraphics[width=\textwidth]{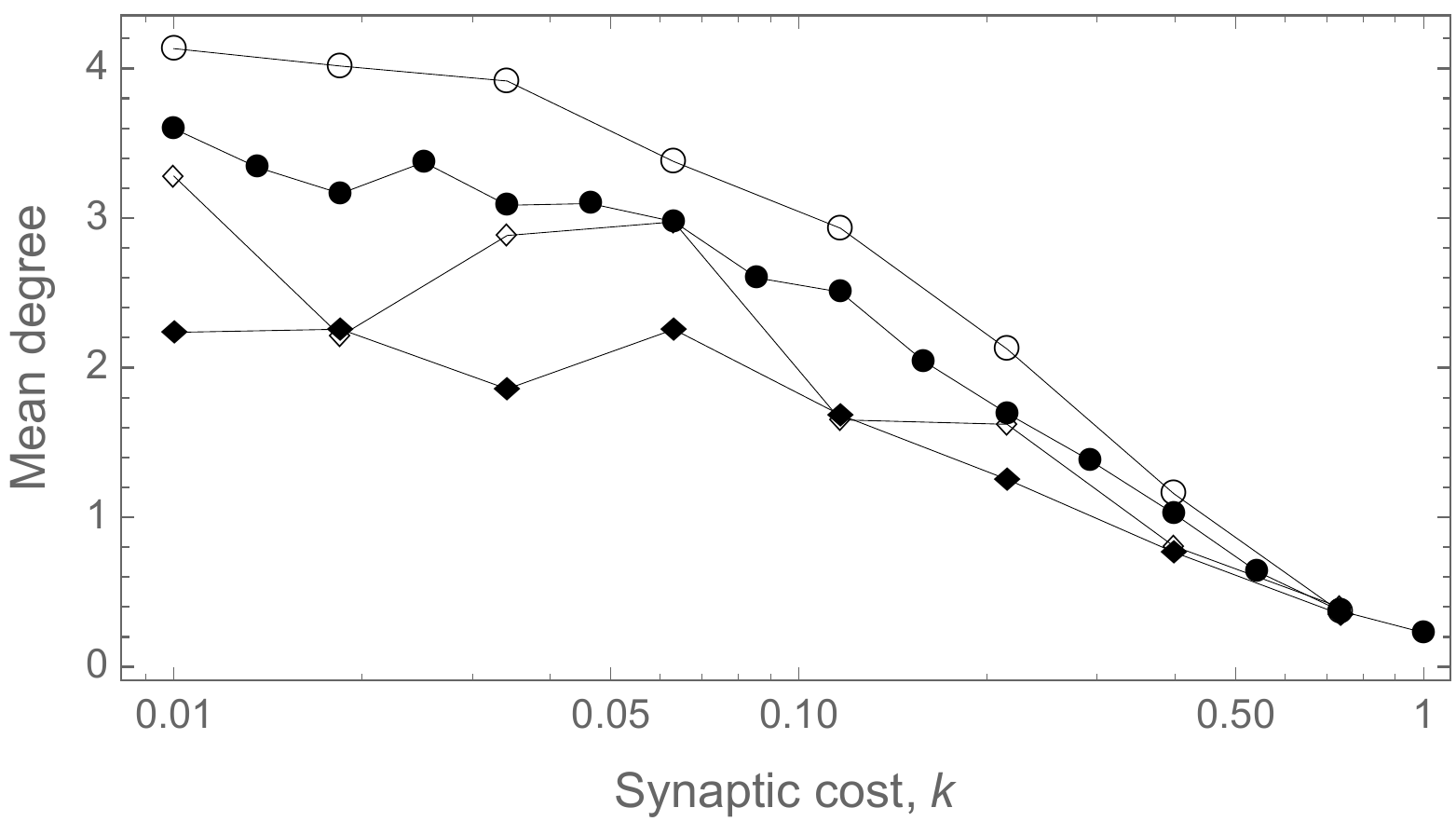}
\caption{\bf Mean degree of the evolved neuronal circuits for different targets and synaptic costs.}
\label{Fig:MeanDeg}
Each point is an average at the stationary values (last 50 time points) and over 16 simulations (except for $T=7$ which is the data in Fig. 7 int he main text). Filled diamonds $T=4$; open diamonds $T=5$; filled bullets $T=7$; open bullets $T=10$. Parameters as in Fig. 7 in the main text.
\end{figure}

Note that there are always zero classes because the target value $T$ determines that the optimal circuit consist of $T$ neurons that are fully connected (particularly with high in-degree) and $n-T$ that are either disconnected or have zero in-degree. Consequently, because this sets a lower bound for the number of unconnected neurons. The distribution of degrees is dependent on both the synaptic cost and the target values (Fig. \ref{Fig:MeanDeg}) .

We tried fitting several distributions: Erdos-Renyi, Watts-Strogatz, Poisson, Binomial, and scale free networks, and none fits well the empirical distributions.

\section{Distribution of synaptic lifetimes}
\label{SI:SynaptilLifetimes}

In this Appendix we present further results on the distribution of synaptic lifetimes. We ran extensive simulations with an alternative target values from that on the text, $T=4$. (Due to prohibitively long computation times we chose to sacrifice analysing more target values in order to focus on one value, but with sufficient statistical power.)

\begin{figure}
\includegraphics[scale=0.35]{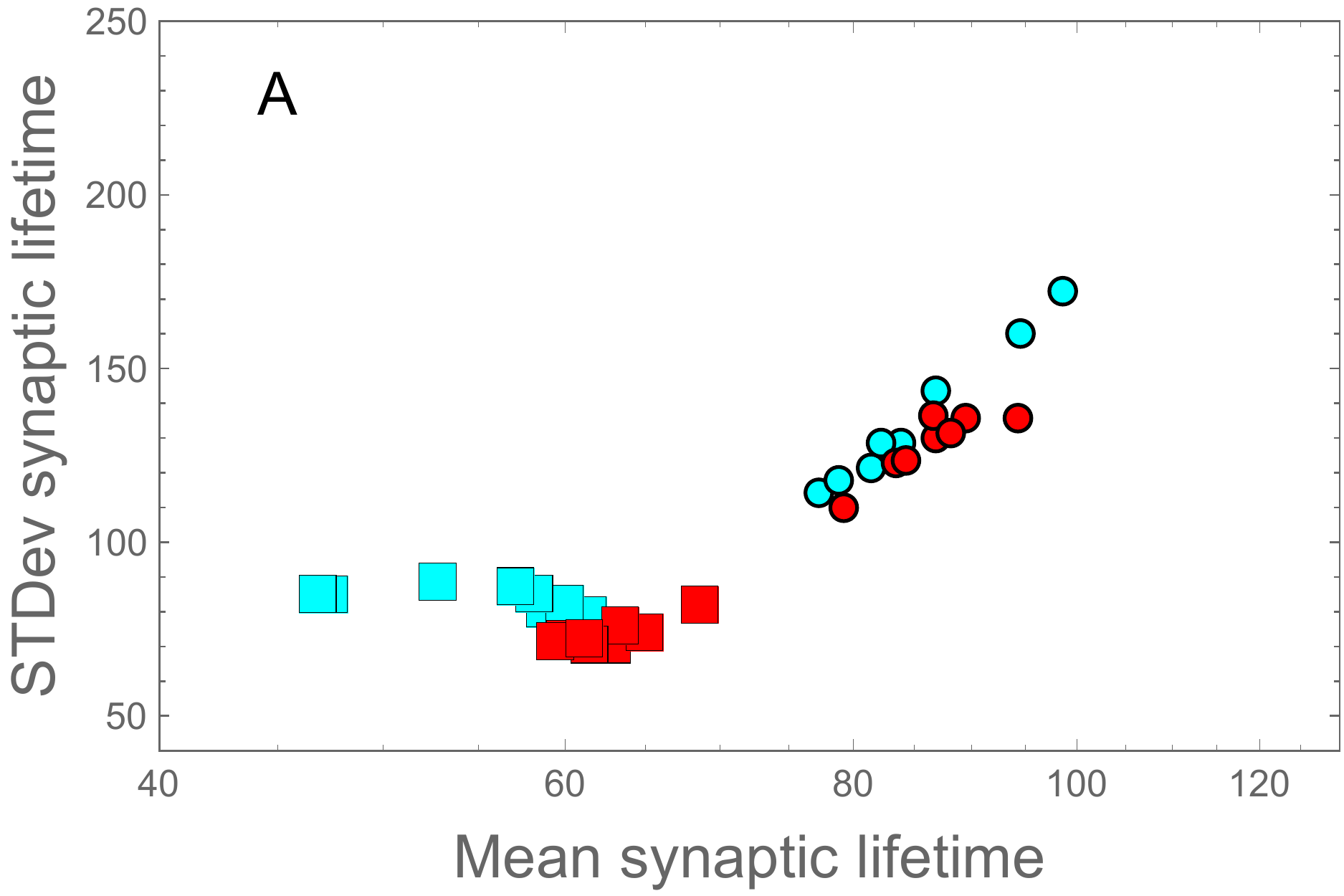}
\includegraphics[scale=0.35]{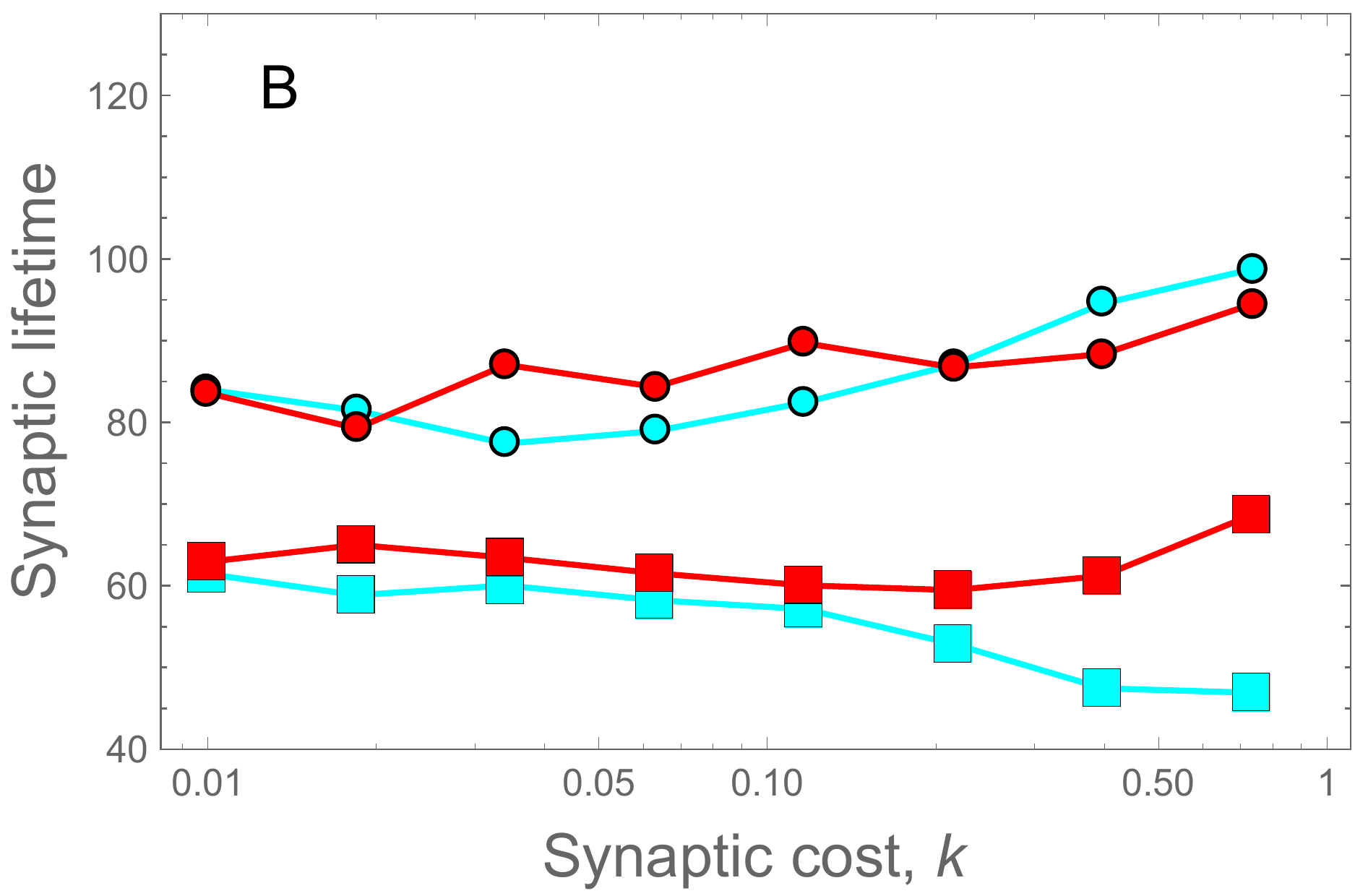}
\caption{\bf Synaptic lifetimes for different costs and targets}
(A) Relationship between the average and standard deviation of the synaptic lifetimes. (B)
In both panels: blue $T=4$, red $T=7$;  rectangles: synapses established during early, learning periods (before 200 iterations); bullets: synapses established in post-learning periods (at or after 200 iterations). The data is pooled from 14 independent runs each with of 1000 iterations. Other parameters as in Fig. 9 in the main text.
\label{Fig:MeanSynLifetimes}
\end{figure}

Figure \ref{Fig:MeanSynLifetimes} shows results for pooled samples of synaptic lifetimes of 14 simulations each lasting 1000 steps. We partition the synaptic lifetimes into  `early (learning) period', and `late (post-learning) period'. At the learning period, there seems to be no relationship between the average lifetime and the standard deviation of the synaptic lifetimes, whereas for the pos-learning stationary period there is a clear trend (Fig. \ref{Fig:MeanSynLifetimes}). This trend seems to be independent of the target values. The average synaptic lifetimes seem to be largely independent of the synaptic costs (Fig.  \ref{Fig:MeanSynLifetimes}B).

\begin{figure}
\includegraphics[width=\columnwidth]{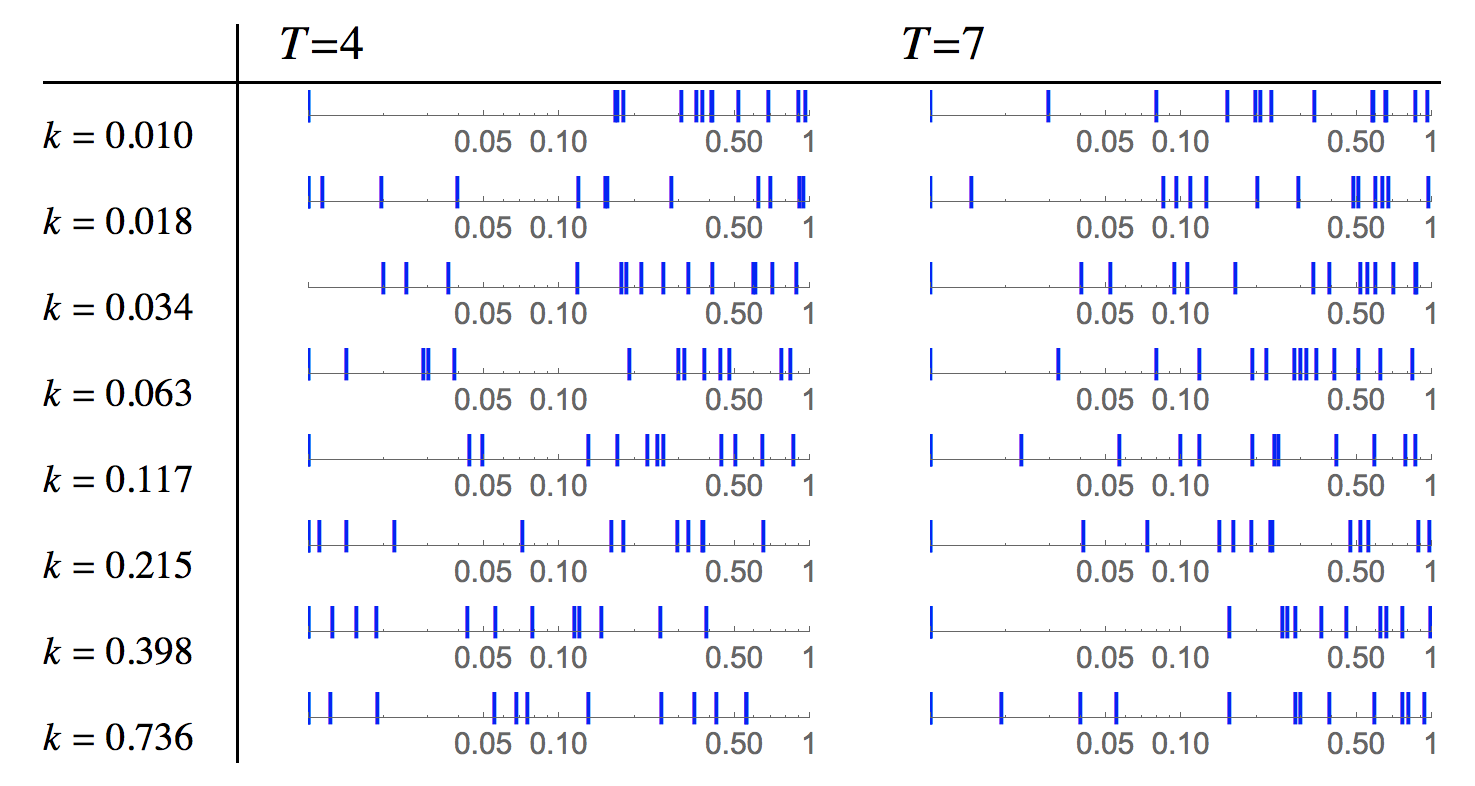}
\caption{\bf Significance of the difference of synaptic lifetimes during and after learning}
The plots show the $p-$values for Mann-Whitney tests (note the log-scale); all with confidence 0.05. Null hypothesis $H_0$: the mean of the distributions of synaptic lifetimes during learning and post-learning periods are different. Alternative hypothesis: $H_a$ the mean of the distributions are different. Marks at $p=0.01$  actually indicate that $p\leq0.01$.
The data corresponds to the simulations of the previous figure (and also Fig. 9 of the main text).
\label{Fig:PValues}
\end{figure}

Despite the fact that there is a  clear difference between the  synapses established early and those established late, Mann-Whitney tests on the each of the different simulations give $p-$values that are largely variable, and very often fail to reject the null hypothesis that the distributions of early and late synapses are distinct (Fig. \ref{Fig:PValues}). However, pooling the replicate simulations leads, in all cases to reject the null hypothesis (Table \ref{Table:Pvals}).

\begin{table}
\caption{
{\bf $p-$values of pooled simulations}}
\begin{tabular}{cll}
\hline
Cost \textbackslash Target	& $T=4$	& $T=7$ \\\hline
$k=0.010$	&	$1.\times 10^{-2} $& $4.\times 10^{-4}$ \\
$k=0.018$	&	$4.\times 10^{-5} $& $2.\times 10^{-2}$ \\
$k=0.034$	&	$4.\times 10^{-4} $& $3.\times 10^{-5}$ \\
$k=0.063$	&	$3.\times 10^{-5} $& $8.\times 10^{-5}$ \\
$k=0.117$	&	$3.\times 10^{-9} $& $5.\times 10^{-8}$ \\
$k=0.215$	&	$2.\times 10^{-11} $& $5.\times 10^{-6}$ \\
$k=0.398$	&	$2.\times 10^{-15} $& $1.\times 10^{-8}$ \\
$k=0.736$	&	$5.\times 10^{-12} $& $1.\times 10^{-6}$ \\\hline
\hline
\end{tabular}
\begin{flushleft} Test as in Fig. \ref{Fig:PValues} but with pooled samples.
\end{flushleft}
\label{Table:Pvals}
\end{table}

\end{document}